\documentclass[preprintnumbers, superscriptaddress, showpacs, nofootinbib, amsfonts, amsmath, amssymb, aps, prd, notitlepage, floatfix]{revtex4-2}

\usepackage{empheq}
\usepackage[colorlinks=true,citecolor=blue,linkcolor=blue,urlcolor=blue]{hyperref}
\usepackage[offset=1.5em]{simpler-wick}
\usepackage{xcolor, graphicx, slashed,  accents,  multirow,caption, subcaption, bm}
\usepackage[most]{tcolorbox}
\tcbset{
  top=-5pt,
}
\usepackage[normalem]{ulem}
\usepackage{simplewick}

\usepackage[export]{adjustbox}

\usepackage{cancel}

\usepackage{ragged2e} 
\DeclareCaptionJustification{justified}{\justifying}

\allowdisplaybreaks

\newcommand{\as}{\alpha_s}

\newcommand{\xo}{x_{10}}

\newcommand{\xti}{x_{21}}
\newcommand{\dprime}{\prime \prime}

\def\eq#1{{Eq.~(\ref{#1})}}
\def\eqs#1{{Eqs.~(\ref{#1})}}
\def\fig#1{{Fig.~\ref{#1}}}

\newcommand{\ben}{\begin{eqnarray*}}
\newcommand{\een}{\end{eqnarray*}}
\newcommand{\un}[1]{\bm{#1}}

\newcommand{\pd}{\partial}

\newcommand{\tr}{\mbox{tr}}
\newcommand{\thalf}{\tfrac{1}{2}}
\newcommand{\llangle}{\Big\langle \!\! \Big\langle}
\newcommand{\rrangle}{\Big\rangle \!\! \Big\rangle}
\newcommand{\cc}{\mbox{c.c.}}

\newcommand{\dhd}{{\textstyle d}
\lower.03ex\hbox{\kern-0.38em$^{\scriptstyle-}$}\kern-0.05em{}}
\newcommand{\dbar}{{\textstyle \delta}
\lower.03ex\hbox{\kern-0.38em$^{\scriptstyle-}$}\kern-0.05em{}}
\newcommand{\half}{{1\over 2}}

\newcommand{\bra}[1]{\left\langle #1 \right|}
\newcommand{\ket}[1]{\left| #1 \right\rangle}

\newcommand{\ul}[1]{\bm{#1}}

\newcommand{\ord}[1]{\mathcal{O}\left( #1 \right)}

\newcommand{\tord}{\textrm{T} \:}
\newcommand{\atord}{\bar{\textrm{T}}\:}

\makeatletter
\DeclareRobustCommand{\cev}[1]{%
  {\mathpalette\do@cev{#1}}%
}
\newcommand{\do@cev}[2]{%
  \vbox{\offinterlineskip
    \sbox\z@{$\m@th#1 x$}%
    \ialign{##\cr
      \hidewidth\reflectbox{$\m@th#1\vec{}\mkern4mu$}\hidewidth\cr
      \noalign{\kern-\ht\z@}
      $\m@th#1#2$\cr
    }%
  }%
}
\makeatother

\newcommand{\iUV}{
    \int \displaylimits^z_{\frac{1}{s \xo^2}} \frac{dz^\prime }{z^\prime} \int \displaylimits^{\xo^2}_{\frac{1}{z^\prime s}} \frac{d\xti^2}{\xti^2}
}
\newcommand{\iIR}{\int \displaylimits^{z}_{\frac{\Lambda^2}{s}} \frac{dz^\prime}{z^\prime} \int \displaylimits^{\mathrm{min}[\frac{z}{z^\prime} \xo^2, \frac{1}{\Lambda^2}]}_{\mathrm{max}[\xo^2, \frac{1}{z^{\prime}s}]} \frac{d \xti^2}{\xti^2}}

\newcommand{\inUV}{\int \displaylimits^{z^\prime}_{\frac{1}{sx_{10}^2}} \frac{dz^{\dprime}}{z^{\dprime}} \int \displaylimits^{\mathrm{min}[\xo^2, \xti^2 \frac{z^\prime}{z^{\dprime}}]}_{\frac{1}{z^{\dprime} s}} \frac{d x_{32}^2}{x_{32}^2}}

\newcommand{\inIR}{\int \displaylimits^{z^\prime \frac{\xti^2}{\xo^2}}_{\frac{\Lambda^2}{s}} \frac{dz^{\dprime}}{z^{\dprime}} 
    \int \displaylimits^{\mathrm{min}[\frac{z^\prime}{z^{\dprime}} \xti^2, \frac{1}{\Lambda^2}]}_{\mathrm{max}[\xo^2, \frac{1}{z^{\dprime}s}]} \frac{d x_{32}^2}{x_{32}^2} }

\newcommand{\tI}{\widetilde{I}}
\newcommand{\tG}{\widetilde{G}}
\newcommand{\tGm}{\widetilde{\Gamma}}

\newcommand{\Sec}[1]{Sec.~\ref{#1}}

\begin{document}

\title{Orbital angular momentum at small $x$ in the large $N_c\&N_f$ limit}

    \makeatletter  
\def\@fnsymbol#1{\ensuremath{\ifcase#1\or *\or \dagger\or \ddagger\or
   \mathsection\or \mathparagraph\or \|\or **\or \dagger\dagger
   \or \ddagger\ddagger \or \mathsection\mathsection \else\@ctrerr\fi}}
    \makeatother

\author{G. Zardo Becker}
\email[Email: ]{zardobecker@gmail.com}
\affiliation{Physics Institute, Federal University of Santa Catarina, Florianópolis, SC 88049-900, Brazil}

\author{Yuri~V.~Kovchegov} 
\email[Email: ]{kovchegov.1@osu.edu}
\affiliation{Department of Physics, The Ohio State University, Columbus, Ohio 43210, USA}

\author{Ming~Li} 
\email[Email: ]{ming.li@hamptonu.edu}
\affiliation{Department of Physics, Hampton University, Hampton, VA 23668, USA}

\author{Brandon Manley}
\email[Email: ]{brandon.manley@uconn.edu}
\affiliation{Department of Physics, The Ohio State University, Columbus, Ohio 43210, USA}
\affiliation{Department of Physics, University of Connecticut, Storrs, Connecticut 06269, USA}
         
\author{Andrey~Tarasov}
\email[Email: ]{ataraso@ncsu.edu}
\affiliation{Department of Physics and Astronomy, North Carolina State University, Raleigh, North Carolina 27695, USA}
\affiliation{Joint BNL-SBU Center for Frontiers in Nuclear Science (CFNS) at Stony Brook University, Stony Brook, New York 11794, USA}

\begin{abstract}
We extend the small-$x$ analysis of the quark and gluon orbital angular momentum (OAM) distributions in the proton from the large-$N_c$ limit considered in our earlier works to the large-$N_c\&N_f$ limit, in which the numbers of quark colors $N_c$ and flavors $N_f$ are large with their ratio held fixed. Working in the double-logarithmic approximation (DLA), summing powers of $\alpha_s \ln^2(1/x)$ with $\alpha_s$ the strong coupling and $x$ the proton's momentum fraction carried by a parton, we correct the small-$x$ operator expression for the quark OAM distribution suggested earlier in \cite{Kovchegov:2024wjs} and relate both the quark and gluon OAM distributions to the impact-parameter moments of the polarized dipole amplitudes. We derive the large-$N_c\&N_f$ evolution equations for these moment amplitudes in the DLA; these include a new equation for the moment amplitude governing the quark OAM. We then solve these evolution equations numerically together with the helicity evolution for $N_f = 2,3,4,5,6$ and $N_c =3$. We find that, similar to the large-$N_c$ case, the OAM distributions share a common small-$x$ intercept with the helicity distributions, 
\begin{align}
    \notag
     L_{q+\bar{q}}(x,Q^2) \sim L_G(x,Q^2) \sim \Delta \Sigma (x,Q^2) \sim \Delta G(x,Q^2) \sim \left(\frac{1}{x} \right)^{\alpha_h},
 \end{align}
with the intercept $\alpha_h \approx {3.48 \sqrt{\alpha_s N_c/2\pi}}$ for $N_f = N_c = 3$ (cf.~\cite{Borden:2025ehe}): this result, along with $\alpha_h$ for other values of $N_f \neq 0$ that we studied, is smaller than the  intercept of $3.66\sqrt{\alpha_s N_c/2\pi}$ found in the large-$N_c$ limit. We also compute the ratios of the OAM distributions to the helicity parton distribution functions as $x\to 0$, obtaining $L_{q+\bar{q}}(x,Q^2)/\Delta\Sigma(x,Q^2) = -1$ and $L_G(x,Q^2)/\Delta G(x,Q^2) =  -2$ at $Q^2=10\, \mathrm{GeV}^2$, with both ratios being independent of $N_f$. 

We also revisit the elastic dijet production cross section for longitudinally polarized electron-proton scattering at small $x$, previously computed in \cite{Kovchegov:2024wjs}. We calculate the contributions coming from the $F^{+-}$ sub-eikonal operator that were previously omitted and show that they generate non-vanishing corrections to the azimuthal harmonics of the longitudinal double-spin asymmetry. 

This paper is the first one in a two-paper sequence: the second paper will employ the formalism developed here to perform a feasibility study of measuring quark and gluon OAM distributions at the future Electron--Ion Collider (EIC). 
\end{abstract}
\maketitle

\tableofcontents

\section{Introduction}\label{sec:intro}

The eventual solution of the proton spin puzzle \cite{EuropeanMuon:1987isl, Jaffe:1989jz, Ji:1996ek, Boer:2011fh, Aidala:2012mv, Accardi:2012qut, Leader:2013jra, Aschenauer:2013woa, Aschenauer:2015eha,  Proceedings:2020eah, Ji:2020ena, AbdulKhalek:2021gbh, Hatta:2026txn} must entail a detailed quantitative understanding of the orbital angular momentum (OAM) distributions of quarks and gluons, $L_{q+\bar{q}}(x,Q^2)$ and $L_G(x,Q^2)$, in the proton. Indeed, the proton spin of 1/2 (in units of $\hbar$) receives contributions from both the quark and gluon spin and their orbital angular momenta. This is usually manifested by the Jaffe-Manohar sum rule \cite{Jaffe:1989jz} (see also the Ji sum rule \cite{Ji:1996ek}), 
\begin{equation}
S_{q} + L_{q + \bar q} + S_G + L_G = \frac{1}{2}\, ,
\label{eqn:JM}
\end{equation}
where $S_q$ and $S_G$ are the quark and gluon spin contributions, respectively, 
\begin{align}\label{spins}
    S_q(Q^2) = \frac{1}{2}\int\limits_0^1\mathrm{d}x \, \Delta\Sigma(x,Q^2)\,, \quad\quad S_G(Q^2) = \int\limits_0^1 \mathrm{d}x \, \Delta G(x,Q^2) \,,
\end{align}
while $L_{q + \bar q}$ and $L_G$ are the OAM contributions, given by \cite{Bashinsky:1998if, Hagler:1998kg, Harindranath:1998ve, Hatta:2012cs, Ji:2012ba}
\begin{align}\label{eqn:LqLG}
L_{q + \bar q}(Q^2) = \sum_f \int\limits_0^1 dx \, L_{q_f + \bar q_f} (x,Q^2)\,, 
\qquad\qquad
L_G (Q^2) = \int\limits_0^1 dx \, L_G (x,Q^2)\,.
\end{align}
Here $x$ is the proton momentum fraction carried by the parton in question, while $Q$ is the momentum scale. $L_{q_f + \bar q_f} (x,Q^2)$ denotes the OAM distribution for the quarks of flavor $f$. The gluon helicity parton distribution function (hPDF) is denoted by $\Delta G(x,Q^2)$, while the quark flavor-singlet hPDF is 
\begin{align}\label{DeltaSigmageneral}
    \Delta \Sigma(x,Q^2) = \sum_{f} \left[\Delta q_f(x,Q^2) + \Delta\overline{q}_f(x,Q^2) \right]\,,
\end{align}
where $\Delta q_f(x,Q^2)$ and $\Delta \overline{q}_f(x,Q^2)$ are the hPDFs for the quarks and antiquarks of flavor $f$. 

Equation~\eqref{eqn:JM} clearly demonstrates that solution of the proton spin puzzle is not possible without understanding, qualitatively and quantitatively, the OAM distributions. At the same time, we know relatively little about the OAM distributions. This is in contrast to the hPDFs, which are much better known, both theoretically and phenomenologically. The $Q^2$ evolution of hPDFs is given by the spin-dependent Dokshitzer-Gribov-Lipatov-Altarelli-Parisi (DGLAP) evolution equations~\cite{Gribov:1972ri, Altarelli:1977zs, Dokshitzer:1977sg} with the splitting functions currently known to three loops~\cite{Altarelli:1977zs, Dokshitzer:1977sg, Zijlstra:1993sh, Mertig:1995ny, Moch:1999eb, vanNeerven:2000uj, Vermaseren:2005qc, Moch:2014sna, Blumlein:2021ryt, Blumlein:2021lmf, Davies:2022ofz, Blumlein:2022gpp}. This, along with existence of several experimental observables coupling to $\Delta q_f, \Delta \bar q_f$, and $\Delta G$, allowed for multiple hPDFs extractions from the experimental data over the years \cite{Gluck:2000dy, Leader:2005ci, deFlorian:2009vb, Leader:2010rb, Jimenez-Delgado:2013boa, Ball:2013lla, Nocera:2014gqa, deFlorian:2014yva, Leader:2014uua, Sato:2016tuz, Ethier:2017zbq, DeFlorian:2019xxt, Aschenauer:2019kzf, Borsa:2020lsz, Zhou:2022wzm, Cocuzza:2022jye, Borsa:2023tqr, Karpie:2023nyg, Borsa:2024mss, Bertone:2024taw, Hunt-Smith:2024khs, Cruz-Martinez:2025ahf, Cocuzza:2025qvf}. Lattice quantum chromodynamics (QCD) calculations have also made progress in determination of hPDFs~\cite{Cichy:2018mum, Karpie:2023nyg, Alexandrou:2021oih, Alexandrou:2020uyt, Delmar:2023agv, HadStruc:2022nay, Good:2025daz, Gao:2026wlz}. (More on hPDFs can be found in the reviews \cite{EuropeanMuon:1987isl, Lampe:1998eu, Boer:2011fh, Aidala:2012mv, Accardi:2012qut, Leader:2013jra, Aschenauer:2013woa, Aschenauer:2015eha, Jimenez-Delgado:2013sma, Proceedings:2020eah, Ji:2020ena, AbdulKhalek:2021gbh, Constantinou:2022yye, Hatta:2026txn}.) In comparison, the $Q^2$ evolution of the OAM distributions is only known at one loop \cite{Hagler:1998kg, Hoodbhoy:1998yb,Hatta:2019csj} and, at the time of writing, there exists not a single extraction of the OAM distributions from the experimental data. (However, see \cite{Alexandrou:2018xnp} for a lattice calculation of all three $x$-integrated contributions to the Ji sum rule.) The lack of OAM distributions extractions from the data is explained by the historical absence of experimental observables that couple to OAM directly. This is beginning to change, with several elastic particle production observables in electron--proton collisions that couple to OAM distributions proposed in the past decade
\cite{Ji:2016jgn, Hatta:2016aoc, Bhattacharya:2022vvo, Bhattacharya:2023hbq, Bhattacharya:2024sck, Bhattacharya:2026qnd, Kovchegov:2024wjs}. Still, it appears that no OAM distributions extraction from the experimental data will be possible until the Electron-Ion Collider (EIC) \cite{Accardi:2012qut,Boer:2011fh,Proceedings:2020eah,AbdulKhalek:2021gbh} comes online. 

A particularly interesting and unexplored corner of the proton spin phase space is the small-$x$ region. There, the uncertainties of hPDF extractions are large \cite{Gluck:2000dy, Leader:2005ci, deFlorian:2009vb, Leader:2010rb, Jimenez-Delgado:2013boa, Ball:2013lla, Nocera:2014gqa, deFlorian:2014yva, Leader:2014uua, Sato:2016tuz, Ethier:2017zbq, DeFlorian:2019xxt, Aschenauer:2019kzf, Borsa:2020lsz, Zhou:2022wzm, Cocuzza:2022jye, Borsa:2023tqr, Karpie:2023nyg, Borsa:2024mss, Bertone:2024taw, Hunt-Smith:2024khs, Cruz-Martinez:2025ahf, Cocuzza:2025qvf} since the existing data in longitudinally polarized scattering does not reach very small values of $x$ (at perturbatively high $Q^2$ values), while accurate lattice QCD calculations at small-$x$ are not possible at this time. This suggests that additional theoretical input on the proton spin carried by the partons at small $x$ is needed and will be important for our understanding of the proton structure. The hPDFs at small $x$ were first tackled theoretically in the pioneering work of Bartels, Ermolaev and Ryskin (BER) \cite{Bartels:1995iu, Bartels:1996wc} (see also the more recent \cite{Ermolaev:2025isl} for a summary). The BER approach employed the infrared evolution equations (IREE) \cite{Gorshkov:1966ht, Kirschner:1983di, Kirschner:1994rq, Kirschner:1994vc, Griffiths:1999dj, Blumlein:1996hb, Ermolaev:2026rkl}.
Phenomenology based on BER formalism can be found in Refs.~\cite{Blumlein:1995jp, Blumlein:1996hb, Ermolaev:1999jx, Ermolaev:2000sg, Ermolaev:2003zx, Ermolaev:2009cq}. BER found that the hPDFs $\Delta \Sigma$ and $\Delta G$ grow as powers of $1/x$ at small $x$. The powers were found to be close to 1, making the small-$x$ contributions in the integrals of Eqs.~\eqref{spins} and \eqref{eqn:LqLG} potentially very important. 

More recently, the hPDFs at small $x$ have been studied in the $s$-channel/shock wave formalism \cite{Mueller:1994rr, Mueller:1994jq, Mueller:1994gb, Balitsky:1995ub, Balitsky:1998ya, Kovchegov:1999yj, Kovchegov:1999ua, Jalilian-Marian:1997ubg, Jalilian-Marian:1997jhx, Weigert:2000gi, Iancu:2001ad, Iancu:2000hn, Ferreiro:2001qy} (see Refs.~\cite{Gribov:1983ivg, Iancu:2003xm, Weigert:2005us, Jalilian-Marian:2005ccm, Gelis:2010nm, Albacete:2014fwa, Kovchegov:2012mbw, Morreale:2021pnn} for reviews extended to describe sub-eikonal scattering (the processes where the cross section is suppressed by one power of $1/s$ compared to the leading eikonal contribution) in \cite{Altinoluk:2014oxa, Balitsky:2015qba, Balitsky:2016dgz, Kovchegov:2017lsr, Kovchegov:2018znm, Chirilli:2018kkw, Jalilian-Marian:2018iui, Jalilian-Marian:2019kaf, Altinoluk:2020oyd, Boussarie:2020vzf, Boussarie:2020fpb, Kovchegov:2021iyc, Altinoluk:2021lvu, Kovchegov:2022kyy, Altinoluk:2022jkk, Altinoluk:2023qfr,Altinoluk:2023dww, Li:2023tlw, Altinoluk:2024dba, Altinoluk:2025ewj, Altinoluk:2025ang, Altinoluk:2025ivn, Kar:2026vzk, Chirilli:2026vij, Chirilli:2026pkv}. Here $s$ is the center-of-mass energy squared. The evolution equations for the so-called polarized dipole amplitudes, knowing which one can construct the quark and gluon hPDFs and the $g_1$ structure function, were derived and analyzed in \cite{Kovchegov:2015pbl, Kovchegov:2016zex, Kovchegov:2016weo, Kovchegov:2017jxc, Kovchegov:2017lsr, Kovchegov:2018znm, Cougoulic:2019aja, Kovchegov:2020hgb, Cougoulic:2020tbc, Chirilli:2021lif, Kovchegov:2021lvz, Cougoulic:2022gbk, Borden:2023ugd, Adamiak:2023okq, Borden:2024bxa, Borden:2025ehe}. Similar to BER, they resum powers of $\as \, \ln^2 (1/x)$ and $\as \, \ln (1/x) \, \ln (Q^2/\Lambda^2)$, with $\as$ the strong coupling constant and $\Lambda$ some infrared (IR) cutoff scale: this is known as the double-logarithmic approximation (DLA). Analogous to the unpolarized eikonal small-$x$ evolution in the $s$-channel/shock wave formalism \cite{Mueller:1994rr, Mueller:1994jq, Mueller:1994gb, Balitsky:1995ub, Balitsky:1998ya, Kovchegov:1999yj, Kovchegov:1999ua, Jalilian-Marian:1997ubg, Jalilian-Marian:1997jhx, Weigert:2000gi, Iancu:2001ad, Iancu:2000hn, Ferreiro:2001qy}, the equations for polarized dipole amplitudes are not closed, forming an infinite hierarchy of equations. The hierarchy for helicity evolution can be truncated, leading to closed evolution equations for polarized dipole amplitudes in the large-$N_c$ \cite{tHooft:1977xjm} and large-$N_c\&N_f$ \cite{Veneziano:1976wm} limits (where $N_c$ and $N_f$ are the number of quark colors and flavors, respectively). The equations in the large-$N_c$ limit are referred to as the KPS-CTT equations \cite{Kovchegov:2015pbl, Kovchegov:2016zex, Kovchegov:2017lsr, Kovchegov:2018znm, Cougoulic:2022gbk}, while the ones in the large-$N_c\&N_f$ limit are referred to as the KPS-CTT-BCL equations \cite{Kovchegov:2015pbl, Kovchegov:2016zex, Kovchegov:2017lsr, Kovchegov:2018znm, Chirilli:2021lif, Cougoulic:2022gbk, Borden:2024bxa}. Let us add that there is a minor disagreement between the BER and shock wave approaches to helicity evolution: the intercepts (powers of $1/x$) of the small-$x$ asymptotics for $\Delta \Sigma$ and $\Delta G$ are different by a few tenths of a percent in the two approaches \cite{Borden:2023ugd, Borden:2025ehe}. The small-$x$ anomalous dimensions in the two approaches differ at four loops \cite{Borden:2023ugd, Borden:2024bxa, Borden:2025ehe}, just outside the precision of the existing finite-order calculations \cite{Altarelli:1977zs, Dokshitzer:1977sg, Zijlstra:1993sh, Mertig:1995ny, Moch:1999eb, vanNeerven:2000uj, Vermaseren:2005qc, Moch:2014sna, Blumlein:2021ryt, Blumlein:2021lmf, Davies:2022ofz, Blumlein:2022gpp}. These differences appear to be small enough to be ignored in phenomenological applications: phenomenology using the KPS-CTT evolution equations was developed in \cite{Adamiak:2021ppq, Adamiak:2023yhz, JAMCollaborationSmall-xAnalysisGroup:2025tfa} (using also \cite{Kovchegov:2024aus}), reducing the extracted hPDF uncertainties. Constraints on the initial conditions for helicity evolution were constructed in \cite{Dumitru:2024pcv, Adamiak:2025mdy}. 

Let us stress the importance of the large-$N_c\&N_f$ limit in the helicity evolution. The unpolarized nonlinear small-$x$ evolution equations \cite{Mueller:1994rr, Mueller:1994jq, Mueller:1994gb, Balitsky:1995ub, Balitsky:1998ya, Kovchegov:1999yj, Kovchegov:1999ua, Jalilian-Marian:1997ubg, Jalilian-Marian:1997jhx, Weigert:2000gi, Iancu:2001ad, Iancu:2000hn, Ferreiro:2001qy} close only in one limit: the large-$N_c$ limit. Quarks do not contribute to the unpolarized small-$x$ evolution at the leading order, in which the powers of $\as \, \ln (1/x)$ are re-summed. In contrast to that, quarks do contribute to the helicity evolution already at the leading order, in DLA. This allows the helicity evolution equations to close in the large-$N_c\&N_f$ limit as well. Importantly, this makes quark helicity distributions at small-$x$ flavor-dependent \cite{Adamiak:2021ppq, Adamiak:2023yhz, JAMCollaborationSmall-xAnalysisGroup:2025tfa}, again largely in contrast to the unpolarized quark distributions at small-$x$, whose small-$x$ asymptotics are entirely gluon-driven and, hence, flavor-independent (modulo quark mass differences, which are small for light flavors). The goal of this paper is to construct and solve the small-$x$ evolution equations for the OAM distributions in the large-$N_c\&N_f$ limit. 

The small-$x$ asymptotics of the OAM distributions $L_{q+\bar{q}}(x,Q^2)$ and $L_G(x,Q^2)$ was first studied in \cite{Hatta:2018itc} using the DGLAP evolution equations: this approach re-sums only powers of $\as \, \ln (1/x) \, \ln (Q^2/\Lambda^2)$ and is insufficient to fully determine the small-$x$ asymptotics. In \cite{Boussarie:2019icw}, the small-$x$ asymptotics of $L_{q+\bar{q}}(x,Q^2)$ and $L_G(x,Q^2)$ were obtained using a generalization of the BER IREE approach. In the $s$-channel/shock-wave formalism, the OAM distributions at small $x$ were first studied in \cite{Kovchegov:2019rrz}: this work was subsequently corrected in \cite{Kovchegov:2023yzd, Manley:2024pcl}, where the small-$x$ asymptotics for these distributions was derived in the large-$N_c$ limit (both numerically \cite{Kovchegov:2023yzd} and analytically \cite{Manley:2024pcl}) and was found to be the same as for hPDFs, 
\begin{align}\label{OAM_LargeNc}
        L_{q+\bar{q}}(x, Q^2) \sim L_{G}(x,Q^2) \sim \Delta \Sigma(x, Q^2) \sim \Delta G(x,Q^2) \sim \left(\frac{1}{x}\right)^{3.66 \, \sqrt{\frac{\as N_c}{2\pi}}}.
\end{align}
The minor differences between BER and the shock wave approaches for hPDFs affected the OAM distributions as well: the intercepts (the power in \eq{OAM_LargeNc}) of \cite{Boussarie:2019icw} and \cite{Kovchegov:2023yzd, Manley:2024pcl} slightly differ as well. The difference is in the next digit, beyond $3.66$ shown in \eq{OAM_LargeNc}.

In this work we generalize the treatment of \cite{Kovchegov:2023yzd} to the large-$N_c \& N_f$ case. As we mentioned above, the large-$N_c \& N_f$ approximation includes quarks and, therefore, is more accurate and realistic than the large-$N_c$ limit. For this reason, it is also a better choice in phenomenological applications, allowing for a non-trivial flavor dependence of quark OAM distributions. It is, therefore, important to construct the large-$N_c \& N_f$ expressions for the quark OAM distributions along with their small-$x$ evolution. This is what is done below. 

In Sec.~\ref{sec:revised}, we review the expressions for the OAM distributions obtained in \cite{Kovchegov:2023yzd, Kovchegov:2024wjs}. In the process, we define a new operator $\widetilde{L}_{10}$ (see \eq{Ltilde}), whose impact parameter moment $\widetilde{I}$ is proportional to the quark OAM distribution. This is different from the conjecture made in \cite{Kovchegov:2024wjs} that the quark OAM distribution is given by the impact-parameter moment of another operator, $\widetilde{Q}_{10}$, introduced in \cite{Borden:2024bxa}. The gluon OAM distribution remains the same as in \cite{Kovchegov:2023yzd}. The OAM distributions are given by \eqs{OAMs_corr} in the same Sec.~\ref{sec:revised}. 

We proceed by constructing the large-$N_c \& N_f$ evolution equations for the moment amplitudes in Sec.~\ref{sec:evolution}. The section begins by accounting for the contribution of the sub-eikonal operator $F^{+-}$ \cite{Chirilli:2018kkw, Chirilli:2021lif, Altinoluk:2021lvu, Li:2023tlw, Mukherjee:2026cte}, which was omitted in the previous OAM distribution analyses in the shock-wave formalism \cite{Kovchegov:2023yzd, Manley:2024pcl}: while this operator does not contribute to helicity evolution at DLA \cite{Kovchegov:2025gcg}, we find that it does contribute to the small-$x$ evolution of OAM distributions. Next, the evolution for $\widetilde{L}_{10}$ (and, hence, $\widetilde{I}$) is constructed in Sec.~\ref{sec:evolution} by analogy to the evolution of $\widetilde{Q}_{10}$ studied earlier in \cite{Borden:2024bxa}. It turns out that the cancellations which apply to the impact-parameter integrated quantity considered in \cite{Borden:2024bxa} (in the DLA) and were conjectured earlier in \cite{Kovchegov:2018znm, Kovchegov:2021iyc, Kovchegov:2022kyy} do not apply for the evolution of the moment amplitude $\widetilde{I}$. This results in two new terms in the integrand in \eq{Itilde_evol2}
as compared to Eq.~(30) of \cite{Kovchegov:2024wjs}, correcting the expression for the quark OAM distribution. The evolution of the remaining moment amplitudes at large $N_c \& N_f$ is constructed similar to that in \cite{Kovchegov:2023yzd}, with the final results given in \eqs{mom_eqn} and \eqref{neigh_mom_eqn}. 

These evolution equations are discretized and solved numerically in Sec.~\ref{sec:numerics}. (While their analytic solution is possible along the lines of \cite{Borden:2025ehe}, it is likely to be rather involved and is left for future work.) The numerical intercepts for various impact-parameter integrated polarized dipole amplitudes along with the moment amplitudes are summarized in Table~\ref{t:intercepts_Nf}. They appear to be consistent, within the statistical error bars, with the hPDFs intercepts found analytically in  \cite{Borden:2025ehe} at large $N_c \& N_f$. Our numerical intercepts are also consistent with those found by BER in \cite{Bartels:1996wc}: however, the intercepts of \cite{Bartels:1996wc} disagree with those in \cite{Borden:2025ehe} at the tenths-of-percent level. While our numerical precision does not allow us to tell the difference, we find it likely that our intercepts will agree with those in \cite{Borden:2025ehe}, since the OAM distributions' intercepts appear to follow those for hPDFs \cite{Manley:2024pcl}. Finally, in Sec.~\ref{sec:ratios}, we calculate the ratios of OAM distributions to hPDFs, $L_{q + \bar q}/\Delta\Sigma$ and $L_G/\Delta G$. The results are summarized in Table~\ref{t:ratio_coeffs}. They are in numerical agreement with the ratios found in \cite{Boussarie:2019icw}. 

Since the sub-eikonal $F^{+-}$ operator affects the evolution of the OAM distributions by mixing with other moment amplitudes, it appears natural to explore whether it contributes to the elastic dijet production in the longitudinally polarized electron--proton collisions studied in \cite{Kovchegov:2024wjs} as a way to measure OAM distributions. This calculation is performed in Sec.~\ref{sec:dijet}, showing that indeed the $F^{+-}$ operator does contribute to the longitudinal double-spin asymmetry (DSA) in the $\vec e + \vec p$ collisions. In Sec.~\ref{sec:dijet} we add these contributions to the expression for the DSA numerator constructed in \cite{Kovchegov:2024wjs}. 
In addition, we perform azimuthal angular integrations over the dipole orientations in our expression for the DSA numerator, making it more amenable for numerical evaluation needed to perform phenomenology.

We summarize our results in Sec.~\ref{sec:conclusions}, in which we also mention the companion paper to this one \cite{JAMCollaborationSmall-xAnalysisGroup:inprep}, where a feasibility study of constraining OAM distributions at the EIC by analyzing the elastic dijet and di-hadron production in the longitudinally polarized $e+p$ collisions is conducted.


\section{Revised evaluation of the OAM distributions at small $x$}

\label{sec:revised}

Scattering of longitudinally polarized particles is naturally described in terms of sub-eikonal operators along the $x^-$ path of the incident projectile. (Our light-cone coordinates are defined by $x^\pm = (t \pm z)/\sqrt{2}$, while the transverse vectors are denoted by bold letters, $\un x = (x^1, x^2)$, such that the 4-vectors are $x^\mu = (x^+, x^-, \un x)$.) Consider a high energy quark moving along the $x^-$ light-cone with large momentum $p^-$ scattering off a high energy proton moving along the $x^+$ light-cone with large momentum $P^+$. The resulting $S$-matrix of this interaction can be expanded in eikonality (that is, in powers of $1/s$), 
\begin{align} \label{Wilsonl}
    V_{\un x, \un y; \sigma^\prime, \sigma} = V_{\un x} \, \delta^2(\un x - \un y) \, \delta_{\sigma, \sigma^\prime} + \Big[ V^{\mathrm{q}[1]}_{\un x} + V^{\mathrm{G}[1]}_{\un x} \Big] \delta^2( \un x- \un y)\, \sigma \delta_{\sigma, \sigma^\prime} + 
     \Big[ V^{\mathrm{q}[2]}_{\un x}\, \delta^2( \un x- \un y)  + V^{\mathrm{G}[2]}_{\un x, \un y} \Big] \, \delta_{\sigma, \sigma^\prime} + 
    \cdots,
\end{align}
where $\un y$ ($\un x$) and $\sigma$ ($\sigma'$) are the incoming (outgoing) transverse position and helicity, respectively, of the quark projectile. The ellipsis in \eq{Wilsonl} denotes terms which are suppressed by two or more powers of center-of-mass-energy squared $s$ (the sub-sub-eikonal terms and beyond). The first term on the right-hand side of \eq{Wilsonl} is proportional to the eikonal light-cone Wilson line, which we define as
\begin{align}\label{Wline_def}
    V_{\un x}[b^-, a^-] = \mathcal{P}\, \exp \left[ ig \int \limits^{b^-}_{a^-} dx^- A^{+}(0^+, x^-, \un x) \right],
\end{align}
where $g$ is the strong coupling constant, $\mathcal{P}$ denotes path-ordering and $A^+ = A^{+a} t^a$ is the gluon field of the target with $t^a$ the generators of $\mathrm{SU}(N)$ in the fundamental representation. We denote infinite Wilson lines via $V_{\un x} = V_{\un x}[\infty, -\infty]$. 

The remaining terms shown explicitly in \eq{Wilsonl} are the sub-eikonal polarized Wilson lines, defined as \cite{Cougoulic:2022gbk} (see \cite{Altinoluk:2014oxa, Balitsky:2015qba, Balitsky:2016dgz, Kovchegov:2017lsr, Kovchegov:2018znm, Chirilli:2018kkw, Jalilian-Marian:2018iui, Jalilian-Marian:2019kaf, Altinoluk:2020oyd, Boussarie:2020vzf, Boussarie:2020fpb, Kovchegov:2021iyc, Altinoluk:2021lvu, Kovchegov:2022kyy, Altinoluk:2022jkk, Altinoluk:2023qfr,Altinoluk:2023dww, Li:2023tlw, Altinoluk:2024dba, Altinoluk:2025ewj, Altinoluk:2025ang, Altinoluk:2025ivn, Chirilli:2026vij, Chirilli:2026pkv} for related sub-eikonal calculations)
\begin{subequations}\label{VqG}
\begin{align}
& V_{\un x}^{\textrm{G} [1]}  = \frac{i \, g \, P^+}{s} \int\limits_{-\infty}^{\infty} d{x}^- V_{\un{x}} [ \infty, x^-] \, F^{12} (x^-, {\un x}) \, \, V_{\un{x}} [ x^-, -\infty]  , \label{VG1} \\
& V_{\un x}^{\textrm{q} [1]}  = \frac{g^2 P^+}{2 \, s} \int\limits_{-\infty}^{\infty} \!\! d{x}_1^- \! \int\limits_{x_1^-}^\infty d x_2^- V_{\un{x}} [ \infty, x_2^-] \, t^b \, \psi_{\beta} (x_2^-,\un{x}) \, U_{\un{x}}^{ba} [x_2^-, x_1^-] \, \left[ \gamma^+ \gamma^5 \right]_{\alpha \beta} \, \bar{\psi}_\alpha (x_1^-,\un{x}) \, t^a \, V_{\un{x}} [ x_1^-, -\infty] , \label{Vq1} \\
& V_{{\un x}, {\un y}}^{\textrm{G} [2]}  = - \frac{i \, P^+}{s} \int\limits_{-\infty}^{\infty} d{z}^- d^2 z \ V_{\un{x}} [ \infty, z^-] \, \delta^2 (\un{x} - \un{z}) \, \cev{D}^i (z^-, {\un z}) \, D^i  (z^-, {\un z}) \, V_{\un{y}} [ z^-, -\infty] \, \delta^2 (\un{y} - \un{z}) , \label{VxyG2} \\
& V_{{\un x}}^{\textrm{q} [2]} = - \frac{g^2 P^+}{2 \, s} \int\limits_{-\infty}^{\infty} \!\! d{x}_1^- \! \int\limits_{x_1^-}^\infty d x_2^- V_{\un{x}} [ \infty, x_2^-] \, t^b \, \psi_{\beta} (x_2^-,\un{x}) \, U_{\un{x}}^{ba} [x_2^-, x_1^-] \, \left[ \gamma^+ \right]_{\alpha \beta} \, \bar{\psi}_\alpha (x_1^-,\un{x}) \, t^a \, V_{\un{x}} [ x_1^-, -\infty] \label{Vq2},
\end{align}
\end{subequations}
where the center-of-mass energy squared of the target-projectile scattering is 
$s = 2P^+ p^-$. Additionally, $F^{12}$ is the (Abelian part of the) gluon field strength tensor, and the left- and right-acting covariant derivatives are defined as $\cev{D}^i = \cev{\partial}^i + ig A^i $ and $D^i = \partial^i - ig A^i $, respectively. Note that Latin indices denote the transverse vector components, $i = 1,2$, and summation over repeated Latin indices is assumed.

Note that there is another sub-eikonal operator \cite{Chirilli:2018kkw, Chirilli:2021lif, Altinoluk:2021lvu, Li:2023tlw}, 
\begin{align}
V_{\un x}^{\textrm{G} [3]}  = \frac{i \, g \, P^+}{s} \int\limits_{-\infty}^{\infty} d{x}^- V_{\un{x}} [ \infty, x^-] \, F^{+-} (x^-, {\un x}) \, \, V_{\un{x}} [ x^-, -\infty]  , \label{VG3} 
\end{align}
which can be obtained by replacing $0^+ \to x^+$ in \eq{Wline_def} and expanding the corresponding infinite light-cone Wilson line $V_{\un x} (x^+)$ in the powers of $x^+$ to the linear order in $A^- =0$ gauge (see \cite{Kovchegov:2025gcg} as well as \cite{Chirilli:2018kkw, Chirilli:2021lif, Altinoluk:2021lvu, Altinoluk:2023lly, Li:2023tlw,Mukherjee:2026cte}),\footnote{Note the sign difference in the exponent and in front of the $V_{\un x}^{\textrm{G} [3]}$ term in \eq{V_expansion}, as compared to Eq.~(60) of \cite{Kovchegov:2025gcg}.} 
\begin{align} \notag
    & \int\limits_{-\infty}^\infty d x^+ \, e^{+ i (p_f^- - p_i^-) \, x^+} \, V_{\un{x}} (x^+) = \int\limits_{-\infty}^\infty d x^+ \, e^{+ i (p_f^- - p_i^-) \, x^+} \, \left[ V_{\un{x}} (0^+) + x^+ \, \pd^- V_{\un{x}} (0^+) + \ldots \right] \\
    & = 2 \pi \, \delta (p_f^- - p_i^-) \, V_{\un{x}} (0^+) +  2 \pi \, i \, \left[ \frac{\pd}{\pd p_f^-} \delta (p_f^- - p_i^-) \right] \, 2 \sqrt{p_f^- \, p_i^-} \, V_{\un x}^{\textrm{G} [3]} + \ldots \, . 
    \label{V_expansion}
\end{align}
Here $p_i^-$ and $p_f^-$ are the large momentum components of the incoming and outgoing quark, respectively, which scatters on a background field to sub-eikonal order. We see that $V_{\un x}^{\textrm{G} [3]}$ contributes to the sub-eikonal quark $S$-matrix Fourier-transformed into the $p^-$-momentum space. Unlike the terms in \eq{Wilsonl}, which all would come into this object multiplying $\delta (p_f^- - p_i^-)$,  $V_{\un x}^{\textrm{G} [3]}$ enters multiplying a derivative of the minus-momentum delta-function. 

This operator ($V_{\un x}^{\textrm{G} [3]}$) does not couple to the proton's longitudinal spin directly at the lowest non-trivial Born level, and, at DLA, its evolution does not mix with other sub-eikonal operators which couple to the proton spin. Similarly, this operator does not seem to contribute to the OAM distributions directly; its contribution to OAM evolution equations was neglected in \cite{Kovchegov:2023yzd, Manley:2024pcl}. Below, we will show that, in fact, the moment dipole amplitude involving $V_{\un x}^{\textrm{G} [3]}$ mixes with the evolution of the other moment amplitudes, which enter the definitions of the quark and gluon OAM distributions. Hence, the large-$N_c$ small-$x$ evolution equations for the moment amplitudes derived in \cite{Kovchegov:2023yzd} receive a correction.

Matrix elements involving longitudinally polarized scattering at small $x$ can often be expressed in terms of the polarized dipole amplitudes \cite{Kovchegov:2015pbl, Kovchegov:2016weo, Kovchegov:2017lsr, Kovchegov:2018znm, Cougoulic:2022gbk, Kovchegov:2019rrz, Kovchegov:2023yzd}
\begin{subequations} \label{pdas}
    \begin{align}
        Q_{10}(s) &= \frac{1}{2N_c} \mathrm{Re}\,  \llangle \tord 
        \tr \left[ V_{\un 0} V_{\un 1}^{\mathrm{pol}[1] \dagger} 
        \right] 
        +
        \tord 
        \tr \left[ V_{\un 1}^{\mathrm{pol}[1]}  V_{\un 0}^\dagger 
        \right] 
        \rrangle(s),
        \\ 
        G^i_{10}(s) &= \frac{1}{2N_c} \mathrm{Re}\,  \llangle \tord 
        \tr \left[ V_{\un 0} V_{\un 1}^{i\mathrm{G}[2] \dagger} 
        \right] 
        +
        \tord 
        \tr \left[ V_{\un 1}^{i\mathrm{G}[2]}  V_{\un 0}^\dagger 
        \right] 
        \rrangle(s),
    \end{align}
\end{subequations}
where $V^{\mathrm{pol}[1]}_{\un x} = V^{\mathrm{G}[1]}_{\un x} + V^{\mathrm{q}[1]}_{\un x}$, $\tord$ is the time-ordering operator, and 
\begin{align} \label{d-d_pw}
    V_{\un z}^{i\mathrm{G}[2]} = \frac{P^+}{2s} \int \displaylimits^\infty_{-\infty} dz^- V_{\un z}[\infty, z^-] \left[ D^i(z^-, \un z) - \cev{D}^i(z^-, \un z) \right] V_{\un z}[z^-, -\infty]
\end{align}
 is related to $V^{\mathrm{G}[2]}_{\un x, \un y}$ in \eq{VxyG2} above \cite{Cougoulic:2022gbk, Kovchegov:2024wjs, Kovchegov:2025gcg}. The double angle brackets in \eqs{pdas} denote averaging in the (polarized) proton state scaled by a factor of $s$ \cite{Kovchegov:2015pbl, Kovchegov:2017lsr} (see also \cite{Kovchegov:2015zha, Wu:2017rry, Kovchegov:2019rrz, Cougoulic:2020tbc, Kovchegov:2013cva, Kovchegov:2026gwb} for the derivation of the relation between the saturation/color glass condensate averaging and the matrix element in the proton state)
\begin{align} \label{double-bracket}
    & \llangle \hat{\cal O} (x_1, x_2) \rrangle(s) = s\,  \Big\langle \hat{\cal O} (x_1, x_2) \Big\rangle(s) \notag \\
    & = s \, \frac{1}{2} \sum_{S_L} S_L \, \frac{1}{2 P^+} \int \frac{d^2 \Delta_\perp \, d \Delta^+}{(2 \pi)^3} \, e^{-i \Delta^+ b^- + i \un \Delta \cdot \un b} \, \bra{P - \Delta, S_L} \hat{\cal O} (x_1 - b, x_2 - b)   \ket{P, S_L} 
\end{align}
where $S_L$ is the proton's helicity and $\hat{\cal O} (x_1, x_2)$ is some bi-local operator dependent on position four-vectors $x_1$ and $x_2$. 
In \eqs{pdas} we have used a shorthand notation $V_{\un 0} \equiv V_{\un x_0}$, $V_{\un 1}^{\mathrm{pol}[1]} = V_{\un x_1}^{\mathrm{pol}[1]}$, etc. 

In the previous works \cite{Kovchegov:2023yzd, Manley:2024pcl, Kovchegov:2024wjs}, to obtain the OAM distributions at small $x$, we needed the impact-parameter integrated polarized dipole amplitudes $Q(x_{10}^2, s)$ and $G_2(x_{10}^2, s)$, defined by 
\begin{subequations} \label{integrated}
    \begin{align}
        \int d^2 x_1 
        \, Q_{10}(s) & \equiv Q(x_{10}^2, s), 
        \\ \label{G2def}
        \int d^2 x_1 
        \, G^i_{10}(s) & \equiv \epsilon^{ij} x_{10}^j G_2(x_{10}^2, s) + \cdots,
    \end{align}
\end{subequations}
as well as their first impact-parameter moments (referred to as ``moment amplitudes'') $I_3, I_4$ and $I_5$, which are defined by
\begin{subequations} \label{moments}
    \begin{align}\label{I3_def}
        \int d^2 x_1 \, x_1^m \, Q_{10}(s) & \equiv x_{10}^m \, I_3(x_{10}^2, s) + \cdots,
        \\ \label{Gi_moms}
        \int d^2 x_1 \, x_1^m \, G^i_{10}(s) & \equiv\epsilon^{mi} x_{10}^2 \, I_4(x_{10}^2, s) + 
        \epsilon^{mk} x_{10}^k x_{10}^i \, I_5(x_{10}^2, s) + 
        \cdots.
    \end{align}
\end{subequations}
The ellipses in \eqs{integrated} and \eqref{moments} indicate terms that come in with other transverse index structures and which do not couple to the helicity PDFs or OAM distributions. Once again, $i,j,k,m = 1,2$, while $\epsilon^{ij}$ is the two-dimensional Levi-Civita symbol. The transverse separations between the partons are defined by vectors $\un x_{ij} \equiv \un x_i - \un x_j$ with the magnitude $x_{ij} \equiv |\un x_{ij}|$, such that $\un x_{10} = \un x_1 - \un x_0$ and $x_{10} = |\un x_{10}|$. The $x_1$-integrals in Eqs.~\eqref{integrated} and \eqref{moments} are assumed to be carried out while keeping $\un x_{10}$ (and, in general, all other relative transverse separations between the partons) constant. 

In terms of the amplitudes in \eqs{integrated} and \eqref{moments}, the OAM distributions were previously found to be~\cite{Kovchegov:2023yzd, Kovchegov:2024wjs}
\begin{subequations}\label{OAMs}
\begin{align}
\notag
    & L_{{q+\bar{q}}}(x, Q^2) = \frac{N_cN_f}{2\pi^3} \int 
    \limits^{1/\Lambda^2}_{1/Q^2} \frac{d x_{10}^2}{x_{10}^2}
    \int \limits^{1/(Q^2 x_{10}^2)}_{x/(Q^2 x_{10}^2) } \frac{dz}{z}  \Big[Q(x_{10}^2,zs) +  3\, G_2(x_{10}^2,zs) - I_{3}(x_{10}^2,zs) 
\\[-0.3cm] \label{final_quark_OAM} 
& \hspace{8.0cm}
    + 2\, I_4(x_{10}^2,zs) - I_5(x_{10}^2,zs) \Big], 
\\
\label{gluon_OAM}
& L_G(x,Q^2) = - \frac{2 \, N_c}{\as \pi^2} 
    \Big[ 2\, I_4 + 3\, I_5\Big]\!\!\left(x_{10}^2 = \frac{1}{Q^2}, \, s = \frac{Q^2}{x} \right).
\end{align}
\end{subequations}
However, in light of the new transition operators introduced in \cite{Borden:2024bxa}, the quark OAM distribution needs to be revised, as follows from the calculation we present now. 

We start with the quark OAM distribution as given in Eq.~(10) of \cite{Kovchegov:2019rrz}, but without the sum over all final states:
\begin{align}\label{qOAM1}
L_q (x, Q^2) = \frac{2 P^+}{(2\pi)^3} \,  \int d^2 k_\perp \, d^{2} x_1 \, d^{2} x_0 \, & \int\limits_{-\infty}^\infty d x_1^- \, \int\limits^{\infty}_{-\infty} d x_0^- \, e^{- i \un k \cdot \un x_{10}} \left( \frac{{\un x}_1 + {\un x}_0}{2} \times {\un k}\right) \,  
\notag 
\\ 
& \qquad\qquad \times
\left\langle \bar\psi (x_0^-, \un x_0) \, V_{\un 0} [x_0^-, \infty] \, \left( \thalf \gamma^+ \right)
 V_{\ul 1} [\infty , x_1^-] \, \psi (x_1^-, \un x_1) \right\rangle .
\end{align}
In arriving at \eq{qOAM1} we have expanded the exponential from Eq.~(10) of \cite{Kovchegov:2019rrz} to the lowest nontrivial order in $x$. Adding the antiquark contribution yields
\begin{align}\notag
& L_{q+\bar{q}}(x, Q^2) 
= \frac{2 P^+}{(2\pi)^3} \,  \int d^2 k_\perp \, d^{2} x_1 \, d^{2} x_0 \, \int\limits_{-\infty}^\infty d x_1^- \, \int\limits^{\infty}_{-\infty} d x_0^- \, e^{- i \un k \cdot \un x_{10}} \left( \frac{{\un x}_1 + {\un x}_0}{2} \times {\un k}\right)   \\ 
& \times
\left\langle \bar\psi (x_0^-, \un x_0) \, V_{\un 0} [x_0^-, \infty] \, \left( \thalf \gamma^+ \right)
V_{\ul 1} [\infty , x_1^-] \, \psi (x_1^-, \un x_1) + \left(  V_{\un 0} [\infty, x_0^-] \, \psi_\beta (x_0^-, \un x_0) \right)^i \, \left( \thalf \gamma^+ \right)_{\alpha\beta} \, \left( \bar\psi_\alpha (x_1^-, \un x_1) \, V_{\un 1} [x_1^-, \infty] \right)^i \right\rangle, 
\label{qOAM2}
\end{align}
where $i$ denotes the color index of the row or column in the second line while $\alpha$ and $\beta$ are the Dirac spinor indices.

By analogy to \cite{Borden:2024bxa} (see Eq.~(44) there), we define
\begin{align} 
\notag 
{\widetilde L}_{10} (s) \equiv & \  \llangle \frac{g^2}{16 \sqrt{k^- \, p_2^-}} \, \int\limits_{-\infty}^\infty d x_1^- \, \int\limits^{\infty}_{-\infty} d x_0^- \, \bigg[ \bar\psi (x_0^-, \un x_0) \, V_{\un 0} [x_0^-, \infty] \, \left( \thalf \gamma^+ \right)
 V_{\ul 1} [\infty , x_1^-] \, \psi (x_1^-, \un x_1)  \\ 
& + \left(  V_{\un 0} [\infty, x_0^-] \, \psi_\beta (x_0^-, \un x_0) \right)^i \, \left( \thalf \gamma^+ \right)_{\alpha\beta} \, \left( \bar\psi_\alpha (x_1^-, \un x_1) \, V_{\un 1} [x_1^-, \infty] \right)^i   \notag \\
& + \bar\psi (x_0^-, \un x_0) \, V_{\un 0} [x_0^-, -\infty] \, \left( \thalf \gamma^+ \right)
 V_{\ul 1} [- \infty , x_1^-] \, \psi (x_1^-, \un x_1)  \notag \\
 &   + \left(  V_{\un 0} [- \infty, x_0^-] \, \psi_\beta (x_0^-, \un x_0) \right)^i \, \left( \thalf \gamma^+ \right)_{\alpha\beta} \, \left( \bar\psi_\alpha (x_1^-, \un x_1) \, V_{\un 1} [x_1^-, - \infty] \right)^i  \bigg] \rrangle (s),  \label{Ltilde}
\end{align}
where now the double angle brackets are defined as
\begin{align}\label{double_def}
\llangle {\hat O}_{10} \rrangle \equiv \sqrt{2 k^- P^+} \, \sqrt{2 p_2^- P^+} \, \Big\langle  {\hat O}_{10} \Big\rangle,
\end{align} 
such that 
\begin{align}
2 P^+ \Big\langle  {\hat O}_{10} \Big\rangle = \llangle \frac{1}{\sqrt{k^- \, p_2^-}} \, {\hat O}_{10} \rrangle 
\end{align}
for an arbitrary operator ${\hat O}_{10}$. The object in \eq{Ltilde} is similar to the object $\widetilde{Q}_{10}$ from \cite{Borden:2024bxa}, whose definition we modify (compared to Eq.~(44) of \cite{Borden:2024bxa}) to 
\begin{align}
 \notag 
{\widetilde Q}_{10} (s) \equiv & \  \llangle \frac{g^2}{16 \sqrt{k^- \, p_2^-}} \, \int\limits_{-\infty}^\infty d x_1^- \, \int\limits^{\infty}_{-\infty} d x_0^- \, \bigg[ \bar\psi (x_0^-, \un x_0) \, V_{\un 0} [x_0^-, \infty] \, \left( \thalf \gamma^+ \gamma^5 \right)
 V_{\ul 1} [\infty , x_1^-] \, \psi (x_1^-, \un x_1)  \\ 
& - \left(  V_{\un 0} [\infty, x_0^-] \, \psi_\beta (x_0^-, \un x_0) \right)^i \, \left( \thalf \gamma^+ \gamma^5 \right)_{\alpha\beta} \, \left( \bar\psi_\alpha (x_1^-, \un x_1) \, V_{\un 1} [x_1^-, \infty] \right)^i   \notag \\
& + \bar\psi (x_0^-, \un x_0) \, V_{\un 0} [x_0^-, -\infty] \, \left( \thalf \gamma^+ \gamma^5 \right)
 V_{\ul 1} [- \infty , x_1^-] \, \psi (x_1^-, \un x_1)   \notag \\
 &  - \left(  V_{\un 0} [- \infty, x_0^-] \, \psi_\beta (x_0^-, \un x_0) \right)^i \, \left( \thalf \gamma^+ \gamma^5 \right)_{\alpha\beta} \, \left( \bar\psi_\alpha (x_1^-, \un x_1) \, V_{\un 1} [x_1^-, - \infty] \right)^i  \bigg] \rrangle (s) . 
 \label{Qtilde}
\end{align}
The minus signs in front of the second and fourth terms in \eq{Qtilde}, as compared to the plus signs in front of the same terms in Eq.~(44) of \cite{Borden:2024bxa}, appear due to moving fermion fields past each other. They are related to the minus sign in the definition of the antiquark hPDF. Analogously, the plus signs in front of the second and fourth terms in \eq{Ltilde} are due to the definition of unpolarized antiquark PDF.

Using \eq{Ltilde} we write the OAM distribution as
\begin{align}\label{qOAM3}
& L_{q+\bar{q}}(x,Q^2) 
= \frac{1}{4 \, \pi^4 \, \as} \, 
\int d^2 k_\perp \, d^{2} x_1 \, d^{2} x_0 \, 
e^{- i \un k \cdot \un x_{10}} 
\left( \frac{{\un x}_1 + {\un x}_0}{2} \times {\un k}
\right)\, 
{\widetilde L}_{10}(s = Q^2/x) .
\end{align}
In doing so we have averaged \eq{qOAM2} over the future- and past-pointing Wilson line staples (that is, over semi-inclusive deep inelastic scattering (SIDIS) and Drell-Yan (DY) staples). This does not change the OAM distribution, since it is a PT-even quantity.

Due to the anti-commutative property of the fermion fields, the object ${\widetilde L}_{10}$ is anti-symmetric, ${\widetilde L}_{10} = - {\widetilde L}_{01}$. Noticing that ${\widetilde L}_{10}^* = {\widetilde L}_{01}$, we conclude that ${\widetilde L}_{10}$ is also purely imaginary. A consequence of ${\widetilde L}_{10}$ being anti-symmetric and carrying no Lorentz indices is (cf. \cite{Kovchegov:2012ga})
\begin{align}\label{b0}
\int d^2 \left( \frac{\un x_0 + \un x_1}{2} \right) \, {\widetilde L}_{10}  (s) = 0 . 
\end{align}
This allows us to replace
\begin{align}
\frac{{\un x}_1 + {\un x}_0}{2} \to {\un x}_1 
\end{align}
in \eq{qOAM3}. In addition, we define the (real) moment amplitude ${\widetilde I}$ by
\begin{align} \label{L_mom}
\int d^2 x_1 \, x_1^i \, {\widetilde L}_{10}  (s)  = i \, \epsilon^{ij} \, x_{10}^j \, {\widetilde I} (x^2_{10},  s) + i \, x_{10}^i \, {\widetilde J} (x^2_{10},  s) . 
\end{align}
(Here ${\widetilde J}$ is an object which does not contribute to OAM distribution.) This allows us to rewrite \eq{qOAM3} as
\begin{align} \label{qOAM4}
    L_{q+\bar{q}}(x,Q^2) 
    = \frac{2\, N_f}{\as \pi^2}\, \widetilde{I} \left(x_{10}^2
    = \frac{1}{Q^2}, s = \frac{Q^2}{x} \right), 
\end{align} 
where we neglected derivatives of ${\widetilde I}$ as not DLA. Furthermore, we have assumed all flavors contribute equally here, giving an overall factor of $N_f$. For phenomenological applications, one should instead furnish $\widetilde{I}$ with a flavor index, and replace $N_f$ in \eq{qOAM4} with a sum over flavors. 

The above discussion demonstrates that, in the DLA, Eqs.~\eqref{OAMs} have to be replaced by
\begin{subequations}\label{OAMs_corr}
\begin{align}
        \label{final_quark_OAM_corr}
    & L_{q+\bar{q}}(x, Q^2) = \frac{2\,N_f}{\as \pi^2} \, \widetilde{I} \left(x_{10}^2 = \frac{1}{Q^2}, s = \frac{Q^2}{x} \right)  ,
    \\  \label{final_gluon_OAM}
    & L_G(x,Q^2) = - \frac{2 \, N_c}{\as \pi^2} 
    \Big[ 2\, I_4 + 3\, I_5\Big]\!\!\left(x_{10}^2 = \frac{1}{Q^2}, \, s = \frac{Q^2}{x} \right).
\end{align}
\end{subequations}
(The gluon OAM in \eq{gluon_OAM} remains unchanged.) Comparing \eqs{OAMs_corr} with \eq{OAMs}, we see both the quark and gluon OAM distribution are completely determined by the moment amplitudes. Furthermore, \eqs{OAMs_corr} mirror the analogous expressions for the helicity PDFs at small $x$, which read
\cite{Kovchegov:2015pbl, Kovchegov:2016zex, Kovchegov:2017lsr, Kovchegov:2018znm, Cougoulic:2022gbk, Borden:2024bxa}
\begin{subequations} \label{pdfs}
    \begin{align}
    \label{quark_hel}
    \Delta \Sigma(x, Q^2) &= \frac{N_f}{\as \pi^2} \, \widetilde{Q}\left(x_{10}^2 = \frac{1}{Q^2}, s = \frac{Q^2}{x} \right) ,
    \\ 
    \label{gluon_hel}
    \Delta G(x,Q^2) &= \frac{2 N_c}{\as \pi^2}\, G_2 \left(x_{10}^2 = \frac{1}{Q^2}, s = \frac{Q^2}{x} \right) ,
    \end{align}
\end{subequations}
where the impact-parameter integrated 
object $\widetilde{Q}_{10} (s)$ from \eq{Qtilde} is \cite{Borden:2024bxa}
\begin{align}\label{Qint_def}
    \int d^2 x_1\, \widetilde{Q}_{10} (s) \equiv \widetilde{Q} (x_{10}^2, s).
\end{align}
The moment amplitudes $I_4, I_5, \widetilde{I}$ only appear in the OAM distributions, not the hPDFs. Furthermore, the moment amplitudes do not appear in any of the observables considered so far in the small-$x$ helicity phenomenology \cite{Adamiak:2021ppq, Adamiak:2023yhz, JAMCollaborationSmall-xAnalysisGroup:2025tfa}. Instead, it was shown in \cite{Kovchegov:2024wjs} that the moment amplitudes $I_{3}, I_4, I_5$ explicitly appear in the DSA in elastic dijet production in electron-proton collisions. In a companion paper, we assess the feasibility of this observable to extract the moment amplitudes $I_{3}, I_4, I_5$ and, via \eqs{OAMs_corr}, the OAM distributions from the future EIC data. However, to do so requires the evolution equations for the moment amplitudes in the large-$N_c\&N_f$ limit, which we turn to next. 


\section{Moment amplitude evolution equations in the large $N_c\&N_f$ limit}
\label{sec:evolution}

\subsection{Contributions from $F^{+-}$ operator and the evolution equation for the moment amplitude $I_E$}

Before we derive the evolution for the moment amplitudes, we first consider the contributions from the type-3 polarized Wilson line in \eq{VG3}, which involves the $F^{+-}$ operator. As we will see below, this operator is not DLA in the helicity evolution and therefore was neglected in the previous works on helicity \cite{Kovchegov:2017lsr, Kovchegov:2018znm, Cougoulic:2022gbk, Borden:2024bxa}, but it does contribute to the moment amplitude evolution (and therefore the OAM distributions) at DLA. Therefore, we first need to derive the type-3 contributions to the evolution of the polarized dipole amplitudes at large-$N_c\&N_f$. 

The relevant dipole amplitudes for helicity evolution at large-$N_c\&N_f$ are $Q_{10}(s), \widetilde{G}_{10}(s),$ and $G^i_{10}(s)$ \cite{Kovchegov:2017lsr, Kovchegov:2018znm, Cougoulic:2022gbk}, along with the object $\widetilde{Q}_{10} (s)$ \cite{Borden:2024bxa}. The polarized dipole amplitude $\widetilde{G}_{10}(s)$ is defined by \cite{Cougoulic:2022gbk, Borden:2024bxa}
\begin{align}\label{G_dfn_NcNf}
& {\widetilde G}_{10} (s) = \frac{1}{2 N_c} \, \mbox{Re} \, \llangle \mbox{T} \, \mbox{tr} \left[ V_{\ul 0} \, W_{{\un 1}}^{\textrm{pol} [1] \,\dagger} \right] + \mbox{T} \, \mbox{tr} \left[ W_{{\un 1}}^{\textrm{pol} [1] } \, V_{\ul 0}^\dagger \right] \rrangle (s)
\end{align}
with 
\begin{align}\label{Wpol_dfn}
W_{{\un x}}^{\textrm{pol} [1] } &= V_{{\un x}}^{\textrm{G} [1]} - \frac{g^2 P^+}{4s} \, \int\limits_{-\infty}^{\infty}dx_1^- \int\limits_{x_1^-}^{\infty}dx_2^- \; V_{{\un x}}[\infty,x_2^-] \, \psi_{\alpha}(x_2^-,{\un x}) \, \left(\frac{1}{2}\gamma^+\gamma_5\right)_{\beta\alpha}\bar{\psi}_{\beta}(x_1^-,{\un x}) \, V_{{\un x}}[x_1^-,-\infty] \, .
\end{align}
It results from the adjoint version of $Q_{10} (s)$, 
\begin{align}\label{G_adj_def}
G^\textrm{adj}_{10} (s) \equiv \frac{1}{2 (N_c^2 -1)} \, \mbox{Re} \, \llangle \mbox{T} \, \mbox{Tr} \left[ U_{\ul 0} \, U_{{\un 1}}^{\textrm{pol} [1] \, \dagger} \right] + \mbox{T} \, \mbox{Tr} \left[ U_{{\un 1}}^{\textrm{pol} [1]} \, U_{\ul 0}^\dagger \right] \rrangle (s) ,
\end{align}
where
\begin{align}
    (U_{\un x}^{\textrm{G} [1]})^{ba} = \frac{2 \, i \, g \, P^+}{s} \int\limits_{-\infty}^{\infty} d{x}^- (U_{\un{x}} [ \infty, x^-])^{bb'} \, ({\cal F}^{12})^{b'a'} (x^-, {\un x}) \, (U_{\un{x}} [ x^-, -\infty])^{a'a}   \label{UG1}
\end{align}
is the adjoint analogue of \eq{VG1}, while $U_{\un x} [b^-, a^-]$ is the adjoint light-cone Wilson line (cf. \eq{Wline_def}). At large-$N_c\&N_f$, the adjoint dipole amplitude ${\widetilde G}_{10} (s)$ cannot be expressed in terms of its fundamental analogue $Q_{10} (s)$ \cite{Kovchegov:2018znm, Cougoulic:2022gbk, Borden:2024bxa}: instead we have
\begin{align}\label{Gadj_Gtilde}
G^{\textrm{adj}}_{10} (s) = 4 \, S_{10} (s) \, {\widetilde G}_{10} (s) ,
\end{align}
where 
\begin{align}\label{Sdef}
S_{10} (s) = \frac{1}{N_c} \, \left\langle \mbox{T} \, \tr \left[ V_{\un 0} \,  V_{\un 1}^{\dagger} \right] \right\rangle (s)
\end{align}
is the unpolarized fundamental dipole $S$-matrix \cite{Iancu:2003xm, Weigert:2005us, Jalilian-Marian:2005ccm, Gelis:2010nm, Albacete:2014fwa, Kovchegov:2012mbw, Morreale:2021pnn}.

We will proceed by assuming that, similar to the helicity case, the small-$x$ evolution of the moment amplitudes $I_4, I_5$, and $\widetilde{I}$ that determine the OAM distributions also results from the evolution of $Q_{10}(s), \widetilde{G}_{10}(s),$ $G^i_{10}(s)$, and $\widetilde{Q}_{10} (s)$. We will see below that this is indeed the case.  

We begin with $G^i_{10}(s)$, whose evolution is calculated in detail in Sec.~IVA of \cite{Cougoulic:2022gbk}. Since the relevant diagrams, given in Fig.~5 of that reference, are unchanged, we will not re-draw them here. However, we note that they are very similar to the diagrams in \fig{FIG:G3_evol} below. The only calculational difference now is due to the contribution of the type-3 terms (the $F^{+-}$ operator insertions from \eq{VG3}) to the sub-eikonal gluon propagators employed in \cite{Cougoulic:2022gbk}. Specifically, to compute the type-3 term's contribution to the evolution of $G^i_{10}(s)$, we need to include it into the following propagator \cite{Kovchegov:2025gcg} (note, again, the sign difference compared to Eqs.~(125) and (130) in \cite{Kovchegov:2025gcg}, here and in \eq{Vi6} below):
\begin{align}\label{Vi5}
\int\limits_{-\infty}^0 dx_{2'}^- \,  
\int\limits_0^\infty dx_2^- \, \Big[ x_{2'}^- {\pd}^i a^{+ \, a} (x_{2'}^- , \ul{x}_1)
\contraction[2ex]
{}
{+}{a^{i \, a}
(x_{2'}^- , \ul{x}_1) \Big] \:}
{a}
\: 
+ a^{i \, a} &(x_{2'}^- , \ul{x}_1) \Big] \:
a^{+ \, b} (x_2^- , \ul{x}_0) 
\notag \\ & 
\supset + \frac{1}{(2 \pi)^3} \int\limits_{0}^{q^-} d k^-\, \int d^2 x_2 \, \frac{x_{20}^j}{x_{20}^2} \, 
\left( \delta^{ij} - 2 \, \frac{x_{21}^i \, x_{21}^j}{x_{21}^2} \right) \, \left( U_{\un 2}^{\textrm{G} [3]} \right)^{ba},
\end{align}
where $U^{\mathrm{G}[3]}_{\un 2}$ is the adjoint version of $V^{\mathrm{G}[3]}_{\un 2}$ defined in \eq{VG3}: 
\begin{align}
 \left( U_{\un x}^{\textrm{G} [3]} \right)^{ba}  = \frac{i \, g \, P^+}{s} \int\limits_{-\infty}^{\infty} d{x}^- \left( U_{\un{x}} [ \infty, x^-] \right)^{bb'} \, \left( \cal{F}^{+-} \right)^{b'a'} (x^-, {\un x}) \, \, \left( U_{\un{x}} [ x^-, -\infty] \right)^{a' a} . \label{UG3} 
\end{align}
The result in \eq{Vi5} was obtained in \cite{Kovchegov:2025gcg} (see Eq.~(125) there) by employing the adjoint version of \eq{V_expansion}. 

Similarly, Eq.~(130) in \cite{Kovchegov:2025gcg} gives
\begin{align}\label{Vi6}
\int\limits_{-\infty}^0 dx_{2'}^- \,  
\int\limits_0^\infty dx_2^- \, \Big[ x_{2'}^- {\pd}^i a^{+ \, b} (x_{2'}^- , \ul{x}_1)
\contraction[2ex]
{}
{+}{a^{i \, b}
(x_{2'}^- , \ul{x}_1) \Big] \:}
{a}
\: 
+ a^{i \, b} &(x_{2'}^- , \ul{x}_1) \Big] \:
a^{+ \, a} (x_2^- , \ul{x}_0) 
\notag \\ & 
\supset + \frac{1}{(2 \pi)^3} \int\limits_{0}^{q^-} d k^-\, \int d^2 x_2 \, \frac{x_{20}^j}{x_{20}^2} \, 
\left( \delta^{ij} - 2 \, \frac{x_{21}^i \, x_{21}^j}{x_{21}^2} \right) \, \left( U_{\un 2}^{\textrm{G} [3]} \right)^{ba}.
\end{align}
Using \eqs{Vi5} and \eqref{Vi6} in the evolution of $G^{i}_{10}$ (specifically, in the diagrams in the second line of Fig.~5 in \cite{Cougoulic:2022gbk}), we get 
\begin{align}\label{Gi_evol1}
G^i_{10} (zs) \supset  - \frac{\as \, N_c}{8 \pi^2} \, \int\limits_\frac{\Lambda^2}{s}^z \frac{d z'}{z'} \, \int d^2 x_2 \, \left[ \frac{x_{20}^j}{x_{20}^2} - \frac{x_{21}^j}{x_{21}^2} \right] \, \left[ \delta^{ij} - 2 \frac{x_{21}^i \, x_{21}^j}{x_{21}^2} \right] \, \frac{1}{N_c^2} \llangle \left( U_{\un 2}^{\textrm{G} [3]} \right)^{ba} \, \tr \left[ t^b \, V_{\un 0} \, t^a \, V^\dagger_{\un 1} \right]  + \cc \rrangle (z's).
\end{align}
Here and below, the dimensionless parameter $z$ can be approximately thought of as the lower minus momentum fraction of some initial projectile carried by the quark and anti-quark in the dipole \cite{Kovchegov:2015pbl}. Hence, $zs$ is the effective center-of-mass energy squared between the dipole and the target proton.  

In the large-$N_c$ or large-$N_c\&N_f$ limit, \eq{Gi_evol1} can simplified to 
\begin{align}\label{Gi_evol2}
G^i_{10} (zs) \supset - \frac{\as \, N_c}{4 \pi^2} \, \int\limits_\frac{\Lambda^2}{s}^z \frac{d z'}{z'} \, \int d^2 x_2 \, \left[ \frac{x_{20}^j}{x_{20}^2} - \frac{x_{21}^j}{x_{21}^2} \right] \, \left[ \delta^{ij} - 2 \frac{x_{21}^i \, x_{21}^j}{x_{21}^2} \right] \, \left[  G^{[3]}_{21} (z's) +  G^{[3]}_{20} (z's)   \right],
\end{align}
where we have defined the polarized dipole amplitude of the third type \cite{Kovchegov:2025gcg}
\begin{align} \label{G3_def}
     G^{[3]}_{10}(s) &= \frac{1}{2N_c} \mathrm{Re}\,  \llangle \tord 
        \tr \left[ V_{\un 0} V_{\un 1}^{\mathrm{G}[3] \dagger} 
        \right] 
        +
        \tord 
        \tr \left[ V_{\un 1}^{\mathrm{G}[3]}  V_{\un 0}^\dagger 
        \right] 
        \rrangle(s).
\end{align}
Additionally, in arriving at \eq{Gi_evol2}, we have set the unpolarized dipole $S$-matrix to 1 in the DLA. 

Restoring the type-1 and type-2 terms from Eq.~(128) of \cite{Cougoulic:2022gbk} (replacing $G\to \widetilde{G}$ and $\Gamma \to \widetilde{\Gamma}$ in the large-$N_c\&N_f$ limit), we get the evolution equation for $G^i_{10}$ in the large $N_c\&N_f$ limit,
\begin{align}\notag
& G^i_{10} (zs) =  G^{i \, (0)}_{10} (zs) + \frac{\as \, N_c}{2 \pi^2} \, \int \limits_{\frac{\Lambda^2}{s}}^z \frac{d z'}{z'} \, \int d^2 x_2 \, \frac{x_{10}^2}{x_{21}^2 \, x_{20}^2} \,  \left[  G^i_{12} (z' s)  - \Gamma^{i \, \textrm{gen}}_{10,21} (z' s)  \right] \\
& + \frac{\as \, N_c}{4 \pi^2} \, \int \limits_{\frac{\Lambda^2}{s}}^z \frac{d z'}{z'} \, \int d^2 x_2 \Bigg\{ 2 \left[ \frac{\epsilon^{ij} x_{21}^j}{x_{21}^2} - \frac{\epsilon^{ij} x_{20}^j}{x_{20}^2} + 2 x_{21}^i \frac{{\un x}_{21} \times {\un x}_{20}}{x_{21}^2 \, x_{20}^2} \right] \, \left[  \widetilde{G}_{21} (z' s) +  \widetilde{\Gamma}^\textrm{gen}_{20,21} (z' s) \right] \notag \\ 
& + \left[ \delta^{ij} \left( \frac{3}{x_{21}^2} -  2 \, \frac{{\un x}_{20} \cdot {\un x}_{21}}{x_{20}^2 \, x_{21}^2} - \frac{1}{x_{20}^2} \right)  - 2 \frac{x_{21}^i \, x_{20}^j}{x_{21}^2 \, x_{20}^2} \left( 2 \frac{{\un x}_{20} \cdot {\un x}_{21}}{x_{20}^2} + 1 \right) + 2 \frac{x_{21}^i \, x_{21}^j}{x_{21}^2 \, x_{20}^2} \left( 2 \frac{{\un x}_{20} \cdot {\un x}_{21}}{x_{21}^2} + 1 \right) + 2 \frac{x_{20}^i \, x_{20}^j}{x_{20}^4} - 2 \frac{x_{21}^i \, x_{21}^j}{x_{21}^4}   \right] \notag \\
& \times \, \left[ G^j_{21} (z' s) +   \Gamma^{j \, \textrm{gen}}_{20,21} (z' s) \right]  
\notag \\ 
& - \left[ \frac{x_{20}^j}{x_{20}^2} - \frac{x_{21}^j}{x_{21}^2} \right] \, \left[ \delta^{ij} - 2 \frac{x_{21}^i \, x_{21}^j}{x_{21}^2} \right] \, \left[ G^{[3]}_{21} (z's) + G^{[3]}_{20} (z's)   \right]
\Bigg\}  \, ,  
\label{Gi_largeNc_evol}
\end{align}
where again we have set the unpolarized $S$ matrix equal to $1$ in the strict DLA, and defined the generalized amplitudes 
\begin{subequations}
\begin{align}
    \tGm^{\mathrm{gen}}_{10, 32}(zs) &\equiv \tG_{10}(zs) \,\theta ( x_{32} - x_{10} ) + \tGm_{10,32}(zs) \, \theta(x_{10}- x_{32}), 
    \\ 
    \Gamma^{i\, \mathrm{gen}}_{10, 32}(zs) &\equiv G^i_{10}(zs) \,\theta ( x_{32} - x_{10} ) + \Gamma^i_{10,32}(zs) \, \theta(x_{10}- x_{32}).
\end{align}
\end{subequations}
Note the $\Gamma$ functions here are the so-called ``neighbor" dipole amplitudes, which are auxiliary functions needed to enforce lifetime ordering in the evolution. Their operator definitions are the same as for the corresponding polarized dipole amplitudes: they differ from those amplitudes only in the constraints on their evolution coming from lifetime ordering — this is also the origin of the dependence on two separate dipole sizes in their definitions (see \cite{Kovchegov:2015pbl} for further details).

Moving on to the evolution of $Q_{10}(s)$ and $\widetilde{G}_{10}(s)$, we need the following type-3 contributions to the relevant propagators: 
\begin{subequations}
    \begin{align}\label{Vi55}
& \int\limits_{-\infty}^0 dx_{2'}^- \,  
\int\limits_0^\infty dx_2^- \, %
\contraction[2ex]
{}
{a}{^{i \, a}
(x_{2'}^- , \ul{x}_1)  \:}
{a}
\: 
a^{i \, a} (x_{2'}^- , \ul{x}_1)  \:
a^{+ \, b} (x_2^- , \ul{x}_0)  
\supset -  \frac{1}{4 \pi^3} \int\limits_0^{p_2^-} d k^-  \int d^2 x_2 \, \frac{x_{20}^i}{x_{20}^2} \, \ln \frac{1}{x_{21} \, \Lambda} \, \left( U_{\un 2}^{\textrm{G} [3]} \right)^{ba} , 
\\  \label{Vi56}
& \int\limits_{-\infty}^0 dx_{2'}^- \,  
\int\limits_0^\infty dx_2^- \, 
\contraction[2ex]
{}
{a}{^{i \, b}
(x_2^- , \ul{x}_1)  \:}
{a}
\: 
a^{i \, b} (x_2^- , \ul{x}_1) \:
a^{+ \, a} (x_{2'}^- , \ul{x}_0) \supset - \frac{1}{4 \pi^3} \int\limits_0^{p_2^-} d k^-  \int d^2 x_2 \, \frac{x_{20}^i}{x_{20}^2} \, \ln \frac{1}{x_{21} \, \Lambda} \, \left( U_{\un 2}^{\textrm{G} [3]} \right)^{ba} ,
\end{align}
\end{subequations}
which are easily obtained by analogy to \eqs{Vi5} and \eqref{Vi6}. Using \eqs{Vi55} and \eqref{Vi56} in the evolution of $Q_{10}(s)$ (where only the propagators in the second line of Fig.~3 in \cite{Cougoulic:2022gbk} are affected, see also \fig{FIG:G3_evol} below), we get 
\begin{align}\label{Q_evol2}
Q_{10} (zs) \supset + \frac{\as \, N_c}{2 \pi^2} \, \int\limits_\frac{\Lambda^2}{s}^z \frac{d z'}{z'} \, \int d^2 x_2 \, \frac{\un x_{21} \times \un x_{20}}{x_{21}^2 \, x_{20}^2} \, \left[  G^{[3]}_{21} (z's) +  G^{[3]}_{20} (z's)   \right] . 
\end{align}
Restoring the type-1 and type-2 terms from Eq.~(141) of \cite{Cougoulic:2022gbk}, we arrive at the evolution equation for $Q_{10}$ in the large-$N_c\&N_f$ limit (while, again, working in the linearized $S=1$ approximation)
\begin{align}\notag
 Q_{10}(zs) &=  Q_{10}^{(0)}(zs) \\
& + \frac{\as \, N_c}{2 \pi^2} \, \int\limits_{\frac{\Lambda^2}{s}}^z \frac{d z'}{z'} \, \int d^2 x_2 \, \Bigg\{ 2 \, \left[ \frac{1}{x_{21}^2} -  \frac{{\un x}_{21}}{x_{21}^2} \cdot \frac{{\un x}_{20}}{x_{20}^2} \right] \, \left(  {\widetilde \Gamma}^{\text{gen}}_{20,21}(z's) +  {\widetilde G}_{21}(z's)\right) \notag \\ 
& + \left[ 2 \frac{\epsilon^{ij} \, x_{21}^j}{x_{21}^4} - \frac{\epsilon^{ij} \, (x_{20}^j + x_{21}^j)}{x_{20}^2 \, x_{21}^2}  - \frac{2 \, {\un x}_{20} \times {\un x}_{21}}{x_{20}^2 \, x_{21}^2} \left( \frac{x_{21}^i}{x_{21}^2} - \frac{x_{20}^i}{x_{20}^2}\right) \right] \left( \Gamma^{i \, \textrm{gen}}_{20,21}(z's) +   G^{i }_{21}(z's)\right)  \Bigg\}  \notag \\
& + \frac{\as N_c}{4 \pi^2 } \, \int\limits_{\frac{\Lambda^2}{s}}^z \frac{d z'}{z'} \, \int \frac{d^2 x_2}{x_{21}^2} \,  \Bigg\{ Q_{21}(z's) +  \frac{2\epsilon^{ij} \, {x}_{21}^j}{x_{21}^2} \, G^{i}_{21}(z's) \Bigg\} \notag \\
& + \frac{\as \, N_c}{2 \pi^2} \, \int\limits_{\frac{\Lambda^2}{s}}^z \frac{d z'}{z'} \, \int d^2 x_2 \, \frac{x_{10}^2}{x_{21}^2 \, x_{20}^2} \,  \Bigg\{ Q_{12}(z's)  - \overline{\Gamma}^{\text{\,gen}}_{10,21}(z's)  \Bigg\} \, 
\notag  \\ & 
+  \frac{\as \, N_c}{2\pi^2} \, \int\limits_\frac{\Lambda^2}{s}^z \frac{d z'}{z'} \, \int d^2 x_2 \, \frac{\un x_{21} \times \un x_{20}}{x_{21}^2 \, x_{20}^2} \, \left\{  G^{[3]}_{21} (z's) +  G^{[3]}_{20} (z's)  \right\} ,
\label{Q_evol_2}
\end{align}
where we have defined another generalized dipole amplitude
\begin{align}
    \overline{\Gamma}^{\mathrm{gen}}_{10, 32}(zs) &\equiv Q_{10}(zs) \,\theta ( x_{32} - x_{10} ) + \overline{\Gamma}_{10,32}(zs) \, \theta(x_{10}- x_{32}).
\end{align}
We have also verified that the contribution of $V^{\mathrm{G}[3]}_{\un 2}$ to evolution of $Q_{10}$ is zero (see diagram III of Fig.~3 in \cite{Cougoulic:2022gbk}, where a soft quark is emitted and absorbed by a dipole). 

Since the gluon operator ($V^{\mathrm{G} [1]}_{\un x}$) parts of $Q_{10} (s)$ and $\widetilde{G}_{10} (s)$ are identical, and, as we mentioned above, type-3 terms do not affect the diagrams with a soft quark emission, the terms in \eq{Q_evol2} also appear in the evolution for $\widetilde{G}_{10}$. Adding \eq{Q_evol2} to Eq.~(150) of \cite{Cougoulic:2022gbk}, we arrive at the evolution equation for $\widetilde{G}_{10}$ (again, for $S=1$):
\begin{align}\notag
 {\widetilde G}_{10} (zs) &=  {\widetilde G}_{10}^{(0)}(zs) \\
& + \frac{\as N_c}{2\pi^2} \, \int\limits_{\frac{\Lambda^2}{s}}^z \frac{d z'}{z'} \, \int d^2 x_2 \, \Bigg\{ 2 \left[ \frac{1}{x_{21}^2} -  \frac{{\un x}_{21}}{x_{21}^2} \cdot \frac{{\un x}_{20}}{x_{20}^2} \right]  \left[ {\widetilde G}_{21}(z's) +  {\widetilde \Gamma}^{\text{gen}}_{20,21}(z's)  \right] \notag  \\ 
& +  \left[ 2 \frac{\epsilon^{ij} \, x_{21}^j}{x_{21}^4} - \frac{\epsilon^{ij} \, (x_{20}^j + x_{21}^j)}{x_{20}^2 \, x_{21}^2}  - \frac{2 \, {\un x}_{20} \times {\un x}_{21}}{x_{20}^2 \, x_{21}^2} \left( \frac{x_{21}^i}{x_{21}^2} - \frac{x_{20}^i}{x_{20}^2}\right) \right] \, \left[ G^{i}_{21} (z's) + \Gamma^{i\,\text{gen}}_{20,21} (z's)  \right]  \Bigg\} \notag \\
& - \frac{\as \, N_f}{8 \pi^2 } \int\limits_{\frac{\Lambda^2}{s}}^z \frac{d z'}{z'}  \int d^2 x_2 \, \Bigg\{ \frac{1}{x_{21}^2} \, \overline{\Gamma}^{\text{gen}}_{20,21}(z's) +   \frac{2\epsilon^{ij} \, {\un x}_{21}^j}{x_{21}^4} \, \Gamma^{i \, \textrm{gen}}_{20,21}(z's)  \Bigg\}  \notag \\
& + \frac{\as N_c}{2 \pi^2} \, \int\limits_{\frac{\Lambda^2}{s}}^z \frac{d z'}{z'} \, \int d^2 x_2 \, \frac{x_{10}^2}{x_{21}^2 \, x_{20}^2}  \, \Bigg\{  {\widetilde G}_{12}(z's)  -  {\widetilde  \Gamma}^{\text{gen}}_{10,21}(z's)  \Bigg\} \,
\notag \\ & 
+ \frac{\as \, N_c}{2 \pi^2} \, \int\limits_\frac{\Lambda^2}{s}^z \frac{d z'}{z'} \, \int d^2 x_2 \, \frac{\un x_{21} \times \un x_{20}}{x_{21}^2 \, x_{20}^2} \, \left[  G^{[3]}_{21} (z's) +  G^{[3]}_{20} (z's)  \right].
\label{G_adj_evol_2.5}
\end{align}
However, this is not our final equation for ${\widetilde G}_{10} (s)$. As we will see in the next Section, \eq{G_adj_evol_2.5} also receives corrections from $\widetilde{Q}_{10}$ and $\widetilde{L}_{10}$, with the latter generating a contribution from the corresponding moment amplitude $\widetilde{I}$. 

\begin{figure}[ht]
\centering
\includegraphics[width= \textwidth]{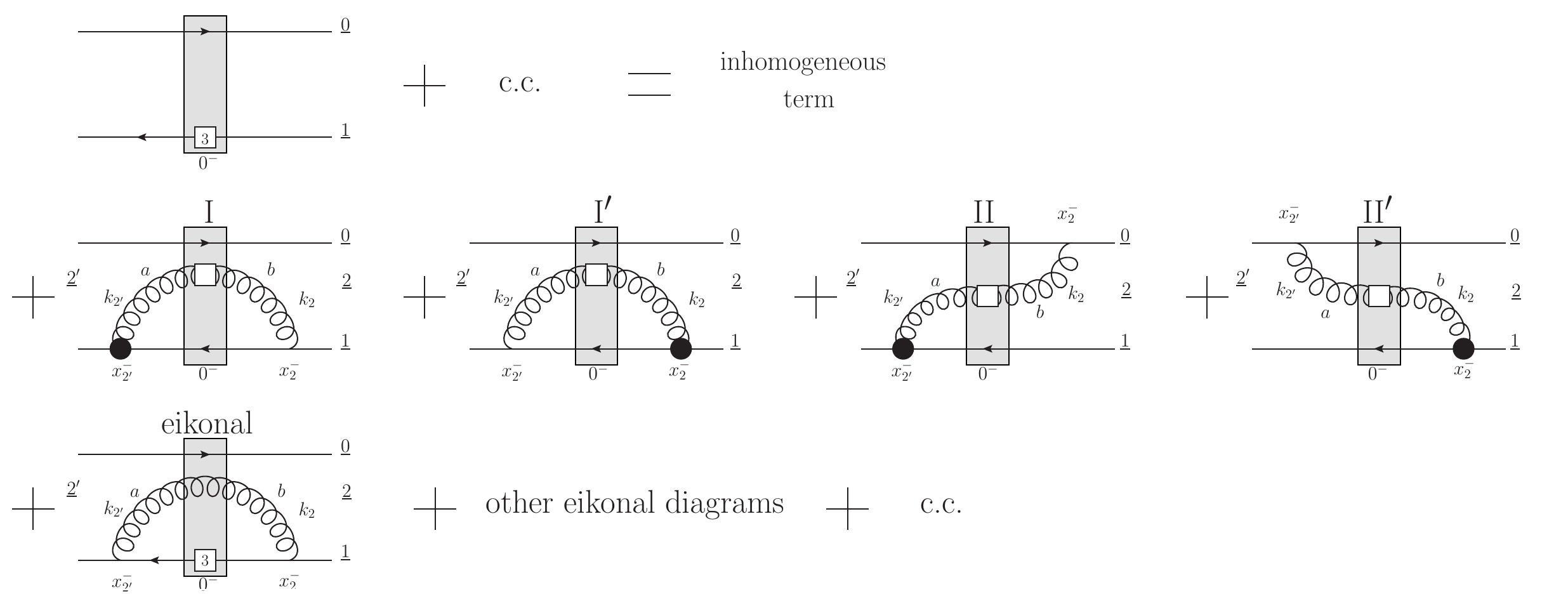}
\caption{Diagrams contributing to the small-$x$ evolution of the polarized dipole amplitude $G^{[3]}_{10} (s)$. The vertical shaded rectangle represents the target (proton) shock wave. The square white box on the gluon and quark lines represents the sub-eikonal interaction with the target: in the diagrams I, I', II, and II' it is given by the propagators in \eqs{dmapap}. All momenta flow to the right.}
\label{FIG:G3_evol}
\end{figure}

Finally, we turn now to the evolution of $G^{[3]}_{10}$ itself. The corresponding one-loop diagrams are shown in \fig{FIG:G3_evol}. They are similar to those in Figs.~3 and 5 of \cite{Cougoulic:2022gbk}, although now some of the gray boxes denote the insertion of the $F^{+-}$ operator from \eqref{VG3}, as indicated by the number 3 in those boxes. 
To compute the evolution of the $F^{+-}$ operator, we need the following propagators (which can be straightforwardly derived following the technique in \cite{Kovchegov:2017lsr, Kovchegov:2018znm, Cougoulic:2022gbk, Li:2023tlw, Kovchegov:2025gcg}):
\begin{subequations}\label{dmapap}
\begin{align}\label{Vi555}
& \int\limits_{-\infty}^0 dx_{2'}^- \,  
\int\limits_0^\infty dx_2^- \, \big[ \pd^- %
\contraction[2ex]
{}
{a}{^{+ \, a}
(x_{2'}^- , \ul{x}_1) \big]  \:}
{a}
\: 
a^{+ \, a} (x_{2'}^- , \ul{x}_1) \big] \:
a^{+ \, b} (x_2^- , \ul{x}_0)  
= \frac{1}{4 \pi^3} \int\limits_0^{p_2^-} d k^-  \int d^2 x_2 \, \left\{ \frac{\un x_{21}}{x_{21}^2} \times \frac{\un x_{20}}{x_{20}^2} \, \left( U_{\un 2}^{\textrm{pol} [1]} \right)^{ba} \right. \notag \\
& 
\hspace{5cm}
\left. - \left[ \left( \pd_{20}^i - \pd_{21}^i \right) \frac{\un x_{21}}{x_{21}^2} \cdot \frac{\un x_{20}}{x_{20}^2} \right] \, \left( U_{\un 2}^{i \, \textrm{G} [2]} \right)^{ba} +  \frac{\un x_{21}}{x_{21}^2} \cdot \frac{\un x_{20}}{x_{20}^2}  \, \left( U_{\un 2}^{\textrm{G} [3]} \right)^{ba} \right\}, \\
\label{Vi666}
& \int\limits_{-\infty}^0 dx_{2'}^- \,  
\int\limits_0^\infty dx_2^- \, \big[ \pd^-
\contraction[2ex]
{}
{a}{^{+ \, b}
(x_2^- , \ul{x}_1) \big] \:}
{a}
\: 
a^{+ \, b} (x_2^- , \ul{x}_1) \big] \:
a^{+ \, a} (x_{2'}^- , \ul{x}_0) = \frac{1}{4 \pi^3} \int\limits_0^{p_2^-} d k^-  \int d^2 x_2 \, \left\{ \frac{\un x_{21}}{x_{21}^2} \times \frac{\un x_{20}}{x_{20}^2} \, \left( U_{\un 2}^{\textrm{pol} [1]} \right)^{ba} \right. \notag \\
& \hspace{5cm} \left. - \left[ \left( \pd_{20}^i - \pd_{21}^i \right) \frac{\un x_{21}}{x_{21}^2} \cdot \frac{\un x_{20}}{x_{20}^2} \right] \, \left( U_{\un 2}^{i \, \textrm{G} [2]} \right)^{ba} +  \frac{\un x_{21}}{x_{21}^2} \cdot \frac{\un x_{20}}{x_{20}^2}  \, \left( U_{\un 2}^{\textrm{G} [3]} \right)^{ba} \right\},
\end{align}
\end{subequations}
where $U^{\mathrm{pol}[1]}_{\un 2} =U^{\mathrm{q}[1]}_{\un 2}  + U^{\mathrm{G}[1]}_{\un 2}  $ and $U^{\mathrm{i\, \mathrm{G}[2]}}_{\un 2}$ are adjoint versions of the polarized Wilson lines in \eqref{VG1}, \eqref{Vq1}, and \eqref{d-d_pw}, respectively. Employing these propagators, we can find the contributions of the diagrams I, I', II, and II' in \fig{FIG:G3_evol}. The standard eikonal propagator \cite{Mueller:1994rr,Mueller:1994jq,Mueller:1994gb,Balitsky:1995ub,Balitsky:1998ya,Kovchegov:1999yj,Kovchegov:1999ua} gives the contribution of the graphs in the last line of \fig{FIG:G3_evol}. Assembling everything together we arrive at the equation for $G^{[3]}_{10}$, 
\begin{align}\label{G3_evol1}
G^{[3]}_{10} (zs) = & \, \frac{\as \, N_c}{2 \pi^2} \, \int\limits_\frac{\Lambda^2}{s}^z \frac{d z'}{z'} \, \int d^2 x_2 \, \left\{ - 2 \, \frac{\un x_{21}}{x_{21}^2} \times \frac{\un x_{20}}{x_{20}^2} \,  \left[ \widetilde{G}_{21} (z's) + \widetilde{G}_{20} (z's)  \right] 
\notag \right.\\  & \left.
+ \left[ \left( \pd_{20}^i - \pd_{21}^i \right) \frac{\un x_{21}}{x_{21}^2} \cdot \frac{\un x_{20}}{x_{20}^2} \right] \, \left[ G^i_{21} (z's) + G^i_{20} (z's) \right] \right. \notag \\
& - \left. \left( \frac{\un x_{21}}{x_{21}^2} \cdot \frac{\un x_{20}}{x_{20}^2}  - \frac{1}{x_{21}^2} \right) \, \left[ G^{[3]}_{21} (z's) + \Gamma^{[3]}_{20, 21} (z's) \right]  + \frac{x_{10}^2}{x_{21}^2 \, x_{20}^2} \, \left[ G^{[3]}_{12} (z's) - \Gamma^{[3]}_{10, 21} (z's) \right] \right\}.
\end{align}
where we again employ the linearized approximation where unpolarized $S$-matrices are set to $1$. In arriving at the terms involving $\widetilde{G}_{10}$ and $G^i_{10}$ in \eq{G3_evol1}, we have used the fact that there are no ultraviolet (UV) divergences in the kernels: this implies that no neighbor amplitudes are needed in those terms. This is not the case for the last two terms on the right-hand side of \eq{G3_evol1}, which is why we keep the neighbor amplitude of $G^{[3]}_{10}$, denoted by $\Gamma^{[3]}_{10, 21}$. Importantly, let us note that \eq{G3_evol1} has no inhomogeneous term on the right: as we will argue below, one can show that the classical initial condition for $G^{[3]}$ is zero, $G^{[3] \, (0)}_{10} (zs) =0$, to all orders in $\as$ at the sub-eikonal level.

We now have equations \eqref{Gi_largeNc_evol}, \eqref{Q_evol_2}, \eqref{G_adj_evol_2.5}, and \eqref{G3_evol1} for the polarized dipole amplitudes $Q_{10}(s), \widetilde{G}_{10}(s),$ and $G^i_{10}(s)$, along with $G^{[3]}_{10} (s)$. While, as we mentioned above, \eq{G_adj_evol_2.5} will receive a correction due to the quark to gluon transition operators from \cite{Borden:2024bxa}, this correction will not affect the mixing between $G^{[3]}_{10} (s)$ and the other dipole amplitudes. This allows us to make an observation here. If we were to integrate \eq{G3_evol1} over all impact parameters (using \eqs{integrated}) and define $G^{[3]} (x^2_{10} , s)$ by
\begin{equation}\label{G3_int_def}
\int d^2 x_1 \, G^{[3]}_{10} (s) \equiv G^{[3]} (x^2_{10} , s) ,  
\end{equation} 
we would see that the first two terms in the curly brackets of \eq{G3_evol1} are not DLA. Therefore, the impact-parameter integrated version of \eq{G3_evol1} only contains terms which couple $G^{[3]} (x^2_{10} , s)$ to itself. Conversely, integrating \eqs{Gi_largeNc_evol}, \eqref{Q_evol_2}, and \eqref{G_adj_evol_2.5} over all impact parameters (while keeping only the $G_2$ term on the left-hand side of \eq{Gi_largeNc_evol}) we would obtain equations in which the contribution of the $G^{[3]} (x^2_{10} , s)$ amplitudes on the right is not DLA and can, therefore, be neglected. Hence $G^{[3]} (x^2_{10} , s)$ does not mix with $Q (x^2_{10} , s)$, $G_2 (x^2_{10} , s)$, and $\widetilde{G} (x^2_{10} , s)$, at least in DLA. The latter amplitude is defined by
\begin{equation}
        \label{Gt_def}
    \int d^2 x_1 \, \tG_{10}(s) \equiv \widetilde{G} (x_{10}^2, s) .
\end{equation}
One can also show that $G^{[3]} (x^2_{10} , s)$ does not mix with $\widetilde{Q} (x^2_{10} , s)$. We thus confirm the observation made in \cite{Kovchegov:2025gcg} that $G^{[3]}$ does not contribute to the DLA helicity evolution of \cite{Kovchegov:2015pbl, Kovchegov:2018znm, Cougoulic:2022gbk, Borden:2024bxa}. 

However, for the moment amplitudes, the situation changes. To take an impact parameter moment, we define $I_E$ and $J_E$ by\footnote{The subscript $E$ refers to the chromo-electric field, whose $z$-component is captured by the $F^{+-}$ operator.}
\begin{align} \label{moment3}
    \int d^2 x_1 \, x_1^m \, G^{[3]}_{10}(zs) \equiv \epsilon^{mk} x_{10}^k \, I_E(x_{10}^2, zs) + x_{10}^m \, J_E(x_{10}^2, zs).
\end{align}
Counting the Levi-Civita symbols in \eqs{Gi_largeNc_evol}, \eqref{Q_evol_2}, and \eqref{G_adj_evol_2.5}, while taking into account the definitions \eqref{moments}, we see that the moment $I_E$ may mix with the other moments ($I_3, I_4$, and $I_5$) in DLA. Focusing on the structures which contribute to the evolution of $I_E$, we multiply \eq{G3_evol1} by $x_1^m$ and integrate over all $x_1$ (keeping inter-partonic separations $\un x_{21}, \un x_{20}, \un x_{10}$ fixed) using \eqs{integrated}, \eqref{moments}, \eqref{Gt_def} and \eqref{moment3}. The first two terms on the right-hand side of \eq{G3_evol1} now have IR divergences ($x_{20} \sim x_{21} \gg x_{10}$), the third term has both IR and UV ($x_{21} \ll x_{20} \sim x_{10}$) divergences, and the last term on the right-hand side of \eq{G3_evol1} still has only a UV divergence. Keeping only the DLA divergent terms, we arrive at the evolution equation for the moment amplitude $I_E$, 
\begin{align} 
\label{IE_evol1}
I_E (x^2_{10}, zs) =  & \, \frac{\as \, N_c}{2 \pi} \, \iIR \,  \bigg\{ - 2 \left[ \widetilde{I}_3 (x^2_{21} , z's) - \widetilde{G} (x^2_{21} , z's) \right] + 2 \, I_4 (x^2_{21} , z's) - I_5 (x^2_{21} , z's) \notag \\ 
&  + 3 \, G_2 (x^2_{21} , z's)  +  I_E (x^2_{21} , z's) \bigg\}.
\end{align}
$\widetilde{I}_3$ is the moment amplitude associated with $\widetilde{G}_{10}$ defined by 
\begin{equation}
        \label{I3t_def}
    \int d^2 x_1 \,x^m_1 \tG_{10}(s) \equiv x_{10}^m \, \widetilde{I}_3(x_{10}^2, s)+ \cdots.
\end{equation}
Note the amplitude $J_E$ and its neighbor dipole amplitude do not contribute to the moment evolution below since they couple to different tensor structures than the moment amplitudes relevant to the OAM distributions. Additionally, in arriving at \eq{IE_evol1}, we have employed $x^-$ lifetime ordering \cite{Kovchegov:2015pbl, Cougoulic:2019aja, Cougoulic:2022gbk} and required that all dipole sizes be smaller than $1/\Lambda$, with $\Lambda$ an IR cutoff. 

In writing down \eqs{G3_evol1} and \eqref{IE_evol1}, we have used the fact that the quasi-classical initial condition for $G^{[3]}_{10}$ is zero to all orders in $\alpha_s$ at the sub-eikonal level. Indeed, for the classical gluon field of the target $\partial^- \sim k^- \sim \perp^2/ (2P^+)$, where $\perp$ denotes a generic transverse momentum scale. Furthermore, for the amplitude $G^{[3]}_{10}$ to be non-zero, one of the $A^+$ fields in its definition must couple to the proton spin. Both of these 
represent a sub-eikonal contribution, and therefore the initial conditions of $G^{[3]}_{10}$ and $I_E$ are sub-sub-eikonal. We neglect such contributions in our sub-eikonal calculation. 

 We see that the new amplitude $I_E$ couples to both the helicity and moment sector through evolution, with its novel DLA evolution equation given in \eq{IE_evol1}. In the next Section, we will also show that $I_E$ enters the evolution of the other moment amplitudes and therefore contributes to the OAM distributions.


\subsection{Evolution equation for $\widetilde{I}$}

We turn now to constructing the evolution equations for the remaining moment amplitudes $I_3, I_4, I_5, \widetilde{I}_3$, and $\widetilde{I}$ in the large-$N_c\&N_f$ limit. The evolution equations for $I_3, I_4,$ and $I_5$ in the large-$N_c$ limit were derived in \cite{Kovchegov:2023yzd} (see also Appendix A of \cite{Kovchegov:2024wjs}). However, note that these earlier derivations did not include contributions from $I_E$, which we include below. 

\begin{figure}[ht]
\centering
\includegraphics[width= 0.95 \textwidth]{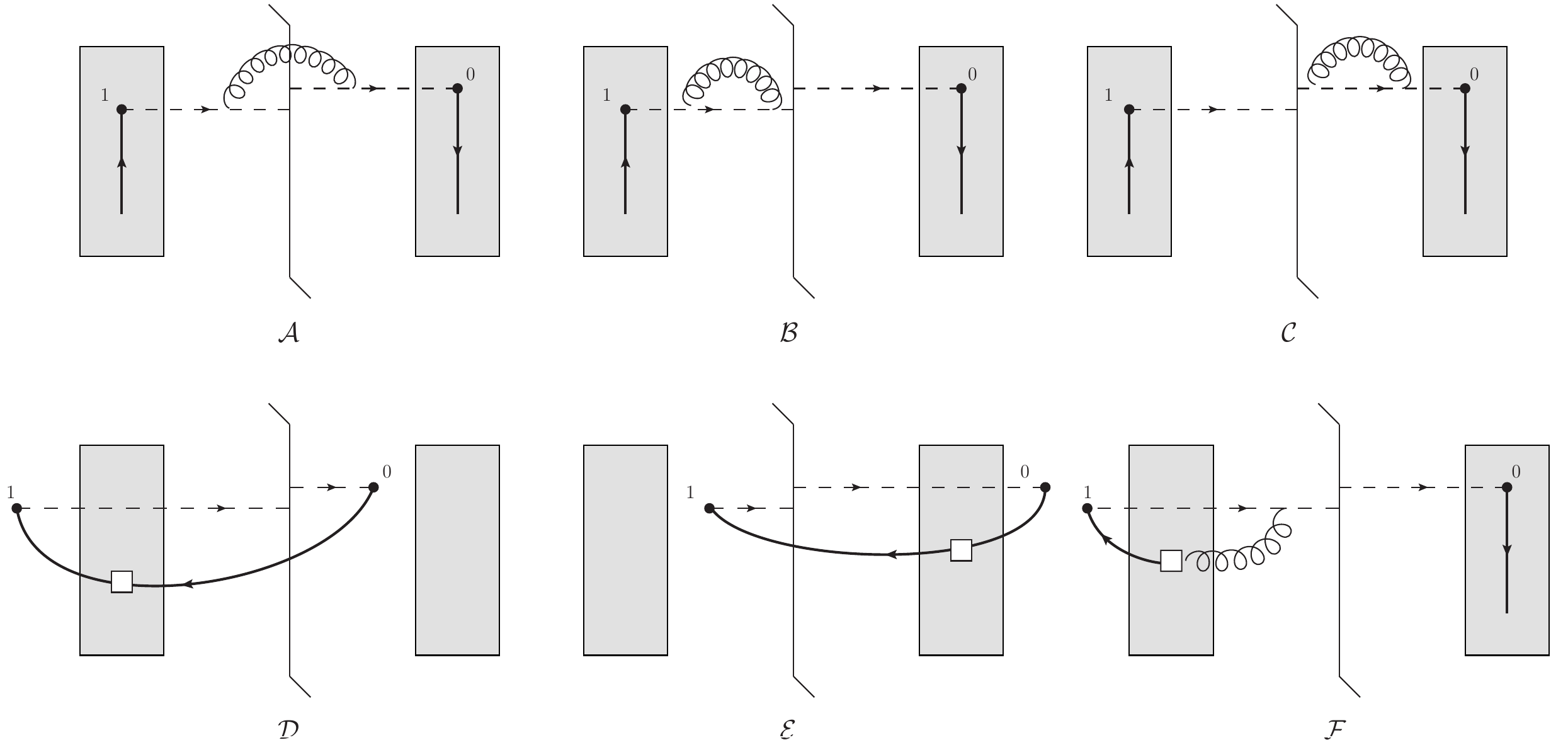} \\~~\\
\hspace*{3mm} \includegraphics[width= 0.95 \textwidth]{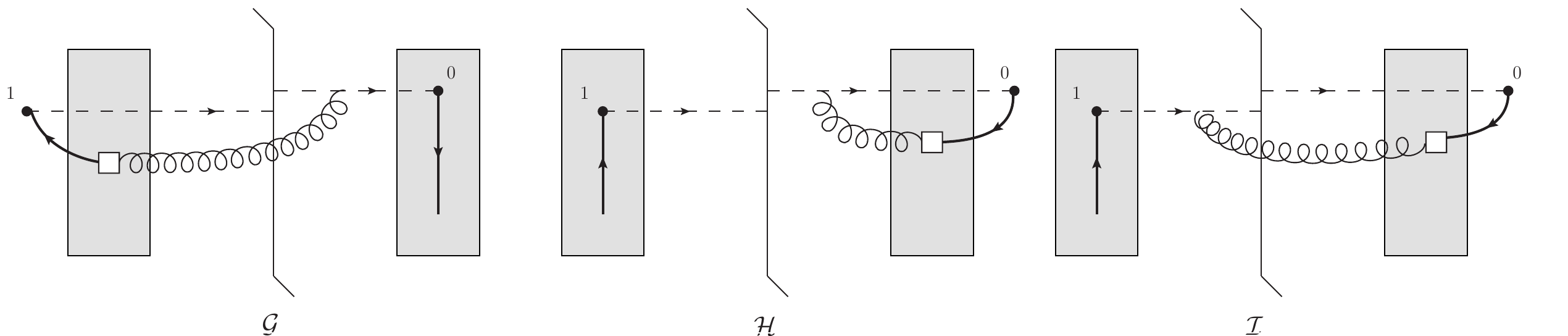}
\caption{Diagrams contributing to the evolution of ${\widetilde L}_{10} (zs)$ calculated here. The shaded rectangles denote the target shockwave, the dashed lines denote the classical Wilson lines, and the gray boxes denote the sub-eikonal quark-to-gluon (or gluon-to-quark) transition operators from \cite{Borden:2024bxa}. We omit some of the proton spin-independent diagrams which do not contribute.}
\label{FIG:Ltilde_evol}
\end{figure}

We begin with the evolution of $\widetilde{I}$. The diagrams contributing to the one-loop evolution of ${\widetilde L}_{10}$ (and therefore $\widetilde{I}$) are shown in \fig{FIG:Ltilde_evol} using the nomenclature of Fig.~3 from  \cite{Borden:2024bxa}. Except now we are taking note of the absence of the time-ordering (T) symbol in \eq{Ltilde} and are including the final state cut explicitly. The diagrams in \fig{FIG:Ltilde_evol} correspond to the evolution of the first term (quark OAM) in \eq{Ltilde}; the second (antiquark) term is obtained from the diagrams shown by reversing the direction of the particle number flow. The diagrams with the Drell-Yan staple (initial state interaction) are not shown either, but are evaluated similarly: their contributions are included into the results quoted below.  

Following \cite{Borden:2024bxa} one can readily calculate (including the contributions of all the terms in \eq{Ltilde}) 
\begin{align}
\notag 
& {\cal A} + {\cal B} + {\cal C} + {\cal F} + {\cal G} + {\cal H} + {\cal I} \\ 
& = - \frac{\as \, N_c}{4 \pi^2} \, \int\limits_\frac{\Lambda^2}{s}^{z} \frac{d z'}{z'} \, \int d^2 x_2 \, \Bigg[ \frac{x_{10}^2}{x_{21}^2 \, x_{20}^2} \, {\widetilde L}_{10} (z' s) - \left( \frac{1}{x_{21}^2} -  \frac{{\un x}_{21} \cdot {\un x}_{20}}{x_{21}^2 \, x_{20}^2} \right)  \, {\widetilde L}_{20} (z' s) + i \, \frac{{\un x}_{21} \times {\un x}_{20}}{x_{21}^2 \, x_{20}^2} \, {\widetilde Q}_{20} (z' s) \notag \\
& \hspace{8cm}
- \left( \frac{1}{x_{20}^2} -  \frac{{\un x}_{21} \cdot {\un x}_{20}}{x_{21}^2 \, x_{20}^2} \right)  \, {\widetilde L}_{12} (z' s)  + i \, \frac{{\un x}_{21} \times {\un x}_{20}}{x_{21}^2 \, x_{20}^2} \, {\widetilde Q}_{12} (z' s)  \Bigg] .
\label{ABCFGHI}
\end{align}
This result is very similar to that in \cite{Borden:2024bxa}, where the cross-product terms were neglected as not DLA. Indeed, none of the terms in \eq{ABCFGHI} are DLA for ${\widetilde L}_{10}$ evolution, though for the moment amplitude $\widetilde{I}$ the situation will change, as we will see shortly.\footnote{The terms multiplying ${\widetilde L}$ on the right-hand side of \eq{ABCFGHI} contain DLA UV contributions: in those UV regions, strictly speaking ${\widetilde L}$ should be replaced by a ``neighbor" amplitude which we could label $\widetilde{\Gamma}_{L}$. However, just as in \cite{Borden:2024bxa}, these UV contributions cancel out in \eq{ABCFGHI}, making such a replacement unnecessary.} Note the type-3 polarized dipole amplitude in \eq{G3_def} does not contribute to \eq{ABCFGHI}. 

Next, we need to calculate diagrams $\cal D$ and $\cal E$ along with their contributions to antiquark OAM, obtained by reversing the particle number flow in all the lines. The latter are labeled $\cal D'$ and $\cal E'$. We obtain
\begin{subequations}\label{D_D_E_E}
\begin{align}
& {\cal D} = - i \, \frac{\as}{8 \pi^2} \,  \int\limits_\frac{\Lambda^2}{s}^{z} \frac{d z'}{z'} \, \int d^2 x_2 \, \Bigg[ \frac{{\un x}_{21} \times {\un x}_{20}}{x_{21}^2 \, x_{20}^2} \, \llangle \tr \left[ V_{\un 1} \, V_{\un 2}^{\textrm{pol} [1] \dagger} \right] \rrangle + \left( \frac{x_{21}^j}{x_{21}^2} \,  \frac{\delta^{ij} \, x_{20}^2 - 2 x_{20}^i \, x_{20}^j}{x_{20}^4}  - (1 \leftrightarrow 0) \right) \llangle \tr \left[ V_{\un 1} \, V_{\un 2}^{i \textrm{G} [2] \dagger} \right] \rrangle \Bigg], \\
& {\cal D}' = - i \, \frac{\as}{8 \pi^2} \,  \int\limits_\frac{\Lambda^2}{s}^{z} \frac{d z'}{z'} \, \int d^2 x_2 \, \Bigg[ \frac{{\un x}_{21} \times {\un x}_{20}}{x_{21}^2 \, x_{20}^2} \, \llangle \tr \left[ V_{\un 1}^\dagger \, V_{\un 2}^{\textrm{pol} [1]} \right] \rrangle + \left( \frac{x_{21}^j}{x_{21}^2} \,  \frac{\delta^{ij} \, x_{20}^2 - 2 x_{20}^i \, x_{20}^j}{x_{20}^4}  - (1 \leftrightarrow 0) \right) \llangle \tr \left[ V_{\un 1}^\dagger \, V_{\un 2}^{i \textrm{G} [2]} \right] \rrangle \Bigg], \\
& {\cal E} = - i \, \frac{\as}{8 \pi^2} \,  \int\limits_\frac{\Lambda^2}{s}^{z} \frac{d z'}{z'} \, \int d^2 x_2 \, \Bigg[ \frac{{\un x}_{21} \times {\un x}_{20}}{x_{21}^2 \, x_{20}^2} \, \llangle \tr \left[ V_{\un 0}^\dagger \, V_{\un 2}^{\textrm{pol} [1]} \right] \rrangle + \left( \frac{x_{21}^j}{x_{21}^2} \,  \frac{\delta^{ij} \, x_{20}^2 - 2 x_{20}^i \, x_{20}^j}{x_{20}^4}  - (1 \leftrightarrow 0) \right) \llangle \tr \left[ V_{\un 0}^\dagger \, V_{\un 2}^{i \textrm{G} [2]} \right] \rrangle \Bigg], \\
& {\cal E}' = - i \, \frac{\as}{8 \pi^2} \,  \int\limits_\frac{\Lambda^2}{s}^{z} \frac{d z'}{z'} \, \int d^2 x_2 \, \Bigg[ \frac{{\un x}_{21} \times {\un x}_{20}}{x_{21}^2 \, x_{20}^2} \, \llangle \tr \left[ V_{\un 0} \, V_{\un 2}^{\textrm{pol} [1] \dagger} \right] \rrangle + \left( \frac{x_{21}^j}{x_{21}^2} \,  \frac{\delta^{ij} \, x_{20}^2 - 2 x_{20}^i \, x_{20}^j}{x_{20}^4}  - (1 \leftrightarrow 0) \right) \llangle \tr \left[ V_{\un 0} \, V_{\un 2}^{i \textrm{G} [2] \dagger} \right] \rrangle \Bigg].
\end{align}
\end{subequations}
We note that we have included the contributions of the DY staples into \eqs{D_D_E_E}: the expression for diagram $\cal D$ includes the contribution of diagram $\cal E'$ with the DY staple, obtained by replacing $V_{\un 0} \to V_{\un 1}$ in the latter (which simply doubles $\cal D$), the expression for diagram $\cal E$ includes the contribution of diagram $\cal D'$ with the DY staple, etc. Adding \eqs{D_D_E_E} up yields (cf. Eq.~(58) in \cite{Borden:2024bxa})
\begin{align}
\notag
 {\cal D} + {\cal D}' + {\cal E} + {\cal E}' = & \,  - i \, \frac{\as \, N_c}{4 \pi^2} \,  \int\limits_\frac{\Lambda^2}{s}^{z} \frac{d z'}{z'} \, \int d^2 x_2 \, \Bigg\{ \frac{{\un x}_{21} \times {\un x}_{20}}{x_{21}^2 \, x_{20}^2} \, \left[ Q_{21} (z' s) + Q_{20} (z' s) \right] \\
&  + \left( \frac{x_{21}^j}{x_{21}^2} \,  \frac{\delta^{ij} \, x_{20}^2 - 2 x_{20}^i \, x_{20}^j}{x_{20}^4}  - \frac{x_{20}^j}{x_{20}^2} \,  \frac{\delta^{ij} \, x_{21}^2 - 2 x_{21}^i \, x_{21}^j}{x_{21}^4} \right) \, \left[ G^i_{21} (z' s) + G^i_{20} (z' s)  \right]  \Bigg\} . 
\label{DDEE}
\end{align}
As one can readily verify, the type-3 operator \eqref{VG3} does not contribute to the diagrams ${\cal D}, {\cal D}', {\cal E}$ and ${\cal E}'$.
Combining \eq{ABCFGHI} with \eq{DDEE} we obtain the evolution equation for ${\widetilde L}_{10} (s)$:
\begin{align}
\notag 
{\widetilde L}_{10} & (z s) =  \, {\widetilde L}_{10}^{(0)} (z s)  - \frac{\as \, N_c}{4 \pi^2} \, \int\limits_\frac{\Lambda^2}{s}^{z} \frac{d z'}{z'} \, \int d^2 x_2 \, \Bigg[ \frac{x_{10}^2}{x_{21}^2 \, x_{20}^2} \, {\widetilde L}_{10} (z' s) - \left( \frac{1}{x_{21}^2} -  \frac{{\un x}_{21} \cdot {\un x}_{20}}{x_{21}^2 \, x_{20}^2} \right)  \, {\widetilde L}_{20} (z' s)  \\
&  - \left( \frac{1}{x_{20}^2} -  \frac{{\un x}_{21} \cdot {\un x}_{20}}{x_{21}^2 \, x_{20}^2} \right)  \, {\widetilde L}_{12} (z' s)  + i \, \frac{{\un x}_{21} \times {\un x}_{20}}{x_{21}^2 \, x_{20}^2} \, \left[ {\widetilde Q}_{20} (z' s) + {\widetilde Q}_{12} (z' s) \right] + i \, \frac{{\un x}_{21} \times {\un x}_{20}}{x_{21}^2 \, x_{20}^2} \, \left[ Q_{21} (z' s) + Q_{20} (z' s) \right]   \notag   \\
&  + i \, \left( \frac{x_{21}^j}{x_{21}^2} \,  \frac{\delta^{ij} \, x_{20}^2 - 2 x_{20}^i \, x_{20}^j}{x_{20}^4}  - \frac{x_{20}^j}{x_{20}^2} \,  \frac{\delta^{ij} \, x_{21}^2 - 2 x_{21}^i \, x_{21}^j}{x_{21}^4} \right) \, \left[ G^i_{21} (z' s) + G^i_{20} (z' s)  \right]   \Bigg] . 
\label{Ltilde_evol}
\end{align}
Here, as with all the DLA evolution, the inhomogeneous term ${\widetilde L}_{10}^{(0)} (z s)$ represents the initial condition for evolution. From \eq{Ltilde_evol} we clearly see that ${\widetilde L}_{10} (s)$ is anti-symmetric and imaginary (if ${\widetilde Q}_{10} (s)$ is real and symmetric, as follows from the definition \eqref{Qtilde}, as we will see shortly). Also, if one simply integrates over impact parameters, none of the terms in \eq{Ltilde_evol} are DLA.

On the other hand, the equation for $\widetilde{I}$ follows from multiplying \eq{Ltilde_evol} by $x_1^m$, integrating over all $x_1$, using \eqs{integrated}, \eqref{moments}, \eqref{L_mom}, and  \eqref{Qint_def}, equating the resulting tensor structures, and extracting the DLA contributions in the kernel. We end up with
\begin{align}
\notag
& {\widetilde I}(x^2_{10},  z s)  = {\widetilde I}^{(0)} (x^2_{10},  z s)  - \frac{\as \, N_c}{4 \pi} \, \iIR \, \Bigg[ - {\widetilde I} (x^2_{21},  z' s) + I_{3} (x^2_{21},  z' s) - Q  (x^2_{21},  z' s) \\
& +  {\widetilde I}' (x^2_{21},  z' s) - {\widetilde Q}(x^2_{21},  z' s) - 2 \, I_4 (x^2_{21},  z' s) + I_5 (x^2_{21},  z' s) - 3 \, G_2 (x^2_{21},  z' s)  \Bigg],
\label{Itilde_evol}
\end{align}
where we have also employed \eq{b0}. ${\widetilde I}^{(0)} (x^2_{10},  z s)$ is determined by the initial conditions. Following \cite{Kovchegov:2024wjs}, we have also defined ${\widetilde I}'$ by
\begin{align}\label{Itilde_def}
        \int d^2 x_1 \, x_1^m \, \widetilde{Q}_{10}(s) = x_{10}^m \, \widetilde{I}' (x_{10}^2, s) + \epsilon^{mn} \, x_{10}^n \, \widetilde{J}' (x_{10}^2, s)
\end{align}
using $\widetilde{Q}_{10}$ from \eq{Qtilde}. Note that the object $\widetilde{Q}_{10}$ is symmetric, $\widetilde{Q}_{10}(s) = \widetilde{Q}_{01}(s)$. This implies that 
\begin{align} \notag 
     x_{10}^m \, \widetilde{I}' (x_{10}^2, s) + \epsilon^{mn} \, x_{10}^n \, \widetilde{J}' (x_{10}^2, s) &= \int d^2 x_1 \, x_1^m \, \widetilde{Q}_{10}(s) 
    \\ \notag &= \int d^2 x_1 \, (x_0^m + x_{10}^m)  \, \widetilde{Q}_{10}(s) 
    \\ \notag &= \int d^2 x_0 \, (x_0^m + x_{10}^m)  \, \widetilde{Q}_{01}(s)
    \\ & = x_{01}^m \, \widetilde{I}' (x_{10}^2, s) + \epsilon^{mn} \, x_{01}^n \, \widetilde{J}' (x_{10}^2, s) + x_{10}^m \, \widetilde{Q} (x_{10}^2, s)\,.
    \label{IJQ2}
\end{align}
Equating different tensor structures in \eq{IJQ2}, we obtain $\widetilde{J}' =0$ and, more importantly, 
\begin{align}\label{half}
    \widetilde{I}' (x_{10}^2, s) = \half \, \widetilde{Q} (x_{10}^2, s). 
\end{align}
Equation \eqref{half} allows us to simplify \eq{Itilde_evol} to 
\begin{align}
\notag
& {\widetilde I} (x^2_{10},  z s)  = {\widetilde I}^{(0)} (x^2_{10},  z s)  - \frac{\as \, N_c}{4 \pi} \, \iIR \, \Bigg[ - {\widetilde I} (x^2_{21},  z' s) + I_{3} (x^2_{21},  z' s) - Q (x^2_{21},  z' s) \\
& \hspace*{7cm}
- \half \, {\widetilde Q} (x^2_{21},  z' s) - 2 \, I_4 (x^2_{21},  z' s) + I_5 (x^2_{21},  z' s) - 3 \, G_2 (x^2_{21},  z' s)  \Bigg] .
\label{Itilde_evol2}
\end{align}
Equation~\eqref{Itilde_evol2} should be compared to Eq.~(30) of Ref.~\cite{Kovchegov:2024wjs}. The difference is the presence of $\widetilde I$ and $\widetilde Q$ on the right-hand side of \eq{Itilde_evol2}. The sign difference between our \eq{qOAM4} and Eq.~(31) from Ref.~\cite{Kovchegov:2024wjs} accounts for $I_{3}, I_4, I_5, Q$ and $G_2$ entering \eq{Itilde_evol2} with the different sign when compared to Eq.~(30) in \cite{Kovchegov:2024wjs}. 

We see that \eq{Itilde_evol2} is a novel evolution equation for the new moment amplitude ${\widetilde I}$. It will have to be combined with the evolution equations for the other moment amplitudes, whose derivation we present next. 


\subsection{Evolution equations for $I_3, \widetilde{I}_3, I_4$, and $I_5$}

For the remaining moment amplitudes $I_3, I_4, I_5$, and $\widetilde{I}_3$, we first recall their definitions from \eqs{moments} and \eqref{I3t_def} above. In addition to the moments $I_4, I_5$ of the dipole amplitude $G^i_{10}$, for the evolution equations we need to define the moments $\Gamma_4, \Gamma_5$ of the neighbor amplitude $\Gamma^i_{10, 21}$ of the second type. We write~\cite{Kovchegov:2023yzd},
\begin{align}\label{neigh_2_defs}
        \int d^2 x_1 \, x_1^m \,\Gamma^i_{10, 21}(zs) \equiv \epsilon^{mi} x_{10}^2 \, \Gamma_4(x_{10}^2, x_{21}^2, zs) + \epsilon^{mk} x_{10}^k x_{10}^i \, \Gamma_5(x_{10}^2, x_{21}^2, zs) + \cdots,
\end{align}
where the ellipsis indicates irrelevant structures which do not appear in the evolution below. 
Similarly, we define the neighbor moment amplitudes $\Gamma_3$ and $\widetilde{\Gamma}_3$ corresponding to $I_3$ and $\widetilde{I}_3$ defined in \eqs{I3_def} and \eqref{I3t_def}, respectively, by
\begin{subequations} \label{neigh_1_defs}
    \begin{align}
        \int d^2 x_1 \,x^m_1 \overline{\Gamma}_{10,21}(s) &\equiv x_{10}^m \, \Gamma_{3} (x_{10}^2, x_{21}^2, s) + \cdots,  \\ 
        \int d^2 x_1 \, x^m_1 \widetilde{\Gamma}_{10, 21}(s) & \equiv x_{10}^m \, \widetilde{\Gamma}_3(x_{10}^2, x_{21}^2, s) + \cdots.
    \end{align}
\end{subequations}

To derive the evolution equations for the moment amplitudes $I_{3}, I_4$, $I_5$, and $\widetilde{I}_3$ along with their neighbor counterparts, we will follow a similar procedure to the derivation of the large-$N_c$ moment evolution equations in \cite{Kovchegov:2023yzd}. We begin with the evolution equation for $Q_{10}$, which is given above in \eq{Q_evol_2} (see also \cite{Cougoulic:2022gbk}). To obtain the moment evolution equations, we multiply both sides of \eq{Q_evol_2} by $x_1^m$ and integrate over all $x_1$, while using \eq{I3_def} and other moment amplitude definitions. We perform the integration so that $\un x_{21}$ and other inter-partonic separations are fixed. Isolating the logarithmically divergent UV ($x_{21} \ll x_{10} \approx x_{20}$) and IR ($x_{21} \approx x_{20} \gg x_{10}$) regions of the $\un x_{2}$ integrals in \eq{Q_evol_2}, we arrive at the DLA evolution equation for $I_3$ in the large-$N_c\&N_f$ limit
\begin{align} 
\notag 
    I_{3} (x_{10}^2, zs) &= I_{3} ^{(0)}(x_{10}^2, zs) + 
    \frac{\as N_c}{2\pi} \iUV \Big[
    2\, \tGm_3  (x_{10}^2, x_{21}^2, z's)
    -\Gamma_{3} (x_{10}^2, x_{21}^2, z's) 
    -2  \, \Gamma_4(x_{10}^2, x_{21}^2, z's)
    \\[-0.3cm] \notag & \hspace{9.5cm}
    + \Gamma_5(x_{10}^2, x_{21}^2, z's)
    - \Gamma_2(x_{10}^2, x_{21}^2, z's)
    \Big]
    \\ \notag 
 &+ \frac{\as N_c}{2\pi} \iIR \Big[2\, \tI_3(x_{21}^2, z's) 
 - 2\, I_4(x_{21}^2,z's) + I_5(x_{21}^2, z's)  - \, I_E (x_{21}^2, z's)
    \\[-0.3cm]  & \hspace{8.5cm} 
 -2 \, \tG(x_{21}^2, z's) - 3 \, G_2(x_{21}^2, z's)
 \Big].
 \label{I3_ee}
\end{align}
Here, $I_{3}^{(0)}$ denotes the inhomogeneous term which represents the initial condition of the evolution. The impact-parameter integral of the neighbor amplitude of the second type is decomposed similarly to \eq{G2def}
\begin{align}
    \int d^2 x_1 \, \Gamma^i_{10, 21} (zs) \equiv \epsilon^{ij} \, x_{10}^j \, \Gamma_2 (x_{10}^2, x_{21}^2, zs) + \cdots .
\end{align}

Similarly, one can obtain the DLA evolution equation for $\tI_3$ by starting with \eq{G_adj_evol_2.5} above for $\widetilde{G}_{10}$. After multiplying by $x_1^m$, integrating over all $x_1$ while keeping inter-partonic distances fixed and identifying the logarithmically divergent parts of the resulting kernels, we obtain
\begin{align}\notag
    \tI_3(x_{10}^2, zs) &= \tI_3^{(0)}(x_{10}^2, zs) + 
    \frac{\as N_c}{2\pi} \iUV \Bigg[
    \tGm_3  (x_{10}^2, x_{21}^2, z's)
    - \frac{N_f}{4 \,N_c}\,\Gamma_{3}  (x_{10}^2, x_{21}^2, z's)
    \\ \notag & 
    + \left(\frac{N_f}{2\, N_c} -2\right)  \, \Gamma_4(x_{10}^2, x_{21}^2, z's)
    + \left(1- \frac{N_f}{4\, N_c}\right) \Gamma_5(x_{10}^2, x_{21}^2, z's)
    + \left(\frac{N_f}{4\, N_c}-1\right)\Gamma_2(x_{10}^2, x_{21}^2, z's)
    \Bigg]
    \\ \notag 
 &+ \frac{\as N_c}{2\pi} \iIR \bigg[2\, \tI_3(x_{21}^2, z's) 
 - \frac{N_f}{4 \, N_c}\, I_{3} (x_{21}^2, z's) 
 + \left(\frac{N_f}{2 N_c} - 2 \right) \, I_4(x_{21}^2,z's) 
 \\  & \hspace{1cm} 
-  I_E(x_{21}^2, z's) 
 + 
 \left(1 - \frac{N_f}{4\, N_c}\right)\, I_5(x_{21}^2, z's) 
 -2 \, \tG(x_{21}^2, z's) +  \left(\frac{N_f}{4\, N_c} - 3 \right)\, G_2(x_{21}^2, z's)
 \bigg]. 
 \label{I3tilde}
\end{align}
However, a correction to Eq.~(150) of \cite{Cougoulic:2022gbk} and, therefore, to our \eq{G_adj_evol_2.5} was found in \cite{Borden:2024bxa} due to the quark-gluon transition operators, see Eqs.~(76) there.\footnote{Strictly speaking, the evolution equation \eqref{Q_evol_2} for $Q_{10} (s)$ also receives a correction due to the transition operators \cite{Borden:2024bxa}: however, this correction is suppressed in the large-$N_c \& N_f$ limit.} To include this correction, we rewrite Eq.~(46) from \cite{Borden:2024bxa} in the form before it was integrated over all impact parameters and the DL approximation was applied to it,
\begin{align}\label{Qt1}
    - \frac{\as }{4 \pi^2} \sum_q \int\limits_{\Lambda^2 /s}^z \frac{d z'}{z'} \, \int d^2 x_2 \, \Bigg\{ \left[ \frac{1}{x_{21}^2} - \frac{\un x_{21} \cdot \un x_{20}}{x_{21}^2 \, x_{20}^2} \right] \, {\widetilde Q}_{12} (z' s) - i \, \frac{{\un x}_{20} \times {\un x}_{21}}{x_{20}^2 \, x_{21}^2} \, {\widetilde L}_{12} (z' s)  \Bigg\} , 
\end{align}
where we have also divided the expression by 4 due to \eq{Gadj_Gtilde} (cf. Eq.~(47) in \cite{Borden:2024bxa}). The term with the cross-product in \eq{Qt1} was discarded as non-DLA in \cite{Borden:2024bxa}, but, as we have seen above, naively non-DLA kernels may still contribute to the DLA evolution for moment amplitudes: hence, we keep that term in \eq{Qt1}. Multiplying \eq{Qt1} by $x_1^m$ and integrating over all $x_1$ while repeating the same steps as above, which includes employing \eq{half}, we obtain its contribution to the moment amplitude evolution equation \eqref{I3tilde}, arriving at the following equation:
\begin{align}
\notag
   & \tI_3(x_{10}^2, zs) = \tI_3^{(0)}(x_{10}^2, zs) + 
    \frac{\as N_c}{2\pi} \iUV \Bigg[
    \tGm_3  (x_{10}^2, x_{21}^2, z's)
    - \frac{N_f}{4 \,N_c} \,  \Gamma_{3}  (x_{10}^2, x_{21}^2, z's)
    \\ \notag & 
    + \left(\frac{N_f}{2\, N_c} -2\right)  \, \Gamma_4(x_{10}^2, x_{21}^2, z's)
    + \left(1- \frac{N_f}{4\, N_c}\right) \Gamma_5(x_{10}^2, x_{21}^2, z's)
    + \left(\frac{N_f}{4\, N_c}-1\right)\Gamma_2(x_{10}^2, x_{21}^2, z's)
    \Bigg]
    \\ \notag 
 &+ \frac{\as N_c}{2\pi} \iIR \bigg[2\, \tI_3(x_{21}^2, z's) 
 - \frac{N_f}{4 \, N_c}\, I_{3} (x_{21}^2, z's) 
 + \left(\frac{N_f}{2 N_c} - 2 \right) \, I_4(x_{21}^2,z's)   -  \, I_E(x_{21}^2, z's) 
 \\ & \hspace*{0cm}
 + 
 \left(1 - \frac{N_f}{4\, N_c}\right)\, I_5(x_{21}^2, z's) 
 -2 \, \tG(x_{21}^2, z's) +  \left(\frac{N_f}{4\, N_c} - 3 \right)\, G_2(x_{21}^2, z's) + \frac{N_f}{8 N_c} \, \widetilde{Q} (x_{21}^2, z's) + \frac{N_f}{4 N_c} \,  \widetilde{I} (x_{21}^2, z's) 
 \bigg].
 \label{I3tilde_2}
\end{align}

Lastly, consider the moment amplitudes $I_4$ and $I_5$ in \eq{Gi_moms}. These amplitudes do not contain a sub-eikonal quark operator. Therefore, the evolution equations for the moment amplitudes $I_4$ and $I_5$ in the large-$N_c\&N_f$ limit are obtained from the ones in the large-$N_c$ limit by simply replacing $G$ by $\tG$ in the latter. The large-$N_c$ moment evolution equations are given in Eq.~(A1) of \cite{Kovchegov:2024wjs} (see also \cite{Kovchegov:2023yzd}). However, as mentioned above, these earlier studies did not include contributions from $I_E$. To obtain these type-3 terms, we multiply \eq{Gi_largeNc_evol} by $x_1^m$ and integrate over $x_1$, concentrating on the last term in the curly brackets. The new terms involving $I_E$ yield 
\begin{subequations}
\begin{align}
& I_4 (x^2_{10}, zs) \supset - \frac{\as \, N_c}{4 \pi} \, \iIR \, I_E (x^2_{21} , z's) , \\
& I_5 (x^2_{10}, zs) \supset  + \frac{\as \, N_c}{4 \pi} \, \iIR \, I_E (x^2_{21} , z's) . 
\end{align}
\end{subequations}
Similarly, the neighbors of $I_4$ and $I_5$ also get contributions from $I_E$ via
\begin{subequations}
\begin{align}
& \Gamma_4 (x^2_{10}, x_{21}^2,  z' s) \supset  - \frac{\as \, N_c}{4 \pi} \, \inIR \, I_E (x^2_{32} , z''s) , \\
& \Gamma_5 (x^2_{10}, x_{21}^2,  z' s)  \supset + \frac{\as \, N_c}{4 \pi} \, \inIR \, I_E (x^2_{32} , z''s) . 
\end{align}
\end{subequations}


\subsection{Evolution equations for moment-amplitudes in the large-$N_c \& N_f$ limit}

\vspace{-2.5mm}

Assembling all of the moment amplitude evolution equations, we may write them in a compact matrix notation as
\begin{tcolorbox}[colback=blue!10!white]
\vspace{-3mm}
\begin{align}     \notag 
    \begin{pmatrix}
        I_{3} \\
        \tI_3\\ 
        I_4 \\
        I_5 \\
        \widetilde{I} \\ 
        I_E 
    \end{pmatrix} (x_{10}^2, zs)
    &= \begin{pmatrix}
        I_{3}^{(0)} \\
        \tI_3^{(0)}  \\ 
        I_4^{(0)} \\
        I_5^{(0)} \\
        \widetilde{I}^{(0)} \\
        0 
    \end{pmatrix} (x_{10}^2, zs) 
    \\ \notag &
    + \frac{\as N_c}{2\pi} \iUV
    \begin{pmatrix}
        -1 & 2 & -2 & 1 & -1   \\
        - \frac{N_f}{4N_c}  & 1 & \frac{N_f}{2N_c} -2  & 1- \frac{N_f}{4N_c} & \frac{N_f}{4N_c} - 1  \\
        0 & 0 & 0 & 0  &  0  \\ 
        0 & 0 & 0 & 0  &  0  \\ 
        0 & 0 & 0 & 0  &  0  \\ 
        0 & 0 & 0 & 0  &  0 
    \end{pmatrix}
    \begin{pmatrix}
        \Gamma_{3} \\ 
        \tGm_3 \\
        \Gamma_4 \\
        \Gamma_5 \\
        \Gamma_2 
    \end{pmatrix}(x_{10}^2, x_{21}^2 ,z^\prime s)
    \\ \notag & 
    + \frac{\as N_c}{4\pi} \iIR
    \begin{pmatrix}
        0 & 4 & -4 & 2 & 0 &  -2\\
        -\frac{N_f}{2N_c}  & 4 & \frac{N_f}{N_c} - 4 & 2 - \frac{N_f}{2N_c}   & \frac{N_f}{2N_c}  &  -2 \\
        0 & 0 & 4 & 2  & 0  &  -1 \\
        0 & -2 & 2 & -1 & 0 &  +1 \\
        -1 & 0 & 2 & -1 & 1 & 0 \\ 
        0 & -4 & 4 & -2 & 0 &  +2
    \end{pmatrix}
    \begin{pmatrix}
        I_{3} \\ 
        \tI_3 \\
        I_4 \\
        I_5 \\
        \widetilde{I} \\ 
        I_E 
    \end{pmatrix}(x_{21}^2 ,z^\prime s)
    \notag
    \\ & 
        + \frac{\as N_c}{4\pi} \iIR
    \begin{pmatrix}
         0 & -4  & -6 & 0   \\
         0 & -4  & \frac{N_f}{2N_c} -6 & \frac{N_f}{4 N_c} \\
         0 &  -2 & -3 & 0 \\
         0 & 4 & 7 & 0 \\
         1 & 0 & 3 & \thalf \\ 
         0 & 4 & 6 & 0 
    \end{pmatrix}
    \begin{pmatrix}
        Q \\
        \tG \\
        G_2 \\
        \widetilde{Q}
    \end{pmatrix}(x_{21}^2 ,z^\prime s).
     \label{mom_eqn}
\end{align}
\vspace{-6mm}
\end{tcolorbox}
Similarly, the evolution equations for the moments of the neighbor amplitudes can be obtained via the existing methods \cite{Cougoulic:2022gbk, Kovchegov:2016zex, Kovchegov:2018znm, Kovchegov:2017lsr, Cougoulic:2019aja, Kovchegov:2021lvz}. They are given by 
\begin{tcolorbox}[colback=blue!10!white]
\vspace{-3mm}
\begin{align}  \notag 
    \begin{pmatrix}
        \Gamma_{3} \\
        \tGm_3\\ 
        \Gamma_4 \\
        \Gamma_5  
    \end{pmatrix} &(x_{10}^2, x_{21}^2, z's)
    = \begin{pmatrix}
        I_{3}^{(0)} \\
        \tI_3^{(0)}  \\ 
        I_4^{(0)} \\
        I_5^{(0)}  
    \end{pmatrix} (x_{21}^2, z's) 
    \\ \notag & \hspace{-1cm}
    + \frac{\as N_c}{2\pi} \inUV
    \begin{pmatrix}
        -1 & 2 & -2 & 1 & -1 \\
        - \frac{N_f}{4N_c}  & 1 & \frac{N_f}{2N_c} -2  & 1- \frac{N_f}{4N_c} & \frac{N_f}{4N_c} - 1 \\
        0 & 0 & 0 & 0  &  0  \\ 
        0 & 0 & 0 & 0  &  0 
    \end{pmatrix}
    \begin{pmatrix}
        \Gamma_{3} \\ 
        \tGm_3 \\
        \Gamma_4 \\
        \Gamma_5 \\
        \Gamma_2 
    \end{pmatrix}(x_{10}^2, x_{32}^2 ,z^{\prime \prime} s) 
    \notag  \\ &
    + \frac{\as N_c}{4\pi}
    \inIR
    \begin{pmatrix}
        0 & 4 & -4 & 2 & 0 & -2 \\
        -\frac{N_f}{2N_c}  & 4 & \frac{N_f}{N_c} - 4 & 2 - \frac{N_f}{2N_c}   & \frac{N_f}{2N_c}  &  -2 \\
        0 & 0 & 4 & 2  & 0  &  -1 \\
        0 & -2 & 2 & -1 & 0 & +1 
    \end{pmatrix}
     \begin{pmatrix}
        I_{3} \\ 
        \tI_3 \\
        I_4 \\
        I_5 \\
        \widetilde{I} \\ 
        I_E
    \end{pmatrix} (x_{32}^2 ,z^{\prime\prime} s)
    \notag 
    \\ &
    + \frac{\as N_c}{4\pi}
    \inIR
    \begin{pmatrix}
         0 & -4  & -6 & 0   \\
         0 & -4  & \frac{N_f}{2N_c} -6 & \frac{N_f}{4 N_c} \\
         0 &  -2 & -3 & 0 \\
         0 & 4 & 7 & 0 
    \end{pmatrix}
    \begin{pmatrix}
        Q \\
        \tG \\
        G_2 \\
        \widetilde{Q}
    \end{pmatrix} (x_{32}^2 ,z^{\prime\prime} s).
    \label{neigh_mom_eqn}
\end{align}
\vspace{-6mm}
\end{tcolorbox}

Equations \eqref{mom_eqn} and \eqref{neigh_mom_eqn} are our main result. Their solution will provide one with $\widetilde{I}, I_4$, and $I_5$, and, through \eqs{OAMs_corr}, with the OAM distributions $L_{q+\bar{q}}(x, Q^2)$ and $L_G(x,Q^2)$ at small $x$ and at large $N_c \& N_f$. 

For the ease of use, let us summarize the ingredients of these equations. For helicity and OAM evolution at small $x$ and at large $N_c \& N_f$, we need the polarized dipole amplitudes $Q_{10}(s), \widetilde{G}_{10}(s)$, $G^{[3]}_{10} (s)$, and $G^i_{10}(s)$, along with the objects $\widetilde{Q}_{10} (s)$ and $\widetilde{L}_{10} (s)$. Using these quantities, we integrate over impact parameters, obtaining
\begin{subequations}
    \begin{align}
        & \int d^2 x_1 
        Q_{10}(s) = Q(x_{10}^2, s), 
        \\ 
        & \int d^2 x_1 \, x_1^m \, Q_{10}(s)  = x_{10}^m \, I_3(x_{10}^2, s) + \cdots,
        \\ 
        & \int d^2 x_1 \, \tG_{10}(s) = \widetilde{G} (x_{10}^2, s) , \\
        & \int d^2 x_1 \, x_1^m \, G^{[3]}_{10}(s) = \epsilon^{mk} x_{10}^k \, I_E(x_{10}^2, s) + \cdots , \\
        & \int d^2 x_1 
        \, G^i_{10}(s) = \epsilon^{ij} x_{10}^j G_2(x_{10}^2, s) + \cdots, \\
        & \int d^2 x_1 \, x_1^m \, G^i_{10}(s) = \epsilon^{mi} x_{10}^2 \, I_4(x_{10}^2, s) + 
        \epsilon^{mk} x_{10}^k x_{10}^i \, I_5(x_{10}^2, s) + 
        \cdots , \\
        & \int d^2 x_1\, \widetilde{Q}_{10} (s) = \widetilde{Q} (x_{10}^2, s) ,  \\
        & \int d^2 x_1 \, x_1^m \, \widetilde{Q}_{10}(s) = x_{10}^m \, \half \, \widetilde{Q} (x_{10}^2, s) + \cdots , \\
        & \int d^2 x_1 \, x_1^i \, {\widetilde L}_{10}  (s)  = i \, \epsilon^{ij} \, x_{10}^j \, {\widetilde I} (x^2_{10},  s) +\cdots . 
    \end{align}
\end{subequations}
We only explicitly show the quantities relevant to either helicity or OAM distributions on the right-hand sides of the above equations. 

The impact parameter integration of the neighbor dipole amplitudes is done similarly, yielding
\begin{subequations}
    \begin{align}
        & \int d^2 x_1 \,x^m_1 \overline{\Gamma}_{10,21}(s) = x_{10}^m \, \Gamma_{3} (x_{10}^2, x_{21}^2, s) + \cdots,  \\ 
        & \int d^2 x_1 \, x^m_1 \widetilde{\Gamma}_{10, 21}(s) = x_{10}^m \, \widetilde{\Gamma}_3(x_{10}^2, x_{21}^2, s) + \cdots, \\
        & \int d^2 x_1 \, \Gamma^i_{10, 21} (s) = \epsilon^{ij} \, x_{10}^j \, \Gamma_2 (x_{10}^2, x_{21}^2, s) + \cdots , \\
        & \int d^2 x_1 \, x_1^m \,\Gamma^i_{10, 21}(s) = \epsilon^{mi} x_{10}^2 \, \Gamma_4(x_{10}^2, x_{21}^2, s) + \epsilon^{mk} x_{10}^k x_{10}^i \, \Gamma_5(x_{10}^2, x_{21}^2, s) + \cdots . 
    \end{align}
\end{subequations}

The impact-parameter integrated quantities $Q(x_{10}^2, s)$, $\widetilde{G} (x_{10}^2, s)$, $G_2 (x_{10}^2, s)$, and $\widetilde{Q} (x_{10}^2, s)$ obey their own evolution equations, which couple them to their own impact-parameter integrated neighbor amplitudes, including $\Gamma_2 (x_{10}^2, x_{21}^2, s)$ which is also present in the above equations. These are the equations~(76) in \cite{Borden:2024bxa} (at large $N_c \& N_f$). Those equations need to be solved first, before solving \eqs{mom_eqn} and \eqref{neigh_mom_eqn}.    

In addition, as we have seen above, the type-3 operator corrections have affected the large-$N_c$ evolution equations for the moment amplitudes as well \cite{Kovchegov:2023yzd, Kovchegov:2024wjs}. For completeness, let us state the corrected version of those equations. First we note that there are two ways of taking the large-$N_c$ limit here: one is to discard the quarks entirely, keeping gluons only. Another way is to keep external quarks, while discarding the internal quark loops. This latter procedure was referred to as the large-$N_c^{+q}$ limit in \cite{JAMCollaborationSmall-xAnalysisGroup:2025tfa}. Here we will consider the former, gluons-only limit: in this limit we, strictly speaking, cannot very reliably obtain the quark OAM distribution. The evolution equations in the large-$N_c^{+q}$ limit will be stated in the companion paper \cite{JAMCollaborationSmall-xAnalysisGroup:inprep}. 

The gluons-only large-$N_c$ evolution equations are obtained from \eqs{mom_eqn} and \eqref{neigh_mom_eqn} by equating $\widetilde{I}_3 = I_3$, $Q = \widetilde{G} \equiv G$, discarding $\widetilde{I}$ and $\widetilde{Q}$ since these come from the quark operators, and by putting $N_f =0$ in the remaining equations. We arrive at 
\begin{subequations}\label{OAM_evol_Large_Nc}
\begin{align} \notag
    \begin{pmatrix}
    I_3 \\ 
    I_4 \\
    I_5 \\ 
    I_E 
    \end{pmatrix}(x_{10}^2, zs) &= \begin{pmatrix}
    I_3^{(0)} \\ 
    I_4^{(0)} \\
    I_5^{(0)} \\
    0
    \end{pmatrix} (x_{10}^2, zs)  
    + \frac{\as N_c}{2\pi} \iUV \,  \begin{pmatrix}
    \Gamma_3 - 2 \,\Gamma_4 + \Gamma_5  - \Gamma_2 \\ 
    0 \\
    0 
    \end{pmatrix}(x_{10}^2, x_{21}^2, z^\prime s) \\ &+ \frac{\as N_c}{4\pi} \iIR \, \begin{pmatrix}
        4 & -4 & 2 &  -2 &  -4 &  - 6 \\
        0 & 4 & 2 &  -1 &  -2 &  -3 \\
        -2 & 2 & -1 &  +1 & 4 &  7 \\
        -4 & 4 & -2 &  +2 & 4 & 6 
    \end{pmatrix} 
    \begin{pmatrix}
    I_3 \\ 
    I_4 \\
    I_5 \\
    I_E \\
    G \\
    G_2
    \end{pmatrix} (x_{21}^2, z^\prime s), 
    \label{Ievol}
    \\
    \notag
    \begin{pmatrix}
    \Gamma_3 \\ 
    \Gamma_4 \\
    \Gamma_5 
    \end{pmatrix}(x_{10}^2, x_{21}^2, z^\prime s) & = \begin{pmatrix}
    I_3^{(0)} \\ 
    I_4^{(0)} \\
    I_5^{(0)} 
    \end{pmatrix} (x_{10}^2, z^\prime s)  \\ \notag 
    & 
    + \frac{\as N_c}{2 \pi} \inUV \, \begin{pmatrix}
    \Gamma_3 - 2 \,\Gamma_4 + \Gamma_5 - \Gamma_2 \\ 
    0 \\
    0 
    \end{pmatrix}(x_{10}^2, x_{32}^2, z^{\prime \prime} s)
    \\ &
    + \frac{\as N_c}{4\pi} \inIR \, \begin{pmatrix}
        4 & -4 & 2 &  -2 &  -4 &  - 6 \\
        0 & 4 & 2 &  -1 &  -2 &  -3 \\
        -2 & 2 & -1 &  1 & 4 &  7 
    \end{pmatrix} 
    \begin{pmatrix}
    I_3 \\ 
    I_4 \\
    I_5 \\
    I_E \\
    G \\ 
    G_2
    \end{pmatrix} (x_{32}^2, z^{\prime \prime} s) .
    \label{Gamma_evol}
\end{align}
\end{subequations}
Once again, the gluon OAM distribution $L_G$ is given by \eq{final_gluon_OAM} with $I_4$ and $I_5$ found from \eqs{OAM_evol_Large_Nc}.


\section{Numerical solution of the moment amplitude evolution in the large $N_c \& N_f$ limit}
\label{sec:numerics}
 
Equations \eqref{mom_eqn} and \eqref{neigh_mom_eqn} do not close on their own: they mix with the polarized dipole amplitudes $Q$, $\tG$, $G_2$, with the object $\widetilde{Q}$, and with the neighbor amplitude $\Gamma_2$. The large-$N_c \& N_f$ helicity evolution for these quantities was derived in
\cite{Kovchegov:2015pbl, Kovchegov:2018znm,  Cougoulic:2022gbk} with important corrections from the $\widetilde{Q}$ amplitude found in \cite{Borden:2024bxa}. The resulting evolution equations were solved numerically in \cite{Adamiak:2023okq} and analytically in \cite{Borden:2025ehe}. While the numerical solution in \cite{Adamiak:2023okq} does not include contributions from $\widetilde{Q}$, the analytic solution in \cite{Borden:2025ehe} does include these contributions. 

Together with the helicity evolution, the moment amplitude evolution equations are closed, and we solve them here numerically, following the procedure developed in \cite{Kovchegov:2016weo, Kovchegov:2020hgb, Cougoulic:2022gbk, Kovchegov:2023yzd, Adamiak:2023okq}. As in \eq{qOAM4}, all the dipole and moment amplitudes are taken to be flavor independent; a flavor decomposition will be needed for phenomenology done in the companion paper \cite{JAMCollaborationSmall-xAnalysisGroup:inprep} along the lines of \cite{Adamiak:2023yhz, JAMCollaborationSmall-xAnalysisGroup:2025tfa}. Our aim in solving \eqs{mom_eqn} and \eqref{neigh_mom_eqn} is to determine the small-$x$ asymptotics of the quark and gluon OAM distributions: we want to obtain their small-$x$ intercepts along with the OAM distributions to hPDF ratios.


\subsection{Discretization and solution}
\label{sec:discretization}
 
Following \cite{Kovchegov:2015pbl, Kovchegov:2016weo, Kovchegov:2020hgb, Kovchegov:2023yzd, Adamiak:2023yhz} we define the following logarithmic variables
\begin{subequations}\label{rescaled}
\begin{align}
  \eta &= \sqrt{\bar{\alpha}_s} \, \ln \frac{zs}{\Lambda^2},
  & \eta' &= \sqrt{\bar{\alpha}_s} \, \ln \frac{z' s}{\Lambda^2},
  & \eta'' &= \sqrt{\bar{\alpha}_s} \, \ln \frac{z'' s}{\Lambda^2},
  \\
  s_{10} &= \sqrt{\bar{\alpha}_s} \, \ln \frac{1}{\xo^2 \Lambda^2},
  & s_{21} &= \sqrt{\bar{\alpha}_s} \, \ln \frac{1}{\xti^2 \Lambda^2},
  & s_{32} &= \sqrt{\bar{\alpha}_s} \, \ln \frac{1}{x_{32}^2 \Lambda^2},
\end{align}
\end{subequations}
where we have defined 
\begin{align}
    \bar{\alpha}_s = \frac{\alpha_s N_c}{2\pi}. 
\end{align}
In terms of the variables in \eqs{rescaled}, \eqs{mom_eqn} and \eqref{neigh_mom_eqn} become
\begin{subequations}\label{rescaled_eqns}
\begin{align}
\notag
    \begin{pmatrix} I_{3} \\ \tI_3 \\ I_4 \\ I_5 \\ \tI  \\ I_E \end{pmatrix}
    (s_{10}, \eta)
    &= \begin{pmatrix} I_{3}^{(0)} \\ \tI_3^{(0)} \\ I_4^{(0)} \\ I_5^{(0)}
       \\ \tI^{(0)}  \\ 0\end{pmatrix} (s_{10}, \eta)
       \notag \\ \notag & 
    + \int\limits_{s_{10}}^{\eta} \! d\eta' \int\limits_{s_{10}}^{\eta'}
      \! ds_{21} \; 
    \begin{pmatrix}
        -1 & 2 & -2 & 1 & -1  \\
        - \frac{N_f}{4N_c}  & 1 & \frac{N_f}{2N_c} -2  & 1- \frac{N_f}{4N_c} & \frac{N_f}{4N_c} - 1 \\
        0 & 0 & 0 & 0  &  0  \\ 
        0 & 0 & 0 & 0  &  0  \\ 
        0 & 0 & 0 & 0  &  0  \\ 
        0 & 0 & 0 & 0  &  0 
    \end{pmatrix}
    \begin{pmatrix} \Gamma_{3} \\ \tGm_3 \\ \Gamma_4 \\ \Gamma_5 \\ \Gamma_2 
    \end{pmatrix} (s_{10}, s_{21}, \eta')
    \\
    & \hspace{0.3cm}
    + \frac{1}{2} \int\limits_{0}^{s_{10}} \! ds_{21}
      \int\limits_{s_{21}}^{\eta - s_{10} + s_{21}} \! d\eta' 
    \begin{pmatrix}
        0 & 4 & -4 & 2 & 0 &  -2 \\
        -\frac{N_f}{2N_c}  & 4 & \frac{N_f}{N_c} - 4 & 2 - \frac{N_f}{2N_c}   & \frac{N_f}{2N_c}  &  -2 \\
        0 & 0 & 4 & 2  & 0  &  -1 \\
        0 & -2 & 2 & -1 & 0 &  1 \\
        -1 & 0 & 2 & -1 & 1 & 0 \\ 
        0 & -4 & 4 & -2 & 0 &  2
    \end{pmatrix}
       \begin{pmatrix} I_{3} \\ \tI_3 \\ I_4 \\ I_5 \\ \tI \\ I_E
      \end{pmatrix} (s_{21}, \eta')
      \notag 
    \\ 
    & \hspace{0.3cm}
    + \frac{1}{2} \int\limits_{0}^{s_{10}} \! ds_{21}
      \int\limits_{s_{21}}^{\eta - s_{10} + s_{21}} \! d\eta' 
    \begin{pmatrix}
         0 & -4  & -6 & 0   \\
         0 & -4  & \frac{N_f}{2N_c} -6 & \frac{N_f}{4 N_c} \\
         0 &  -2 & -3 & 0 \\
         0 & 4 & 7 & 0 \\
         1 & 0 & 3 & \thalf \\ 
         0 & 4 & 6 & 0 
    \end{pmatrix}
    \begin{pmatrix} Q \\ \tG \\ G_2 \\ \widetilde{Q}
      \end{pmatrix} (s_{21}, \eta')  ,
    \label{mom_rescaled}
\\[0.2cm]
\notag
    \begin{pmatrix} \Gamma_{3} \\ \tGm_3 \\ \Gamma_4 \\ \Gamma_5
    \end{pmatrix}
    (s_{10}, s_{21}, \eta')
    &= \begin{pmatrix} I_{3}^{(0)} \\ \tI_3^{(0)} \\ I_4^{(0)} \\ I_5^{(0)} 
       \end{pmatrix} (s_{10}, \eta')
    \\ \notag & 
    + \int\limits_{s_{10}}^{\eta'} \! d\eta''
      \int\limits_{\mathrm{max}[s_{10},\, s_{21} + \eta'' - \eta']}^{\eta''}
      \! ds_{32} \; 
    \begin{pmatrix}
        -1 & 2 & -2 & 1 & -1 &  \\
        - \frac{N_f}{4N_c}  & 1 & \frac{N_f}{2N_c} -2  & 1- \frac{N_f}{4N_c} & \frac{N_f}{4N_c} - 1 \\
        0 & 0 & 0 & 0  &  0  \\ 
        0 & 0 & 0 & 0  &  0  
    \end{pmatrix}
    \begin{pmatrix} \Gamma_{3} \\ \tGm_3 \\ \Gamma_4 \\ \Gamma_5 \\ \Gamma_2 
    \end{pmatrix} (s_{10}, s_{32}, \eta'')
    \\
    & 
    + \frac{1}{2} \int\limits_{0}^{s_{10}} \! ds_{32}
      \int\limits_{s_{32}}^{\eta' - s_{21} + s_{32}} \! d\eta'' 
    \begin{pmatrix}
        0 & 4 & -4 & 2 & 0 &  -2 \\
        -\frac{N_f}{2N_c}  & 4 & \frac{N_f}{N_c} - 4 & 2 - \frac{N_f}{2N_c}   & \frac{N_f}{2N_c}  &  -2 \\
        0 & 0 & 4 & 2  & 0  &  -1 \\
        0 & -2 & 2 & -1 & 0 &  1
    \end{pmatrix}
      \begin{pmatrix} I_{3} \\ \tI_3 \\ I_4 \\ I_5 \\ \tI \\ I_E
      \end{pmatrix} (s_{32}, \eta'')
      \notag
    \\
    & 
    + \frac{1}{2} \int\limits_{0}^{s_{10}} \! ds_{32}
      \int\limits_{s_{32}}^{\eta' - s_{21} + s_{32}} \! d\eta'' 
    \begin{pmatrix}
         0 & -4  & -6 & 0   \\
         0 & -4  & \frac{N_f}{2N_c} -6 & \frac{N_f}{4 N_c} \\
         0 &  -2 & -3 & 0 \\
         0 & 4 & 7 & 0 
    \end{pmatrix}
      \begin{pmatrix} Q \\ \tG \\ G_2 \\ \widetilde{Q}
      \end{pmatrix} (s_{32}, \eta'')  ,
    \label{neigh_rescaled}
\end{align}
\end{subequations}
where the ordering $0 \le s_{10} \le \eta$ is assumed in \eq{mom_rescaled}, while $0 \le s_{10} \le s_{21} \le \eta'$ is assumed in \eq{neigh_rescaled}. (Outside of these orderings the moment amplitudes and their neighbor amplitudes are ill-defined, since these are not the physical regions.) We do not present the helicity equations for $Q$, $\tG$, $G_2$, $\widetilde{Q}$ and the neighbor amplitudes $\overline{\Gamma}$, $\tGm$ and $\Gamma_2$ here for brevity: as we mentioned above, they are given by Eqs.~(76) in \cite{Borden:2024bxa}. However, since \eqs{rescaled_eqns} close only when those equations are included, we solve them numerically as well, in exactly the same way as described in \cite{Adamiak:2023okq}. In solving the helicity evolution, we also include the contributions from $\widetilde{Q}$, which are given explicitly in Eqs.~(76) of \cite{Borden:2024bxa}. 
 
We now discretize the integrals in \eqs{rescaled_eqns} with a step size $\delta$ in both the $\eta$ and $s_{10}$ directions. The discretized form of the dipole and moment amplitudes, along with their corresponding neighbor amplitudes, is denoted via
\begin{align}\label{disc_amps}
\begin{aligned}
  Q_{ij} &= Q(i \delta, j \delta), \\
  \tG_{ij} &= \tG(i \delta, j \delta), \\
  G_{2,ij} &= G_2(i \delta, j \delta), \\
  \widetilde{Q}_{ij} &= \widetilde{Q}(i \delta, j \delta),
\end{aligned}
\qquad
\begin{aligned}
  I_{3,ij} &= I_{3}(i \delta, j \delta), \\
  \tI_{3,ij} &= \tI_3(i \delta, j \delta), \\
  I_{4,ij} &= I_4(i \delta, j \delta), \\
  I_{5,ij} &= I_5(i \delta, j \delta), \\
  \tI_{ij} &= \tI(i \delta, j \delta), \\ 
  I_{E, ij} &= I_E(i \delta, j \delta), 
\end{aligned}
\qquad
\begin{aligned}
  \Gamma_{2,ikj} &= \Gamma_2(i \delta, k \delta, j \delta), \\
  \Gamma_{3,ikj} &= \Gamma_{3}(i \delta, k \delta, j \delta), \\
  \tGm_{3,ikj} &= \tGm_3(i \delta, k \delta, j \delta), \\
  \Gamma_{4,ikj} &= \Gamma_4(i \delta, k \delta, j \delta), \\
  \Gamma_{5,ikj} &= \Gamma_5(i \delta, k \delta, j \delta). 
\end{aligned}
\end{align}
The integrals in \eqs{rescaled_eqns} are replaced by left-handed Riemann sums: 
\begin{subequations}\label{disc_eqns}
\begin{align}
\notag
    &\begin{pmatrix} I_{3,ij} \\ \tI_{3,ij} \\ I_{4,ij} \\ I_{5,ij} \\ \tI_{ij} \\ I_{E,ij}
    \end{pmatrix}
    = \begin{pmatrix} I_{3,ij}^{(0)} \\ \tI_{3,ij}^{(0)} \\ I_{4,ij}^{(0)}
       \\ I_{5,ij}^{(0)} \\ \tI_{ij}^{(0)} \\ 0\end{pmatrix}
    + \delta^2 \sum_{j'=i}^{j-1} \; \sum_{i'=i}^{j'} 
    \begin{pmatrix}
        -1 & 2 & -2 & 1 & -1   \\
        - \frac{N_f}{4N_c}  & 1 & \frac{N_f}{2N_c} -2  & 1- \frac{N_f}{4N_c} & \frac{N_f}{4N_c} - 1  \\
        0 & 0 & 0 & 0  &  0  \\ 
         0 & 0 & 0 & 0  &  0  \\ 
        0 & 0 & 0 & 0  &  0  \\ 
        0 & 0 & 0 & 0  &  0 
    \end{pmatrix}
    \begin{pmatrix} \Gamma_{3,ii'j'} \\ \tGm_{3,ii'j'} \\ \Gamma_{4,ii'j'}
    \\ \Gamma_{5,ii'j'} \\ \Gamma_{2,ii'j'} 
    \end{pmatrix}
    \\
    &
    + \frac{\delta^2}{2} \sum_{i'=0}^{i-1} \; \sum_{j'=i'}^{j-i+i'} \left[
    \begin{pmatrix}
        0 & 4 & -4 & 2 & 0 &  -2 \\
        -\frac{N_f}{2N_c}  & 4 & \frac{N_f}{N_c} - 4 & 2 - \frac{N_f}{2N_c}   & \frac{N_f}{2N_c}  &  -2 \\
        0 & 0 & 4 & 2  & 0  &  -1 \\
        0 & -2 & 2 & -1 & 0 &  1 \\
        -1 & 0 & 2 & -1 & 1 & 0 \\ 
        0 & -4 & 4 & -2 & 0 &  2
    \end{pmatrix}
      \begin{pmatrix} I_{3,i'j'} \\ \tI_{3,i'j'} \\ I_{4,i'j'}
      \\ I_{5,i'j'} \\ \tI_{i'j'} \\ I_{E, i'j'} \end{pmatrix}
    +       \begin{pmatrix}
         0 & -4  & -6 & 0   \\
         0 & -4  & \frac{N_f}{2N_c} -6 & \frac{N_f}{4 N_c} \\
         0 &  -2 & -3 & 0 \\
         0 & 4 & 7 & 0 \\
         1 & 0 & 3 & \thalf \\ 
         0 & 4 & 6 & 0 
    \end{pmatrix}
    \begin{pmatrix} Q_{i'j'} \\ \tG_{i'j'} \\ G_{2,i'j'}
      \\ \widetilde{Q}_{i'j'} \end{pmatrix} \right] ,
    \label{mom_disc}
\\
\notag
    &\begin{pmatrix} \Gamma_{3,ikj} \\ \tGm_{3,ikj} \\ \Gamma_{4,ikj}
    \\ \Gamma_{5,ikj} 
    \end{pmatrix}
    = \begin{pmatrix} I_{3,ij}^{(0)} \\ \tI_{3,ij}^{(0)} \\ I_{4,ij}^{(0)}
       \\ I_{5,ij}^{(0)} \end{pmatrix}
    + \delta^2 \sum_{j'=i}^{j-1} \; \sum_{i'=\mathrm{max}[i,\, k+j'-j]}^{j'}
    \begin{pmatrix}
        -1 & 2 & -2 & 1 & -1   \\
        - \frac{N_f}{4N_c}  & 1 & \frac{N_f}{2N_c} -2  & 1- \frac{N_f}{4N_c} & \frac{N_f}{4N_c} - 1 \\
        0 & 0 & 0 & 0  &  0 \\ 
         0 & 0 & 0 & 0  &  0 
    \end{pmatrix}
    \begin{pmatrix} \Gamma_{3,ii'j'} \\ \tGm_{3,ii'j'} \\ \Gamma_{4,ii'j'}
    \\ \Gamma_{5,ii'j'} \\ \Gamma_{2,ii'j'}  
    \end{pmatrix}
    \\
    &
    + \frac{\delta^2}{2} \sum_{i'=0}^{i-1} \; \sum_{j'=i'}^{j-k+i'} \left[
    \begin{pmatrix}
        0 & 4 & -4 & 2 & 0 &  -2 \\
        -\frac{N_f}{2N_c}  & 4 & \frac{N_f}{N_c} - 4 & 2 - \frac{N_f}{2N_c}   & \frac{N_f}{2N_c}  &  -2 \\
        0 & 0 & 4 & 2  & 0  &  -1 \\
        0 & -2 & 2 & -1 & 0 & 1
    \end{pmatrix}
    \begin{pmatrix} I_{3,i'j'} \\ \tI_{3,i'j'} \\ I_{4,i'j'}
      \\ I_{5,i'j'} \\ \tI_{i'j'}  \\ I_{E,i'j'} \end{pmatrix}
    +  \begin{pmatrix}
         0 & -4  & -6 & 0   \\
         0 & -4  & \frac{N_f}{2N_c} -6 & \frac{N_f}{4 N_c} \\
         0 &  -2 & -3 & 0 \\
         0 & 4 & 7 & 0 
    \end{pmatrix}
    \begin{pmatrix} Q_{i'j'} \\ \tG_{i'j'} \\ G_{2,i'j'}
      \\ \widetilde{Q}_{i'j'} \end{pmatrix} \right] .
    \label{neigh_disc}
\end{align}
\end{subequations}
In order to speed up the numerical computation, we follow the method of \cite{Kovchegov:2020hgb} and write recursive relations for each of the equations in \eqs{disc_eqns}. Subtracting $I_{p,i(j-1)}$ from $I_{p,ij}$ and $\Gamma_{p,i(k-1)(j-1)}$ from $\Gamma_{p,ikj}$ for $p = 3, \tilde 3, 4, 5, E$ and for $\tI$, we obtain
\begin{subequations}\label{recursions}
\begin{align}
   &\begin{pmatrix} I_{3,ij} \\ \tI_{3,ij} \\ I_{4,ij} \\ I_{5,ij} \\ \tI_{ij} \\ I_{E,ij}
    \end{pmatrix}
    = \begin{pmatrix}
       I_{3,ij}^{(0)} - I_{3,i(j-1)}^{(0)} + I_{3,i(j-1)} \\
       \tI_{3,ij}^{(0)} - \tI_{3,i(j-1)}^{(0)} + \tI_{3,i(j-1)} \\
       I_{4,ij}^{(0)} - I_{4,i(j-1)}^{(0)} + I_{4,i(j-1)} \\
       I_{5,ij}^{(0)} - I_{5,i(j-1)}^{(0)} + I_{5,i(j-1)} \\
       \tI_{ij}^{(0)} - \tI_{i(j-1)}^{(0)} + \tI_{i(j-1)} \\ 
       I_{E,i(j-1)}
       \end{pmatrix}
    + \delta^2 \sum_{i'=i}^{j-1} 
    \begin{pmatrix}
        -1 & 2 & -2 & 1 & -1  \\
        - \frac{N_f}{4N_c}  & 1 & \frac{N_f}{2N_c} -2  & 1- \frac{N_f}{4N_c} & \frac{N_f}{4N_c} - 1 \\
        0 & 0 & 0 & 0  &  0  \\ 
        0 & 0 & 0 & 0  &  0  \\ 
        0 & 0 & 0 & 0  &  0 \\ 
        0 & 0 & 0 & 0  &  0 
    \end{pmatrix}
    \begin{pmatrix} \Gamma_{3,ii'(j-1)} \\ \tGm_{3,ii'(j-1)}
    \\ \Gamma_{4,ii'(j-1)} \\ \Gamma_{5,ii'(j-1)} \\ \Gamma_{2,ii'(j-1)} 
    \end{pmatrix}
     \label{mom_recur}
    \\
    & 
    + \frac{\delta^2}{2} \sum_{i'=0}^{i-1} \left[
         \begin{pmatrix}
        0 & 4 & -4 & 2 & 0 &  -2 \\
        -\frac{N_f}{2N_c}  & 4 & \frac{N_f}{N_c} - 4 & 2 - \frac{N_f}{2N_c}   & \frac{N_f}{2N_c}  &  -2 \\
        0 & 0 & 4 & 2  & 0  & -1 \\
        0 & -2 & 2 & -1 & 0 & 1 \\
        -1 & 0 & 2 & -1 & 1 & 0 \\ 
        0 & -4 & 4 & -2 & 0 &  2
    \end{pmatrix}
    \begin{pmatrix} I_{3,i'(i'+j-i)} \\ \tI_{3,i'(i'+j-i)}
      \\ I_{4,i'(i'+j-i)} \\ I_{5,i'(i'+j-i)} \\ \tI_{i'(i'+j-i)}
      \\ I_{E,i'(i'+j-i)}
      \end{pmatrix}
    +  \begin{pmatrix}
         0 & -4  & -6 & 0   \\
         0 & -4  & \frac{N_f}{2N_c} -6 & \frac{N_f}{4 N_c} \\
         0 &  -2 & -3 & 0 \\
         0 & 4 & 7 & 0 \\
         1 & 0 & 3 & \thalf \\ 
         0 & 4 & 6 & 0 
    \end{pmatrix}  \begin{pmatrix} Q_{i'(i'+j-i)} \\ \tG_{i'(i'+j-i)}
      \\ G_{2,i'(i'+j-i)} \\ \widetilde{Q}_{i'(i'+j-i)} \end{pmatrix} \right] , \notag
\\
    &\begin{pmatrix} \Gamma_{3,ikj} \\ \tGm_{3,ikj} \\ \Gamma_{4,ikj}
    \\ \Gamma_{5,ikj} 
    \end{pmatrix}
    = \begin{pmatrix}
       I_{3,ij}^{(0)} - I_{3,i(j-1)}^{(0)} + \Gamma_{3,i(k-1)(j-1)} \\
       \tI_{3,ij}^{(0)} - \tI_{3,i(j-1)}^{(0)} + \tGm_{3,i(k-1)(j-1)} \\
       I_{4,ij}^{(0)} - I_{4,i(j-1)}^{(0)} + \Gamma_{4,i(k-1)(j-1)} \\
       I_{5,ij}^{(0)} - I_{5,i(j-1)}^{(0)} + \Gamma_{5,i(k-1)(j-1)}
       \end{pmatrix}
    + \delta^2 \sum_{i'=k-1}^{j-1} 
     \begin{pmatrix}
        -1 & 2 & -2 & 1 & -1   \\
        - \frac{N_f}{4N_c}  & 1 & \frac{N_f}{2N_c} -2  & 1- \frac{N_f}{4N_c} & \frac{N_f}{4N_c} - 1  \\
        0 & 0 & 0 & 0  &  0  \\ 
         0 & 0 & 0 & 0  &  0 
    \end{pmatrix}
    \begin{pmatrix} \Gamma_{3,ii'(j-1)} \\ \tGm_{3,ii'(j-1)}
    \\ \Gamma_{4,ii'(j-1)} \\ \Gamma_{5,ii'(j-1)} \\ \Gamma_{2,ii'(j-1)} 
    \end{pmatrix} ,
    \label{neigh_recur}
\end{align}
\end{subequations}
where $0 \le i < j$ and $0 \le i < k \le j$ for $i \le i_{\rm max}$, $j \le j_{\rm max}$, as can be seen from \eqs{disc_eqns}. For $i = j$ and $i = k$, the amplitudes are
\begin{align}\label{bcs}
\begin{aligned}
  I_{3,ii} &= I_{3,ii}^{(0)}, \quad
  \tI_{3,ii} = \tI_{3,ii}^{(0)}, \quad
  I_{4,ii} = I_{4,ii}^{(0)}, \quad
  I_{5,ii} = I_{5,ii}^{(0)}, \quad
  \tI_{ii} = \tI_{ii}^{(0)}, \quad 
  I_{E,ii} = 0
  \\
  \Gamma_{3,iij} &= I_{3,ij}, \quad
  \tGm_{3,iij} = \tI_{3,ij}, \quad
  \Gamma_{4,iij} = I_{4,ij}, \quad
  \Gamma_{5,iij} = I_{5,ij}. 
\end{aligned}
\end{align}
The physical regions where the evolution in \eqs{disc_eqns} applies are $j \ge i \ge 0$ and $j \ge k \ge i \ge 0$, corresponding to the $0 \le s_{10} \le \eta$ and $0 \le s_{10} \le s_{21} \le \eta'$ conditions above. Following \cite{Cougoulic:2022gbk, Kovchegov:2020hgb, Adamiak:2023okq, Kovchegov:2023yzd}, the dipole and moment amplitudes are left equal to the inhomogeneous terms outside those regions.
 
\begin{figure}[ht!]
\centering
\includegraphics[scale=0.7]{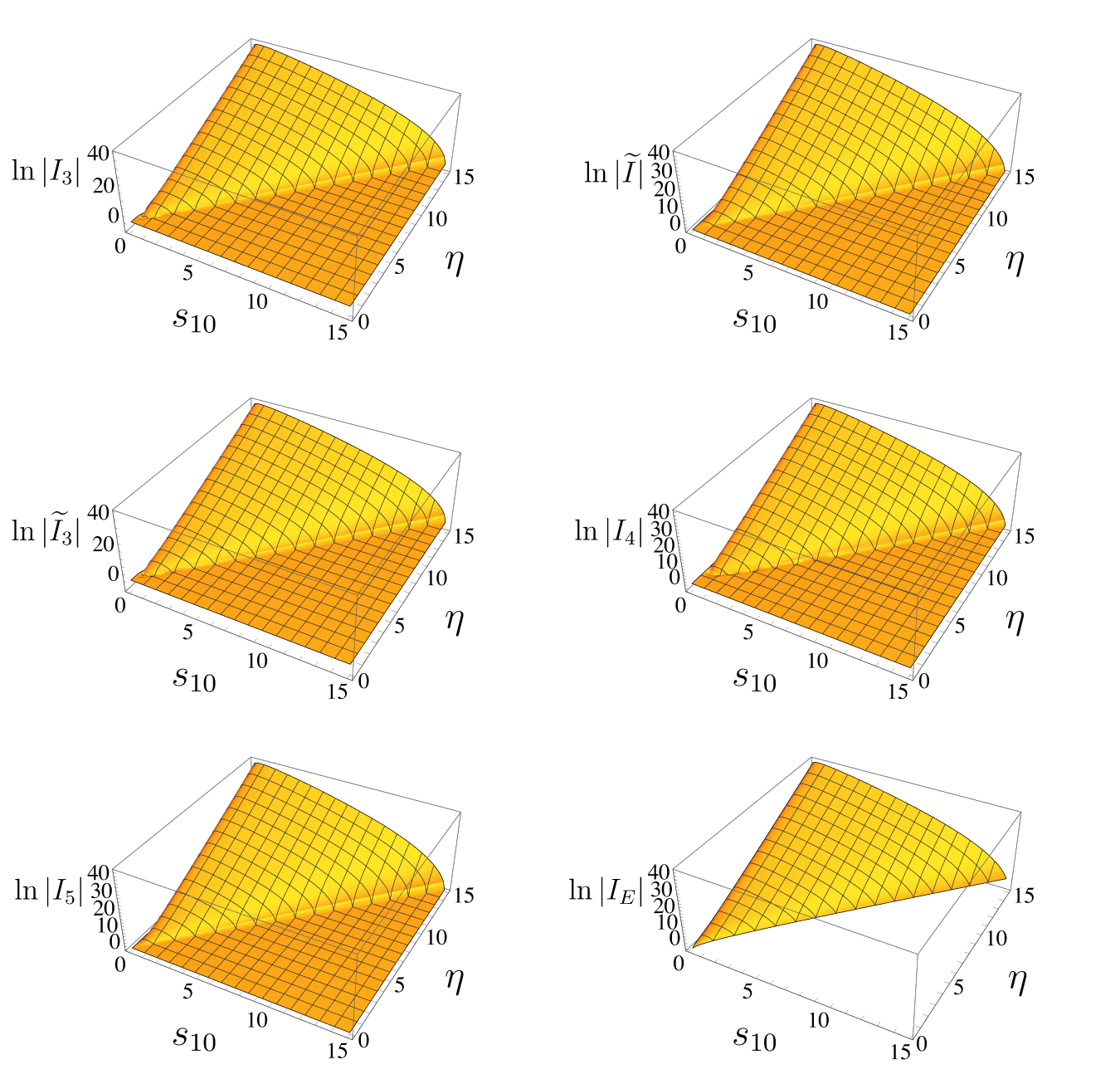}
\caption{Logarithms of the absolute values of the moment amplitudes (starting in the upper left panel and moving clockwise: $I_3$, $\tI$, $I_4$, $I_E$, $I_5$, $\tI_3$) as functions of $s_{10}$ and $\eta$, resulting from our numerical solution of \eqs{recursions} and \eqref{bcs} at $N_f = 3$ with a step size of $\delta = 0.025$. All the inhomogeneous terms are set to $1$ except for the one for $I_E$, which is zero. We use $\alpha_s = 0.25$, $N_c=3$, and $\Lambda=1\, \mathrm{GeV}$. Note in the $I_E$ plot, we have omitted the $s_{10}= 0$ and $s_{10} \geq \eta$ regions since $I_E(s_{10}, \eta) = 0$ there.}
\label{fig:amp3d}
\end{figure}

For simplicity, we set the inhomogeneous terms to $1$ for all dipole and moment amplitudes (except for $I_E$, for which it is 0). We have solved \eqs{recursions} and \eqref{bcs}, along with their counterparts for $Q$, $\tG$, $G_2$, $\widetilde{Q}$ and $\Gamma_2$, for $N_f = 2, 3, 4, 5, 6$. We plot our solution for the moment amplitude evolution with unit initial conditions in \fig{fig:amp3d}.
 
\begin{figure}[ht!]
\centering
\includegraphics[scale=0.6]{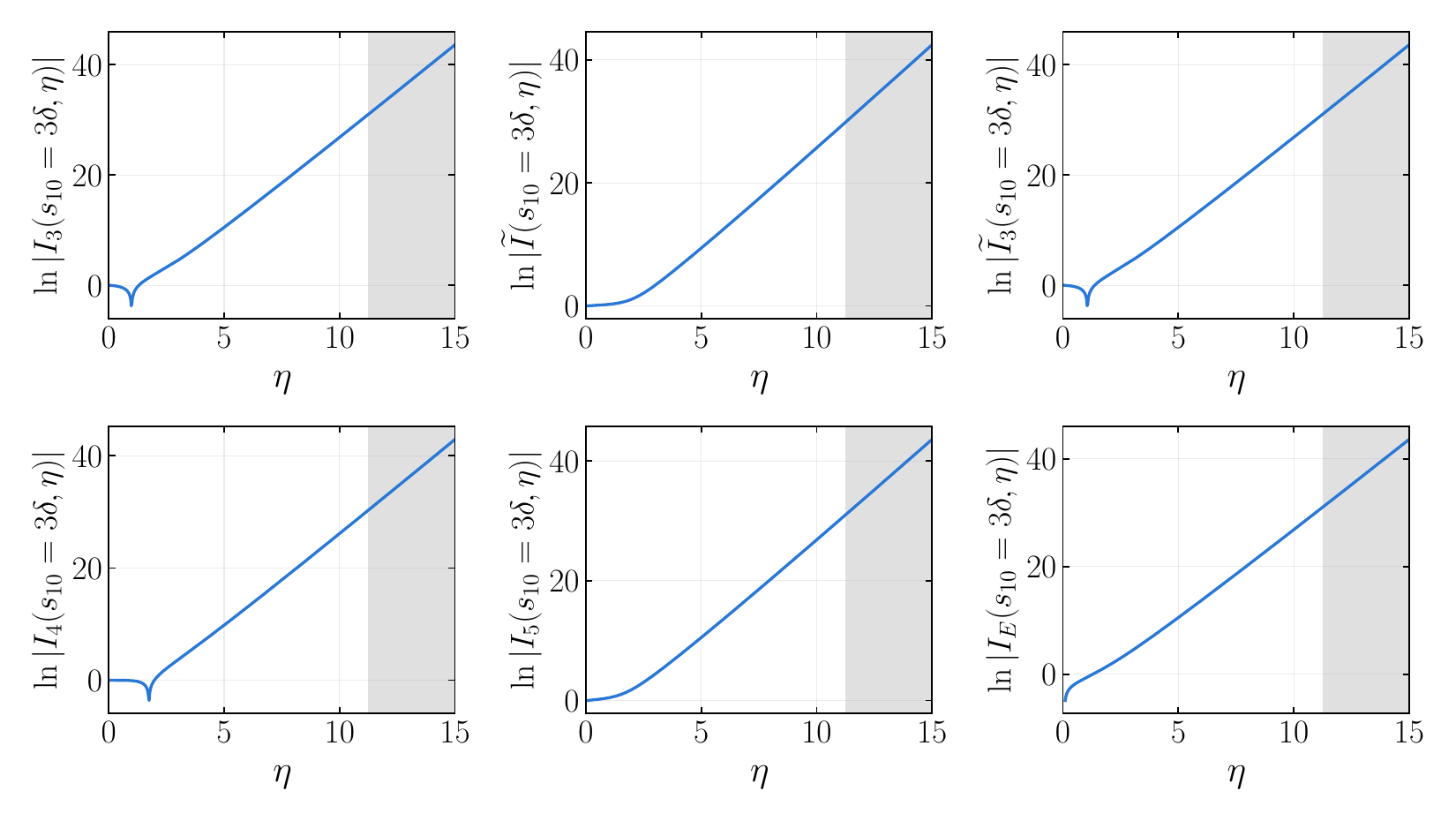}
\caption{Logarithms of the absolute values of the moment amplitudes as functions of $\eta$ for $s_{10} = 3 \delta$, at $N_f = 3$ and with a step size of $\delta = 0.025$. The amplitudes in top row plots are, from left to right, $I_3$, $\widetilde{I}$, and $\widetilde{I}_3$. The bottom row shows $I_4, I_5,$ and $I_E$. All the inhomogeneous terms are set to $1$ except for the one for $I_E$ which is 0. The shaded band marks the fitting region $0.75 \, \eta_{\rm max} \le \eta \le \eta_{\rm max}$ used to extract the intercepts as described in the main text.}
\label{fig:amp_vs_eta}
\end{figure}

\subsection{Small-$x$ asymptotics of the OAM distributions}
\label{sec:asymptotics}
 
From \fig{fig:amp3d} we see that the moment amplitudes grow exponentially with $\eta$ for small values of $s_{10}$, similar to the polarized dipole amplitudes \cite{Adamiak:2023okq}. This corresponds to a power-law growth in the center-of-mass energy squared $zs$. To extract the powers, also known as intercepts, we plot the logarithms of the absolute values of the moment amplitudes as functions of $\eta$ at fixed $s_{10}$ in \fig{fig:amp_vs_eta}. (For brevity, we omit the plots for the dipole amplitudes $Q, \widetilde{Q}, \widetilde{G}, G_2$ as they are very similar to the moment amplitude plots.) All the curves increase linearly with $\eta$ in the region sufficiently far away from $\eta = 0$: at small $\eta$ values the initial conditions and the discretization errors likely play a dominant role, so we want to avoid that region. We therefore have the following ansatz as $\eta \to \infty$ \cite{Kovchegov:2016weo, Kovchegov:2020hgb, Adamiak:2023okq, Kovchegov:2023yzd},
\begin{align}\label{intercept_ansatz}
  A(s_{10}, \eta) \sim \exp \left[ 
 \frac{\alpha_A}{\sqrt{\bar{\alpha}_s}} \, \eta\right] ,
\end{align}
where $A$ runs over $Q, \widetilde{Q}, \widetilde{G}, G_2, I_3, \widetilde{I}, \widetilde{I}_3, I_4, I_5$ and $I_E$ and the intercepts $\alpha_A$ are quoted below in units of $\sqrt{\bar{\alpha}_s}$. In our analysis below, we evaluate each amplitude at $s_{10} = 3 \delta$ rather than at $s_{10} = 0$ to avoid the following feature of the evolution kernel on the right-hand sides of \eqs{rescaled_eqns}: in the evolution equations for $I_4, I_5, \widetilde{I}, I_E$ the kernel vanishes for $s_{10} = 0$ exactly, leaving those amplitudes given by their inhomogeneous terms. Additionally, in \fig{fig:amp3d}, we have omitted plotting the $s_{10}=0$ and $s_{10} \geq \eta$ regions of $\ln|I_E(s_{10}, \eta)|$ since in these regions $I_E(s_{10}, \eta) = I_E^{(0)}(s_{10}, \eta) = 0$. 
 
Since this exponential growth is dominant for larger $\eta$, we deduce the $\alpha_A$'s by regressing the logarithms of the amplitudes over $0.75 \, \eta_{\rm max} \le \eta \le \eta_{\rm max}$ for a given step size $\delta$, where $\eta_{\rm max}$ is the maximum value of $\eta$ in our simulation \cite{Kovchegov:2016weo, Kovchegov:2020hgb, Adamiak:2023okq, Kovchegov:2023yzd}. We can take the continuum limit ($\delta \to 0$, $\eta_{\rm max} \to \infty$) by repeating this regression for other choices of $\delta$ and $\eta_{\rm max}$, fitting the results for $\alpha_A$ in the $(\delta, 1/\eta_{\rm max})$ space with a polynomial-model surface, and extrapolating this model to $\delta = 0$, $1/\eta_{\rm max} = 0$. We detail each of these step sizes and the range of $\eta_{\rm max}$ at a particular step size, given by $\eta_{\rm max} \in \{10, 20, \ldots, M(\delta)\}$, in Table~\ref{t:grids}.

\begin{table}[h!]
\centering
\begin{tabular}{c|c|c|c|c|c|c|c|c}
 $\delta$ & \hspace{0.15cm} 0.025 \hspace{0.15cm} & \hspace{0.15cm} 0.032 \hspace{0.15cm} & \hspace{0.15cm} 0.0375 \hspace{0.15cm} & \hspace{0.15cm} 0.05 \hspace{0.15cm} & \hspace{0.15cm} 0.0625 \hspace{0.15cm} & \hspace{0.15cm} 0.075 \hspace{0.15cm} & \hspace{0.15cm} 0.08 \hspace{0.15cm} &\hspace{0.15cm} 0.1 \hspace{0.15cm} \\ \hline
$M(\delta)$ & 15 & 20 & 25 & 30 & 35 & 50 & 60 & 70
\end{tabular}
\caption{Maximum values of $\eta_{\rm max}$, $M(\delta)$, for each step size $\delta$.}
\label{t:grids}
\end{table}
 
Similar to \cite{Kovchegov:2016weo, Kovchegov:2020hgb, Adamiak:2023okq, Kovchegov:2023yzd}, we model the intercepts $\alpha_A$ using $\delta$ and $1/\eta_{\rm max}$ as independent variables. We use polynomial regression models of various degrees, employing polynomials of different orders in $\delta$ and $1/\eta_{\rm max}$, weighted by the uncertainties of the intercepts, which we determine through regression at the 95\% confidence level. Once we have the best models, we can give an estimate for the intercepts in the continuum limit by extrapolating to $\delta = 0$, $1/\eta_{\rm max} = 0$. In particular we consider the following models with increasing polynomial degree:
\begin{itemize}
  \item Model 1: $\alpha_A = a_1$,
  \item Model 2: $\alpha_A = a_1 + a_2 \delta + \frac{a_3}{\eta_{\rm max}}$,
  \item Model 3: $\alpha_A = a_1 + a_2 \delta + \frac{a_3}{\eta_{\rm max}}
        + a_4 \delta^2 + \frac{a_5 \delta}{\eta_{\rm max}}
        + \frac{a_6}{\eta_{\rm max}^2}$,
  \item Model 4: $\alpha_A = a_1 + a_2 \delta + \frac{a_3}{\eta_{\rm max}}
        + a_4 \delta^2 + \frac{a_5 \delta}{\eta_{\rm max}}
        + \frac{a_6}{\eta_{\rm max}^2} + a_7 \delta^3
        + \frac{a_8 \delta^2}{\eta_{\rm max}}
        + \frac{a_9 \delta}{\eta_{\rm max}^2}
        + \frac{a_{10}}{\eta_{\rm max}^3}$.
\end{itemize}
Once we fit and evaluate all four models for each $\alpha_A$, we observe the Akaike information criterion (AIC) \cite{Akaike:1974vps} decrease significantly from model 1 to 2 and from model 2 to 3. The AIC has a minimum at model 3 for all intercepts and all values of $N_f$ we consider, while examining model 4 reveals a higher AIC and insignificant parameters. Therefore, we decide to use model 3, the quadratic model, for all of our intercepts, as was also done in the large $N_c$ limit in \cite{Kovchegov:2023yzd} (see also \cite{Kovchegov:2020hgb, Cougoulic:2022gbk, Adamiak:2023okq}). 
  
Taking into account the residuals of the quadratic model and the uncertainties at each data point, we arrive at the following estimates for the amplitude intercepts at $N_f = 3$ and $N_c =3$,
\begin{subequations} \label{intercepts_res}
\begin{align}
 \alpha_{I_3} &= 3.477(3) \, \sqrt{\bar{\alpha}_s}, \\
 \alpha_{\widetilde{I}}  &= 3.477(3) \, \sqrt{\bar{\alpha}_s}, \\
 \alpha_{\widetilde{I}_3} &= 3.477(3) \, \sqrt{\bar{\alpha}_s}, \\
 \alpha_{I_4} &= 3.477(3) \, \sqrt{\bar{\alpha}_s}, \\
 \alpha_{I_5} &= 3.477(3) \, \sqrt{\bar{\alpha}_s}, \\
 \alpha_{I_E} &= 3.477(3) \, \sqrt{\bar{\alpha}_s}, \\
 \alpha_{Q} &= 3.479(3) \, \sqrt{\bar{\alpha}_s}, \\
 \alpha_{\widetilde{Q}} &= 3.477(3) \, \sqrt{\bar{\alpha}_s}, \\
 \alpha_{\widetilde{G}} &= 3.480(3) \, \sqrt{\bar{\alpha}_s}, \\
 \alpha_{G_2} &= 3.477(3) \, \sqrt{\bar{\alpha}_s},
\end{align}
\end{subequations}
with the values for $N_f=2,3,4,5,6$ collected in Table~\ref{t:intercepts_Nf}. Comparing Table~\ref{t:intercepts_Nf} with the numerical results of \cite{Adamiak:2023okq}, we observe the contributions from $\widetilde{Q}$, not accounted for in \cite{Adamiak:2023okq}, slightly raise the intercepts. This is consistent with the analytic result of \cite{Borden:2025ehe}. Furthermore, our intercepts themselves, given in Table~\ref{t:intercepts_Nf}, are consistent with the analytic results in \cite{Borden:2025ehe}. However, our numerical solution is not accurate enough to verify the discrepancy with BER reported in \cite{Borden:2025ehe} for hPDFs. We also observe that the moment amplitudes share the same intercept as the helicity amplitudes, exactly as in the large-$N_c$ limit \cite{Kovchegov:2023yzd, Manley:2024pcl}.
\begin{table}[ht]
\centering
\begin{tabular}{c|ccccc}
$N_f$ &  2  & 3 & 4 & 5 & 6 \\ \hline
$\alpha_{I_3}$ & \hspace{0.15cm}$3.544(3) $\hspace{0.15cm} & \hspace{0.15cm}$3.477(3) $\hspace{0.15cm} & \hspace{0.15cm}$3.404(3)$\hspace{0.15cm} & \hspace{0.15cm}$3.319(3)$\hspace{0.15cm}  & \hspace{0.15cm}$3.218(3)$\hspace{0.15cm} \\
$\alpha_{\widetilde{I}}$   & $3.544(3)$ & $3.477(3) $ & $3.404(3)$ & $3.319(3)   $ & $3.218(3) $ \\
$\alpha_{\widetilde{I}_3}$   & $3.544(3)$ & $3.477(3) $ & $3.404(3)$ & $3.319(3)   $ & $3.218(3) $ \\
$\alpha_{I_4}$   & $3.544(3)$ & $3.477(3) $ & $3.404(3)$ & $3.319(3)   $ & $3.218(3) $ \\
$\alpha_{I_5}$   & $3.544(3)$ & $3.477(3) $ & $3.404(3)$ & $3.319(3)   $ & $3.218(3) $ \\
$\alpha_{I_E}$   & $3.544(3)$ & $3.477(3) $ & $3.404(3)$ & $3.319(3) $ & $3.218(3) $ \\
$\alpha_{Q}$   & $3.545(3)$ & $3.479(3) $ & $3.405(3) $ & $3.320(3) $ & $3.219(3)$ \\
$\alpha_{\widetilde{Q}}$  & $3.544(3)$ & $3.477(3)$ & $3.403(3)$ & $3.319(3)$ & $3.218(3)$ \\
$\alpha_{\widetilde{G}}$  & $3.546(3)$ & $3.480(3)$ & $3.406(3)$ & $3.321(3)$ & $3.220(3)$ \\
$\alpha_{G_2}$   & $3.544(3)$ & $3.477(3) $ & $3.404(3)$ & $3.319(3)$ & $3.218(3)$
\end{tabular}
\caption{Continuum limit intercepts, in units of $\sqrt{\bar{\alpha}_s}$, of the moment and dipole amplitudes for $N_f=2,3,4,5,6$ and $N_c =3$.}
\label{t:intercepts_Nf}
\end{table}

Finally, employing \eqs{intercept_ansatz} and \eqref{intercepts_res} in \eqs{OAMs_corr}, we obtain the following large-$N_c \& N_f$, small-$x$ asymptotics for the quark and gluon OAM distributions,
\begin{align}\label{oam_asym}
  L_{q+\bar{q}}(x, Q^2) \sim L_G(x, Q^2) \sim \Delta \Sigma(x, Q^2)
  \sim \Delta G(x, Q^2)
  \sim \left( \frac{1}{x} \right)^{3.48 \sqrt{\bar{\alpha}_s}},
\end{align}
quoted here for $N_f = 3$ and $N_c=3$. As in the large-$N_c$ case, the OAM distributions and the helicity PDFs share the same intercept. Comparing \eq{oam_asym} to the analogous result in the large-$N_c$ limit from \cite{Kovchegov:2023yzd}, we see that including the quarks reduces the intercept below its large-$N_c$ value of $3.66$, following the same trend already established for the helicity PDFs \cite{Adamiak:2023okq, Borden:2025ehe}. Our result \eqref{oam_asym} is also in agreement with the analysis of the OAM asymptotics in \cite{Boussarie:2019icw} using IREE technique of \cite{Bartels:1996wc} — any potential discrepancies with the result in \cite{Boussarie:2019icw}, as expected on the basis of \cite{Manley:2024pcl}, are beyond the precision of our numerical approximation.


\subsection{Ratio of the OAM distributions to the helicity PDFs}
\label{sec:ratios}
 
Another important asymptotic quantity to investigate beyond the small-$x$ intercepts of the OAM distributions is the ratio of the OAM distributions to the helicity PDFs at very small $x$, which was previously studied in \cite{Boussarie:2019icw, Hatta:2016aoc, Hatta:2018itc, Kovchegov:2023yzd}. In
the large-$N_c$ analysis of \cite{Kovchegov:2023yzd}, constructing the quark ratio required integrating the amplitudes over both $s_{10}$ and $\eta$. The revised expressions \eqref{OAMs_corr} and \eqref{pdfs} simplify the task considerably, with the ratios now given by 
\begin{subequations}\label{ratios_def}
\begin{align}
  \frac{L_{q+\bar{q}}(x, Q^2)}{\Delta \Sigma (x, Q^2)}
  &=  \frac{2\,\tI}{\widetilde{Q}}
     \left( \xo^2 = \frac{1}{Q^2}, \, s = \frac{Q^2}{x} \right) ,
  \\
  \frac{L_G(x, Q^2)}{\Delta G(x, Q^2)}
  &= - \, \frac{2 \, I_4 + 3 \, I_5}{G_2}
     \left( \xo^2 = \frac{1}{Q^2}, \, s = \frac{Q^2}{x} \right) .
\end{align}
\end{subequations}
In terms of the variables \eqref{rescaled}, \eqs{ratios_def} become
\begin{subequations}\label{ratios_def_scal}
\begin{align}
  \frac{L_{q+\bar{q}}(x, Q^2)}{\Delta \Sigma (x, Q^2)}
  &=  \frac{2\,\tI}{\widetilde{Q}}
     \left( s_{10} = \sqrt{\bar{\alpha}_s} \ln \frac{Q^2}{\Lambda^2}, \, \eta = \sqrt{\bar{\alpha}_s} \left[ \ln \frac{Q^2}{\Lambda^2} + \ln \frac{1}{x} \right] \right),
  \\
  \frac{L_G(x, Q^2)}{\Delta G(x, Q^2)}
  &= - \, \frac{2 \, I_4 + 3 \, I_5}{G_2}
      \left( s_{10} = \sqrt{\bar{\alpha}_s} \ln \frac{Q^2}{\Lambda^2}, \, \eta = \sqrt{\bar{\alpha}_s} \left[ \ln \frac{Q^2}{\Lambda^2} + \ln \frac{1}{x} \right] \right).
\end{align}
\end{subequations}

\begin{figure}[ht!]
\centering
\includegraphics[width=1\textwidth]{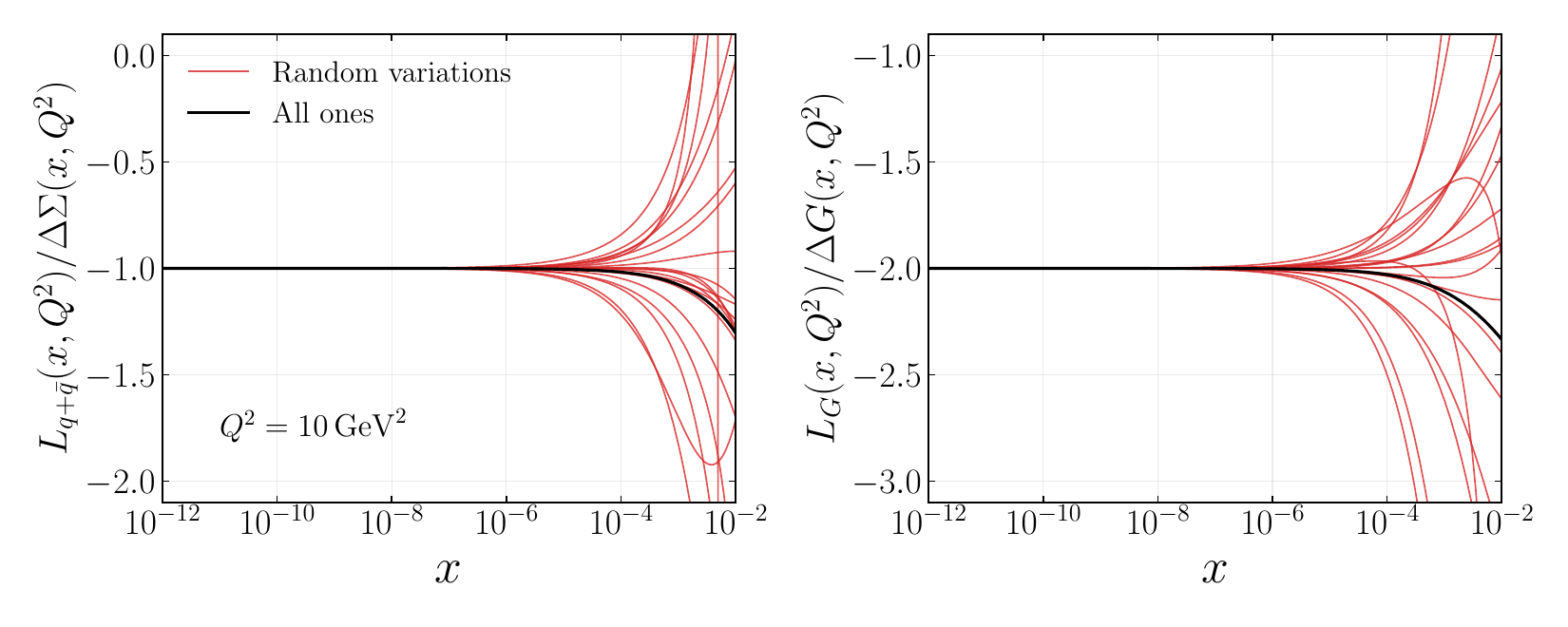}
\caption{Ratios of the OAM distributions to the helicity PDFs for quarks (left
panel) and gluons (right panel) as functions of $x$ at
$Q^2 = 10 \, \mathrm{GeV}^2$, for $N_f = 3$, $N_c =3$, $\as = 0.25$, and $\delta = 0.025$. The black curve is obtained via setting the initial conditions of all polarized dipole and moment amplitudes to one. The red curves are obtained by using the linear ansatz in \eq{lin_ans} along with random numbers for the coefficients for the initial conditions.}
\label{fig:ratios_Y}
\end{figure}

The OAM to hPDF ratios are plotted in \fig{fig:ratios_Y}, for both the ``all ones'' initial conditions used in the previous Section, and variations where the initial conditions for each amplitude are chosen randomly. That is, we use the linear ansatz of \cite{Adamiak:2021ppq, Adamiak:2023yhz, JAMCollaborationSmall-xAnalysisGroup:2025tfa} 
\begin{align} \label{lin_ans}
    A^{(0)}(s_{10}, \eta) = a_A \, \eta + b_A\, s_{10} + c_A,
\end{align}
and choose the parameters $a_A, b_A, c_A$ randomly in the range $[-20, 20]$ (for all amplitudes and moment amplitudes except $I_E$, for which the initial conditions are zero). This range is consistent with the results of the small-$x$ helicity phenomenology program in \cite{Adamiak:2021ppq, Adamiak:2023yhz, JAMCollaborationSmall-xAnalysisGroup:2025tfa}. The ratios are illustrated as a function of $x$ at $Q^2 = 10 \, \mathrm{GeV}^2$ with $\Lambda = 1 \, \mathrm{GeV}$. As demonstrated in \fig{fig:ratios_Y}, the OAM to hPDF ratios are strongly dependent on the initial conditions of the evolution all the way down to $x \sim 10^{-7}$. At this small of an $x$, even at $Q^2 = 10$~GeV$^2$, saturation effects will likely become important, even in the proton, and the linear approximation (via the DLA) we are using here will become invalid. Therefore, the asymptotic ratios we extract below are mostly of academic value, and a useful cross check of our formalism. Nevertheless, an interesting physical interpretation of these ratios has been proposed in \cite{Bhattacharya:2024sck, Bhattacharya:2024sno} based on the idea that small $x$ partons are maximally entangled.  
 
Both the quark and gluon ratios in \fig{fig:ratios_Y} appear to approach a constant with decreasing $x$. Following \cite{Kovchegov:2023yzd}, we parametrize the asymptotic form ($x\to 0$) of these ratios via
\begin{subequations}\label{m2_ansatz}
\begin{align}
  \frac{L_{q+\bar{q}}(x, Q^2)}{\Delta \Sigma(x, Q^2)} &= A_q + \frac{B_q}{\ln(1/x)} + \mathcal{O}\left( \frac{1}{\ln^{2}(1/x)} \right),
  \\ 
  \frac{L_G(x, Q^2)}{\Delta G(x, Q^2)} &= A_G + \frac{B_G}{\ln (1/x)}
  + \mathcal{O}\left( \frac{1}{\ln^{2}(1/x)} \right).
\end{align}
\end{subequations}
Similar to the analysis in \cite{Manley:2024pcl} (see also Appendix B of \cite{Kovchegov:2023yzd} and \cite{Borden:2025ehe}), one can probably adapt the method in \cite{Borden:2025ehe} to obtain an analytic solution of \eqs{rescaled_eqns}, and subsequently justify the form of the ansätze in \eqs{m2_ansatz}. We leave such an exercise for future work.  

 Analogous to our method of extracting the intercepts in the previous Section, we can fit the ratios directly to extract $A$ and $B$ for a set of $\delta$ values and maximum values of $\ln(1/x)$. Then we can repeat the procedure as described in the previous Section to get an estimate for the parameters in the continuum limit ($\delta \to 0$ and $1/\ln(1/x)_{\rm max} \to 0$). The set of $\delta$ values along with their associated maximum values of $\ln(1/x)$ are shown in Table~\ref{t:grids_ymax}. Note we use more $\delta$ values when analyzing the ratios as compared to the intercepts since the ratios are inherently sensitive to the sub-asymptotic behavior of the OAM distributions and helicity PDFs and therefore require slightly higher numerical precision. 
 \begin{table}[ht!]
\centering
\setlength{\tabcolsep}{4pt}
\begin{tabular}{c|c|c|c|c|c|c|c|c|c|c|c|c|c|c|c|c}
 $\delta$ &  0.025 & 0.03 & 0.035 & 0.04 & 0.045 & 0.05 & 0.055 & 0.06 & 0.065 &  0.07 &  0.075   &  0.08  &  0.085  &  0.09  &  0.095 &  0.1 
 \\ \hline
$M_{\ln(1/x)}(\delta)$ & 63 & 76 & 89 & 102 & 115 & 128 & 141 & 154 & 167 & 180 & 193 & 206 & 219 & 232 & 245 & 258 
\end{tabular}
\caption{Maximum values of $\ln(1/x)_{\rm max}$, $M_{\ln(1/x)}(\delta)$, for each step size $\delta$.}
\label{t:grids_ymax}
\end{table}

 To avoid dependence on the specific initial conditions we choose, we set the minimum value of $\ln(1/x)_{\mathrm{min}} \approx 20$, corresponding to $x \approx 10^{-9}$. Similar to what we observed above for the intercepts, quadratic models work best for all the parameters in \eq{m2_ansatz}. The resulting continuum limit parameters we extract from our numerical solution give the following ratios,
\begin{subequations}\label{m2_res}
\begin{align}
  \frac{L_{q+\bar{q}}(x, Q^2 = 10 \, \mathrm{GeV}^2)}
       {\Delta \Sigma(x, Q^2 = 10 \, \mathrm{GeV}^2)}
  &= - 0.9999995(9)  - \frac{0.00003(4)}{\ln (1/x)} ,
  \\
  \frac{L_G(x, Q^2 = 10 \, \mathrm{GeV}^2)}
       {\Delta G(x, Q^2 = 10 \, \mathrm{GeV}^2)}
  &= -1.999999(2) - \frac{0.00006(7)}{\ln(1/x)},
\end{align}
\end{subequations}
for $N_f=3$ and $N_c=3$. The ratios for $N_f=2,3,4,5,6$ are collected in Table~\ref{t:ratio_coeffs}. 

\begin{table}[ht]
\centering
\begin{tabular}{c|cc|cc}
$N_f$ & $A_q$ & $B_q$ & $A_G$ & $B_G$ \\ \hline
2 & $-0.999999(1)$ & $-0.00004(4)$ & $-1.999999(2)$ & $-0.00008(9)$ \\
3 & $-0.9999995(9)$ & $-0.00003(4)$ & $-1.999999(2)$ & $-0.00006(7)$ \\
4 & $-0.9999997(7)$ & $-0.00002(3)$ & $-2.000000(1)$  & $-0.00004(5)$ \\
5 & $-0.9999999(4)$ & $-0.00001(2)$ & $-1.9999998(8)$  & $-0.00002(3)$ \\
6 & $-1.0000000(2)$ & $-0.000004(9)$ & $-2.0000001(3)$  & $-0.00001(1)$
\end{tabular}
\caption{Continuum-limit coefficients of the ratios in \eqs{m2_ansatz},
for each $N_f=2,3,4,5,6$, at $N_c=3$, $Q^2 = 10\,\mathrm{GeV}^2$, $\alpha_s = 0.25$, and $\Lambda=1\, \mathrm{GeV}$.}
\label{t:ratio_coeffs}
\end{table}

While the intercepts of Table~\ref{t:intercepts_Nf} vary appreciably with $N_f$, that variation is common to the OAM distributions and the helicity PDFs and cancels in the ratios. This is reflected in Table~\ref{t:ratio_coeffs}, where we see that all the coefficients are numerically consistent with each other for $2\leq N_f \leq 6$.
 
The results in \eqs{m2_res} and Table~\ref{t:ratio_coeffs} should be compared to the predictions from \cite{Boussarie:2019icw}. Taking Eqs.~(6) and~(7) from \cite{Boussarie:2019icw} while assuming that the parameter $\alpha$ from \cite{Boussarie:2019icw} is perturbatively small, we get
\begin{align}\label{BHY}
  \frac{L_{q+\bar{q}}(x, Q^2)}{\Delta \Sigma(x, Q^2)} = -1 ,
  \qquad
  \frac{L_G(x, Q^2)}{\Delta G(x, Q^2)} = -2 .
\end{align}
Comparing \eq{BHY} to \eqs{m2_res} and Table~\ref{t:ratio_coeffs}, we see that we are in exact numerical agreement with \cite{Boussarie:2019icw} in both the quark and the gluon sectors. Furthermore, our ratios remain close to the ratios obtained in the large-$N_c$ limit in \cite{Kovchegov:2023yzd}. However, we note that, in \cite{Kovchegov:2023yzd}, the $B$-coefficient of both the quark and gluon ratios, defined in \eqs{m2_ansatz}, were found to be non-zero. By contrast, both the quark and gluon $B$-coefficients found above are numerically consistent with zero for all $N_f$: this difference is attributable to the $I_E$-terms, which are included here, but not in \cite{Kovchegov:2023yzd}. Nevertheless, we conclude that the quark contributions retained here (along with the type-3 $I_E$ contributions) affect the ratios far less than they affect the intercept itself.


\section{$F^{+-}$ contribution to elastic dijet production in electron-proton collisions}

\label{sec:dijet}

Having established in \Sec{sec:evolution} that the type-3 amplitude $G^{[3]}_{10}$ and its moment $I_E$ enter the DLA evolution of the moment amplitudes, we must now revisit the elastic dijet production cross section computed in \cite{Kovchegov:2024wjs}. That calculation employed the sub-eikonal quark $S$-matrix in  \eqref{Wilsonl} (and the analogous antiquark one given explicitly in \cite{Kovchegov:2024wjs}), which was constructed from the type-1 and type-2 polarized Wilson lines only. The operator $V^{\mathrm{G}[3]}_{\un x}$ of \eq{VG3} was not included there, since it arises from the $x^+$-dependence of the light-cone Wilson line and was believed not to couple to the proton's longitudinal spin at the order considered. Although this is true for DLA helicity evolution, as we saw above, $G^{[3]}_{10}$ does couple to the moment amplitudes and, hence, to the OAM distributions, through evolution. Consistency therefore demands that we also compute its direct contribution to the dijet observable. 

\subsection{$F^{+-}$ contributions to the elastic dijet production amplitude}
\label{sec:t3_amplitude}

\begin{figure}[ht!]
    \centering
    \includegraphics[scale=0.6]{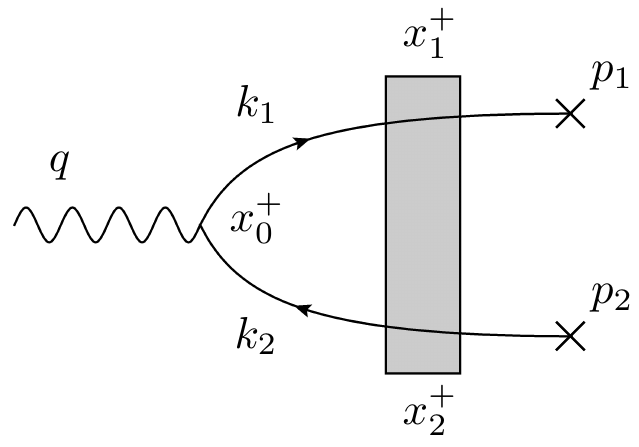}
    \caption{Diagram contributing to elastic dijet production. The shaded rectangle represents the target shockwave, and the crosses denote the measured quark and antiquark.}
    \label{fig:dijet}
\end{figure}

Consider the elastic dijet production amplitude depicted in \fig{fig:dijet}. We work in the dipole/Trento frame \cite{Bacchetta:2004jz, Braun:2005rg, Mantysaari:2020lhf, Bhattacharya:2022vvo,Bhattacharya:2024sck,Hatta:2016aoc, Kovchegov:2024wjs, Kovchegov:2024aus}, where the virtual photon has zero transverse momentum and the photon ($q$) and the incoming proton ($P$) momenta are
\begin{subequations}
    \begin{align}
        q^\mu &= \left( -\frac{Q^2}{2q^-}, q^- , \un 0_\perp \right),
        \\ 
        P^\mu &\approx (P^+, 0^-, \un 0_\perp). \label{P}
    \end{align}
\end{subequations}
The scattering is elastic: the outgoing proton has momentum $P'$. 

As follows from \eq{V_expansion}, to account for the type-3 operators, we need to carefully treat the minus-momentum dependence of the scattering amplitude, or, equivalently, $x^+$-dependence of the operators.  Suppressing the helicity and color structures, which play a passive role here, the reduced scattering amplitude ($A=M/(2 s)$ with $M$ the standard scattering amplitude) reads
\begin{align} \label{t3_a1}
    i A_\lambda \, & \, (2\pi) \, \delta (p_1^- + p_2^- - q^- - \Delta^-) = 
    \int \limits_{0}^{\infty} \frac{dk_1^-}{2\pi 2k_1^-} \frac{dk_2^-}{2\pi 2k_2^-} \,
    \int \limits^{\infty}_{-\infty} dx_1^+ \, \int \limits^{\infty}_{-\infty} dx_2^+ \,  \int \limits^{\infty}_{-\infty} dx_0^+ \,
     e^{i x_1^+ \left( p_1^- - k_1^- \right) + i x_2^+ \left( p_2^- - k_2^- \right)}  e^{i x_0^+ \left( k_1^- + k_2^-  - q^- \right)}
     \notag
     \\  & \times
      \int d^2 x_1 \, d^2 x_2 \, e^{-i \un p_1 \cdot \un x_1} \, e^{- i \un p_2 \cdot \un x_2} \,
     \Psi^{\gamma^* \to q\bar{q}} \left( \un x_{12}, \frac{k_1^-}{q^-}, \frac{k_2^-}{q^-} \right) \,
     \Big \langle \tord \tr \left[
     V_{\un 1}(x_1^+) V^\dagger_{\un 2} (x_2^+) -1 \right]
     \Big \rangle 2 \sqrt{p_1^- k_1^-} 2 \sqrt{p_2^- k_2^-},
\end{align}
where $k_1^-$ and $k_2^-$ are the minus momenta of the quark and anti-quark lines adjacent to the photon vertex, while $p_1^-$ and $p_2^-$ are the minus momenta of the measured jets. Minus momentum components are not conserved yet, this should result from the integrations over the vertex positions $x_0^+$, $x_1^+$, and $x_2^+$, as indicated in \fig{fig:dijet}. Due to this (temporary) momentum non-conservation, we have to keep both the $k_1^-$ and $k_2^-$ arguments explicitly in the wave function $\Psi^{\gamma^* \to q\bar{q}}$ in \eq{t3_a1}. 

Our calculation broadly follows that in \cite{Altinoluk:2023lly}. We change the integration variables to 
\begin{align}
    y_1^+ \equiv x_1^+ - x_0^+, \ \ \ y_2^+ \equiv x_2^+ - x_0^+ 
\end{align}
and notice that since the angle brackets averaging is related to the non-forward (momentum) matrix element by 
\cite{Kovchegov:2015zha, Wu:2017rry, Kovchegov:2019rrz, Cougoulic:2020tbc, Kovchegov:2013cva, Kovchegov:2026gwb} (cf. \eq{double-bracket})
\begin{align}\label{sat_ave}
  \bra{P' = P - \Delta, S_L} \hat{\cal O} \left( x_1 - b , x_2 - b \right) \ket{P, S_L}  = 2 P^+ \, \int db^- \, d^2 b_\perp \, e^{i \Delta^+ \, b^- - i {\un \Delta} \cdot {\un b}} \, \left\langle \hat{\cal O} \left( {x}_1, {x}_2 \right) \right\rangle  
\end{align}
for some bi-local operator $\hat{\cal O} \left( {x}_1, {x}_2 \right)$ with $b= (0^+, b^-, \un b)$ the impact parameter, we can integrate out $x_0^+$ and write 
\begin{align} \label{t3_a2}
    i A_\lambda = & \,
    \int \limits_{0}^{\infty} \frac{dk_1^-}{2\pi} \frac{dk_2^-}{2\pi} \,
    \int \limits^{\infty}_{-\infty} dy_1^+ \, \int \limits^{\infty}_{-\infty} dy_2^+ \, 
     e^{i y_1^+ \left( p_1^- - k_1^- \right) + i y_2^+ \left( p_2^- - k_2^- \right)} 
     \notag
     \\  & \times
      \int d^2 x_1 \, d^2 x_2 \, e^{-i \un p_1 \cdot \un x_1} \, e^{- i \un p_2 \cdot \un x_2} \, 
     \Psi^{\gamma^* \to q\bar{q}} \left( \un x_{12}, \frac{k_1^-}{q^-}, \frac{k_2^-}{q^-} \right) \,
        \sqrt{\frac{p_1^-}{k_1^-}} \sqrt{\frac{p_2^-}{k_2^-}}
     \Big \langle \tord \tr \left[
     V_{\un 1}(y_1^+) V^\dagger_{\un 2} (y_2^+) -1 \right]
     \Big \rangle .
\end{align}
We have dropped the overall factor of $(2\pi) \, \delta (p_1^- + p_2^- - q^- - \Delta^-)$ on both sides of \eq{t3_a2}: however, now $p_1^- + p_2^- = q^- + \Delta^- \approx q^-$ since $\Delta^-$ is negligibly small,
\begin{align}\label{Delta_minus}
    \frac{\Delta^-}{q^-}  = \frac{P^{\prime 2}_\perp}{2 P^{\prime +} q^-}
    \approx \frac{\Delta_\perp^2}{s},
\end{align}
leading to sub-sub-eikonal corrections to the spin-dependent terms we will keep. 

Expanding the Wilson lines,
\begin{align}
    V_{\un 1}(y_1^+) V^\dagger_{\un 2} (y_2^+) = V_{\un 1}(0^+) V^\dagger_{\un 2} (0^+) + y_1^+ \, \left( \pd^- V_{\un 1}(0^+) \right) V^\dagger_{\un 2} (0^+) + V_{\un 1}(0^+) \, y_2^+ \left( \pd^- V^\dagger_{\un 2} (0^+) \right) + {\cal O} \left( (y_1^+)^2, (y_2^+)^2 , y_1^+ \, y_2^+ \right),
\end{align}
we can integrate out $y_1^+$, $y_2^+$, restoring minus momentum conservation, and then integrate over $k_1^-, k_2^-$, obtaining 
\begin{align}\label{t3_a3}
   i A_\lambda = & \, 
      \int d^2 x_1 \, d^2 x_2 \, e^{-i \un p_1 \cdot \un x_1} \, e^{- i \un p_2 \cdot \un x_2} \, \Bigg\langle \Psi^{\gamma^* \to q\bar{q}}( \un x_{12}, z, 1- z) \, \tord \tr \left[
     V_{\un 1} V^\dagger_{\un 2}  -1 \right] \notag \\
      & + 2 i \, 
     \Bigg\{ z \, \left[ \frac{\pd}{\pd z} \Psi^{\gamma^* \to q\bar{q}}( \un x_{12}, z, \bar z) \right] \, \tord \tr \left[
     V_{\un 1}^{\mathrm{G} [3]} V^\dagger_{\un 2}  \right] +  (1-z) \, \left[ \frac{\pd}{\pd \bar z} \Psi^{\gamma^* \to q\bar{q}}( \un x_{12}, z, \bar z) \right] \, \tord \tr \left[
     V_{\un 1} V^{\mathrm{G} [3] \, \dagger}_{\un 2}  \right] 
     \notag \\ 
     & 
      - \frac{1}{2} \Psi^{\gamma^* \to q\bar{q}}(\un x_{12}, z, \bar{z}) \Bigg( \tord \tr \left[ V_{\un 1}^{\mathrm{G}[3]} V_{\un 2}^\dagger \right] 
      + \tord \tr \left[ V_{\un 1} V_{\un 2}^{\mathrm{G}[3] \dagger} \right]
      \Bigg) 
     \Bigg\} \Bigg\rangle ,
\end{align}
where we have dropped $0^+$ in the arguments of the Wilson lines for brevity. We have defined
\begin{align}
z \equiv \frac{k_1^-}{q^-} = \frac{p_1^-}{q^-}, \ \ \ \bar z \equiv \frac{k_2^-}{q^-} = \frac{p_2^-}{q^-}
\end{align}
as two independent variables. Since now the minus momentum is conserved, after differentiation we put $\bar z = 1-z$ in \eq{t3_a3}. In arriving at \eq{t3_a3}, we have also discarded the terms in which the $k_1^-$ and $k_2^-$ derivatives act on the correlators, since these remove a logarithm of energy and are therefore sub-sub-eikonal. Switching to the double angle bracket notation, we write 
\begin{align} \notag 
    iA_\lambda = & \, 
      \int d^2 x_1 \, d^2 x_2 \, e^{-i \un p_1 \cdot \un x_1} \, e^{- i \un p_2 \cdot \un x_2} \, \Bigg\{ \Psi^{\gamma^* \to q\bar{q}}( \un x_{12}, z, 1- z) \, \left\langle \tord \tr \left[
     V_{\un 1} V^\dagger_{\un 2}  -1 \right] \right\rangle \\
      & + \frac{2 \, i}{s} \, 
     \left[ \left( \frac{\pd}{\pd z} \Psi^{\gamma^* \to q\bar{q}}( \un x_{12}, z, \bar z) \right) \, \llangle \tord \tr \left[
     V_{\un 1}^{\mathrm{G} [3]} V^\dagger_{\un 2}  \right] \rrangle (s) +   \left( \frac{\pd}{\pd \bar z} \Psi^{\gamma^* \to q\bar{q}}( \un x_{12}, z, \bar z) \right) \, \llangle \tord \tr \left[
     V_{\un 1} V^{\mathrm{G} [3] \, \dagger}_{\un 2}  \right] \rrangle (s)
     \right] 
     \notag  \\ 
     &
        - \frac{i}{s} \Psi^{\gamma^* \to q\bar{q}}(\un x_{12}, z, \bar{z}) \Bigg[ 
        \frac{1}{z} \llangle \tord \tr \left[
     V_{\un 1}^{\mathrm{G} [3]} V^\dagger_{\un 2}  \right] \rrangle (s) + \frac{1}{1-z} \llangle \tord \tr \left[
     V_{\un 1} V^{\mathrm{G} [3] \, \dagger}_{\un 2}  \right] \rrangle (s)
        \Bigg]
     \Bigg\}.
     \label{t3_a4}
\end{align}
Following \cite{Kovchegov:2024wjs}, we have put the arguments of the sub-eikonal correlators to be $zs \sim (1-z)s \sim s$ since we are assuming $z\sim 1-z \sim \mathcal{O}(1)$. Otherwise, we would need to include small-$x$ evolution corrections in the rapidity interval between the quark and antiquark jets. 

While $\Delta^-$ is energy-suppressed, per \eq{Delta_minus}, this does not necessarily imply that the minus momentum transfers to the quark ($\Delta_1^-$) and anti-quark ($\Delta_2^-$) lines in \fig{fig:dijet} are also small: we only know that their sum is small, $\Delta^- = \Delta_1^- + \Delta_2^- = {\cal O} (1/P^+)$. Since
\begin{align}
    \Delta^- = \Delta_1^- + \Delta_2^- \sim (\pd^-_1 + \pd^-_2) \, \left\langle \tord \tr \left[  V_{\un 1}(x_1^+) V^\dagger_{\un 2} (x_2^+) \right] \right\rangle \approx \left\langle \tord \tr \left[  \left( \pd^- V_{\un 1} \right) \, V^\dagger_{\un 2} \right] \right\rangle + \left\langle \tord \tr \left[  V_{\un 1} \, \left( \pd^- V^\dagger_{\un 2} \right) \right] \right\rangle
\end{align}
(where we suppress the $0^+$ arguments of the Wilson lines on the right-hand side) and $\Delta^-$ is small, we conclude that 
\begin{align}\label{zero_tr}
    \left\langle \tord \tr \left[  \left( \pd^- V_{\un 1} \right) \, V^\dagger_{\un 2} \right] \right\rangle + \left\langle \tord \tr \left[  V_{\un 1} \, \left( \pd^- V^\dagger_{\un 2} \right) \right] \right\rangle = {\cal O} \left( \frac{1}{P^+} \right) \approx 0. 
\end{align}
This result, in turn, can be rewritten as
\begin{align}\label{zero_tr2}
    \llangle \tord \tr \left[  V_{\un 1}^{\mathrm{G} [3]}  \, V^\dagger_{\un 2} \right] \rrangle + \llangle \tord \tr \left[  V_{\un 1} \, V^{\mathrm{G} [3] \, \dagger}_{\un 2} \right] \rrangle = {\cal O} \left( \frac{1}{P^+} \right) \approx 0. 
\end{align}
Employing \eq{zero_tr2}, we rewrite \eq{t3_a4} as
\begin{align}\label{t3_a5}
   & i A_\lambda =  \, 
      \int d^2 x_1 \, d^2 x_2 \, e^{-i \un p_1 \cdot \un x_1} \, e^{- i \un p_2 \cdot \un x_2} \, \Bigg\{ \Psi^{\gamma^* \to q\bar{q}}( \un x_{12}, z, 1- z) \, \left\langle \tord \tr \left[
     V_{\un 1} V^\dagger_{\un 2}  -1 \right] \right\rangle \\
      & + \frac{i}{s} \, 
    \left[ \frac{\pd}{\pd z} \Psi^{\gamma^* \to q\bar{q}}( \un x_{12}, z, 1- z)  + \frac{z- \frac{1}{2}}{z(1-z)}
    \Psi^{\gamma^* \to q\bar{q}}( \un x_{12}, z, 1- z) 
    \right]\,  \Bigg[  \llangle \tord \tr 
    \left[
     V_{\un 1}^{\mathrm{G} [3]} V^\dagger_{\un 2}  \right] \rrangle (s) - \llangle \tord \tr \left[
     V_{\un 1} V^{\mathrm{G} [3] \, \dagger}_{\un 2}  \right] \rrangle (s)
     \Bigg] 
     \Bigg\},     \notag 
\end{align}
where now the $z$-derivative acts on both the $z$ and $1-z$ arguments of the light-cone wave function. Returning to the conventional notation for the latter, 
\begin{align}
    \Psi^{\gamma^* \to q\bar{q}}( \un x_{12}, z, 1- z) \to \Psi^{\gamma^* \to q\bar{q}}( \un x_{12}, z),
\end{align}
and defining
\begin{align} \label{Psi01}
    \frac{\partial}{\partial z}
    \Psi^{\gamma^* \to q\bar{q}}(\un x_{12}, z)
    \equiv  \Psi^{\gamma^* \to q\bar{q}\, (01)}(\un x_{12}, z) ,
\end{align}
we finally write 
\begin{align} \notag 
    &i A_\lambda =  \, 
      \int d^2 x_1 \, d^2 x_2 \, e^{-i \un p_1 \cdot \un x_1} \, e^{- i \un p_2 \cdot \un x_2} \, \Bigg\{ \Psi^{\gamma^* \to q\bar{q}}( \un x_{12}, z) \, \left\langle \tord \tr \left[
     V_{\un 1} V^\dagger_{\un 2}  -1 \right] \right\rangle \\
      & + \frac{i}{s} \, \Bigg[
    \Psi^{\gamma^* \to q\bar{q}\, (01)}(\un x_{12}, z) + \frac{z- \frac{1}{2}}{z(1-z)}
    \Psi^{\gamma^* \to q\bar{q}}( \un x_{12}, z) 
    \Bigg]  \Bigg[ \llangle \tord \tr \left[
     V_{\un 1}^{\mathrm{G} [3]} V^\dagger_{\un 2}  \right] \rrangle (s) - \llangle \tord \tr \left[
     V_{\un 1} V^{\mathrm{G} [3] \, \dagger}_{\un 2}  \right] \rrangle (s)
     \Bigg] 
     \Bigg\}.
     \label{t3_a6}
\end{align}


\subsection{$F^{+-}$ contributions to the double spin asymmetry}
\label{sec:t3_dsa}

We now construct the type-3 contribution to the elastic dijet cross section. In terms of the photon-proton cross section, $\sigma_{\lambda \lambda'}^{\gamma^* p}$ with $\lambda (\lambda')$ the polarization of the photon in the (complex conjugate) amplitude, the elastic dijet DSA can be written as 
\begin{align} \label{DSA}
        E_{k^\prime} \frac{d \sigma^{DSA}}{d^3 k^\prime} &= 
        \frac{\alpha_{EM}}{4\pi^2 \, Q^2}
         \half \, \sum_{S_L} S_L 
        \Bigg\{ 
            (2-y) \, \sum_{\lambda = \pm 1} \lambda \, \sigma^{\gamma^*p}_{\lambda \lambda} +  \sqrt{2(1-y)} \sum_{\lambda = \pm 1} 
            \Big[
            e^{i \lambda \phi_k} \sigma^{\gamma^*p}_{0 \lambda} + \mathrm{c.c.} 
            \Big] 
        \Bigg\}.
\end{align}
Following \cite{Kovchegov:2024wjs}, we denote the first term on the right-hand-side of \eq{DSA} as corresponding to the ``TT" channel and the second term corresponding to the ``LT" channel, where T and L label the transverse ($\lambda = \pm 1$) and longitudinal ($\lambda =0$) polarizations of the virtual photon. Here $\alpha_{EM}$ is the fine structure constant, $y=P\cdot q/P\cdot k$ is the inelasticity of the collision, $k$ is the 4-vector of the incoming electron momentum, while $k'$ is the momentum of the outgoing electron. In the dipole/Trento frame we are working in, $\un k = \un k' = k_\perp (\cos \phi_{k}, \sin \phi_{k})$. 

Before we consider the specific TT and LT channels, it is helpful first to keep the photon polarizations general. Squaring \eq{t3_a6} and, at the sub-eikonal order, keeping only the interference between the type-3 term in the amplitude and the eikonal term in the complex conjugate amplitude (and vice versa), in analogy with Eq.~(54) of \cite{Kovchegov:2024wjs}, we obtain the contribution of the type-3 operator to elastic dijet production for scattering on a polarized proton,
\begin{align}\notag
 & z (1-z) \, \frac{1}{2} \! \sum_{S_L = \pm 1} \!\! S_L \,
 \frac{d\sigma_{\lambda' \lambda}^{\gamma^* p \to q {\bar q} p'\, \mathrm{G}[3]}}
 {d^2 p_1 \, d^2 p_2 \, d z}
 = \frac{1}{2 (2 \pi)^5} \,  \frac{1}{2} \sum_{S_L= \pm 1} S_L\, A_{\lambda} \, A^*_{\lambda'}
 \\
& = \frac{i}{2 (2 \pi)^5 \, s} \, \int d^2 x_1 \, d^2 x_{1'} \, d^2 x_2 \, d^2 x_{2'} \,
e^{- i {\un p}_1 \cdot {\un x}_{11'} - i {\un p}_2 \cdot {\un x}_{22'}} \,
\sum_{\sigma_1, \sigma_2, i , j}
\notag
\\ &
\times \Bigg\{ \Bigg[ \Psi_{\lambda, \sigma_1, \sigma_2; i, i}^{\gamma^* \to q {\bar q}\, (01)} ({\un x}_{12}, z)+ \frac{z- \frac{1}{2}}{z(1-z)} \Psi_{\lambda, \sigma_1, \sigma_2; i, i}^{\gamma^* \to q {\bar q}}
({\un x}_{12}, z) \Bigg] 
\,
\left[ \Psi_{\lambda', \sigma_1, \sigma_2; j, j}^{\gamma^* \to q {\bar q}}
({\un x}_{1'2'}, z) \right]^* \, 
\notag \\ & \hspace{4cm} \times 
\frac{1}{N_c^2}
\llangle \tord \tr \left[ V^\textrm{G[3]}_{\un 1} \, V_{\un 2}^\dagger
- V_{\un 1} V^{\mathrm{G}[3] \,\dagger}_{\un 2} \right] \rrangle(s) \,
\left\langle \atord \tr \left[ V_{{\un 2'}} \, V_{{\un 1'}}^\dagger - 1 \right]  \right\rangle
\notag
 \\&
 -  \Psi_{\lambda, \sigma_1, \sigma_2; i, i}^{\gamma^* \to q {\bar q}} ({\un x}_{12}, z) \,
 \left[ \Psi_{\lambda', \sigma_1, \sigma_2; j, j}^{\gamma^* \to q {\bar q}\,(01)}
 ({\un x}_{1'2'}, z)  + \frac{z-\frac{1}{2}}{z(1-z)}
  \Psi_{\lambda', \sigma_1, \sigma_2; j, j}^{\gamma^* \to q {\bar q}}
 ({\un x}_{1'2'}, z)
  \right]^* 
 \notag \\ & \hspace{4cm} \times 
 \frac{1}{N_c^2}
 \left\langle \tord \tr \left[   V_{{\un 1}} \, V_{{\un 2}}^\dagger - 1 \right]  \right\rangle
 \, 
\llangle \atord \tr \left[ V_{\un 2'} V^{\textrm{G[3]} \, \dagger}_{\un 1'} \, 
- V^{\mathrm{G}[3]}_{\un 2'} V^\dagger_{\un 1'}  \right] \rrangle(s)
 \Bigg\} .
 \label{t3_XS1}
\end{align}
Here, and in what follows, $\lambda$ denotes the photon polarization in the amplitude and $\lambda'$ denotes that in the complex conjugate amplitude. 

Since the derivative with respect to $z$ does not alter the helicity structure of the light-cone wave function, the decomposition of the wave function overlaps is exactly analogous to the type-2 case in Eq.~(65) of \cite{Kovchegov:2024wjs}. We define
\begin{subequations} \label{type3_overlaps}
\begin{align}
\notag
& \textrm{\bf TT \& T, -T:} \sum_{\sigma_1, i , j} \,
\Psi_{\lambda =T, \sigma_1, - \sigma_1; i, i}^{\gamma^* \to q {\bar q}\,(01)} ({\un x}_{12}, z) \,
\left[ \Psi_{\lambda' =T, \sigma_1, - \sigma_1; j, j}^{\gamma^* \to q {\bar q}}
({\un x}_{1'2'}, z) \right]^*
\\ &
\equiv \delta_{\lambda, \lambda^\prime} \Big[ \lambda \, i \,
\Phi^{[3]}_{\mathrm{TT}}(\un x_{12}, \un x_{1^\prime 2^\prime},z)
+ \Phi^{\prime [3]}_{\mathrm{TT}}(\un x_{12}, \un x_{1^\prime 2^\prime}, z) \Big]
+ \delta_{\lambda, -\lambda^\prime} \,
e^{- i \lambda (\phi_{12} +\phi_{1'2'} )} \,
\Phi^{[3]}_{\mathrm{T,-T}} (\un x_{12}, \un x_{1^\prime 2^\prime}, z),
\label{TT3}
\\
& \textrm{\bf LT:}  \sum_{\sigma_1, i , j} \,
\Psi_{\lambda =0, \sigma_1, - \sigma_1; i, i}^{\gamma^* \to q {\bar q}\, (01)} ({\un x}_{12}, z) \,
\left[ \Psi_{\lambda' =T, \sigma_1, - \sigma_1; j, j}^{\gamma^* \to q {\bar q}}
({\un x}_{1'2'}, z) \right]^*
\equiv -i \, \frac{\un \epsilon_{\lambda'}^* \cdot \un x_{1'2'}}{x_{1'2'}} \,
\Phi^{[3]}_{\mathrm{LT}} (\un x_{12}, \un x_{1^\prime 2^\prime}, z) ,\\
& \textrm{\bf TL:} \sum_{\sigma_1, i , j} \,
\Psi_{\lambda =T, \sigma_1, - \sigma_1; i, i}^{\gamma^* \to q {\bar q}\,(01)} ({\un x}_{12}, z) \,
\left[ \Psi_{\lambda' =0 , \sigma_1, - \sigma_1; j, j}^{\gamma^* \to q {\bar q}}
({\un x}_{1'2'}, z) \right]^* \equiv  i \, \frac{\un \epsilon_{\lambda} \cdot \un x_{12}}{x_{12}} \,
\Phi^{[3]}_{\mathrm{TL}} (\un x_{12}, \un x_{1' 2'}, z) ,
\\
& \textrm{\bf LL:} \sum_{\sigma_1, i , j} \,
\Psi_{\lambda =0, \sigma_1, - \sigma_1; i, i}^{\gamma^* \to q {\bar q}\,(01)} ({\un x}_{12}, z) \,
\left[ \Psi_{\lambda' =0 , \sigma_1, - \sigma_1; j, j}^{\gamma^* \to q {\bar q}}
({\un x}_{1'2'}, z) \right]^*
 \equiv \Phi^{[3]}_{\mathrm{LL}}(\un x_{12}, \un x_{1^\prime 2^\prime}, z).
\end{align}
\end{subequations}
The wavefunction overlaps in \eqs{type3_overlaps} can be explicitly computed given the light-cone wavefunctions of the photon, which are to leading order (see \cite{Kovchegov:2012mbw} and references therein) 
\begin{subequations} \label{psis}
\begin{align} \label{t-psi}
    \Psi^{\gamma^* \to q\bar{q}}_{\lambda=\pm 1, \sigma_1, \sigma_2; i, j}({\un x}_{12}, z) &= \frac{e Z_f}{2\pi} \delta_{ij} \delta_{\sigma_1, - \sigma_2} z (1-z) (1- 2z + \sigma_1 \lambda) \, i Q \, \frac{\un{\epsilon}_{\lambda} \cdot \un x_{12}}{x_{12}} \, K_1 \left( x_{12} Q \sqrt{z(1-z)}\right),
    \\ 
    \Psi^{\gamma^* \to q\bar{q}}_{\lambda=0 , \sigma_1, \sigma_2; i, j}({\un x}_{12}, z) &=   
    -\frac{e Z_f}{2\pi} \delta_{ij} \delta_{\sigma_1, - \sigma_2} \big[ z (1-z) \big]^{3/2} \, 2 Q \, K_0 \left( x_{12} Q \sqrt{z(1-z)}\right).
\end{align}
\end{subequations}
where $Z_f$ is the fraction of electric charge carried by a quark of flavor $f$ (in units of the electron electric charge $e$). Furthermore $i,j$ denote the color indices of the quark and antiquark, respectively, while $\sigma_1, \sigma_2$ label the helicities of the quark and antiquark. 

Employing the explicit wave functions in \eqs{psis} together with the definition \eqref{Psi01}, and using
\begin{align}
{\un \epsilon}_\lambda \cdot {\un v} \ {\un \epsilon}^*_\lambda \cdot {\un w}
= \frac{1}{2} \, {\un v} \cdot {\un w} + \frac{i}{2} \, \lambda \, {\un v} \times {\un w}
\end{align}
for ${\un \epsilon}_\lambda = - (1/\sqrt{2}) (-\lambda, i)$, we see that the wave function overlaps defined in \eqs{type3_overlaps} are given by
\begin{subequations} \label{type3_phis}
\begin{align}
    \Phi^{[3]}_{\mathrm{TT}}(\un x_{12}, \un x_{1'2'}, z) &=
    \frac{\alpha_{EM} Z_f^2 N_c^2 Q^2}{2\pi} \,
    \frac{\un x_{12} \times \un x_{1'2'}}{x_{12} \, x_{1'2'}} \,
    z(1-z)(1-2z) \, K_1\!\left( x_{1'2'} Q \sqrt{z(1-z)} \right)
    \notag \\ & \times
    \Bigg\{
    \Big[ (1-2z)^2 + 1 \Big]
    \bigg[ K_1\!\left( x_{12} Q \sqrt{z(1-z)} \right)
    - x_{12} Q \sqrt{z(1-z)} \, K_0\!\left( x_{12} Q \sqrt{z(1-z)} \right) \bigg]
    \notag \\ & \hspace{1cm}
    - 4 \, z(1-z) \, K_1\!\left( x_{12} Q \sqrt{z(1-z)} \right)
    \Bigg\},
    \\[2mm]
    \Phi^{\prime [3]}_{\mathrm{TT}}(\un x_{12}, \un x_{1'2'}, z) &=
    \frac{\alpha_{EM} Z_f^2 N_c^2 Q^2}{2\pi} \,
    \frac{\un x_{12} \cdot \un x_{1'2'}}{x_{12} \, x_{1'2'}} \,
    z(1-z)(1-2z) \, K_1\!\left( x_{1'2'} Q \sqrt{z(1-z)} \right)
    \notag \\ & \times
    \Bigg\{
    \Big[ (1-2z)^2 + 1 \Big]
    \bigg[ K_1\!\left( x_{12} Q \sqrt{z(1-z)} \right)
    - x_{12} Q \sqrt{z(1-z)} \, K_0\!\left( x_{12} Q \sqrt{z(1-z)} \right) \bigg]
    \notag \\ & \hspace{1cm}
    - 4 \, z(1-z) \, K_1\!\left( x_{12} Q \sqrt{z(1-z)} \right)
    \Bigg\},
    \\[2mm]
    \Phi^{[3]}_{\mathrm{T,-T}}(\un x_{12}, \un x_{1'2'}, z) &=
    \frac{2 \alpha_{EM} Z_f^2 N_c^2 Q^2}{\pi} \,
    z^2 (1-z)^2 (1-2z) \, K_1\!\left( x_{1'2'} Q \sqrt{z(1-z)} \right)
    \notag \\ & \times
    \bigg[ 2 K_1\!\left( x_{12} Q \sqrt{z(1-z)} \right)
    - x_{12} Q \sqrt{z(1-z)} \, K_0\!\left( x_{12} Q \sqrt{z(1-z)} \right) \bigg],
    \\[2mm]
    \Phi^{[3]}_{\mathrm{LT}}(\un x_{12}, \un x_{1'2'}, z) &=
    - \frac{2 \alpha_{EM} Z_f^2 N_c^2 Q^2}{\pi} \,
    (1-2z)^2 \big[ z(1-z) \big]^{3/2} \, K_1\!\left( x_{1'2'} Q \sqrt{z(1-z)} \right)
    \notag \\ & \times
    \bigg[ 3 K_0\!\left( x_{12} Q \sqrt{z(1-z)} \right)
    - x_{12} Q \sqrt{z(1-z)} \, K_1\!\left( x_{12} Q \sqrt{z(1-z)} \right) \bigg],
    \\[2mm]
    \Phi^{[3]}_{\mathrm{TL}}(\un x_{12}, \un x_{1'2'}, z) &=
    - \frac{2 \alpha_{EM} Z_f^2 N_c^2 Q^2}{\pi} \,
    \big[ z(1-z) \big]^{3/2} \, K_0\!\left( x_{1'2'} Q \sqrt{z(1-z)} \right)
    \notag \\ & \times
    \Bigg\{
    (1-2z)^2
    \bigg[ K_1\!\left( x_{12} Q \sqrt{z(1-z)} \right)
    - x_{12} Q \sqrt{z(1-z)} \, K_0\!\left( x_{12} Q \sqrt{z(1-z)} \right) \bigg]
    \notag \\ & \hspace{1cm}
    - 4 \, z(1-z) \, K_1\!\left( x_{12} Q \sqrt{z(1-z)} \right)
    \Bigg\},
    \\[2mm]
    \Phi^{[3]}_{\mathrm{LL}}(\un x_{12}, \un x_{1'2'}, z) &=
    \frac{4 \alpha_{EM} Z_f^2 N_c^2 Q^2}{\pi} \,
    (1-2z) \big[ z(1-z) \big]^{2} \, K_0\!\left( x_{1'2'} Q \sqrt{z(1-z)} \right)
    \notag \\ & \times
    \bigg[ 3 K_0\!\left( x_{12} Q \sqrt{z(1-z)} \right)
    - x_{12} Q \sqrt{z(1-z)} \, K_1\!\left( x_{12} Q \sqrt{z(1-z)} \right) \bigg].
\end{align}
\end{subequations}
We note that all $\Phi$'s are real. In addition to the type-3 wavefunction overlaps, we will also need the type-2 ones, defined explicitly in Eq.~(65) of \cite{Kovchegov:2024wjs}. With \eq{t3_XS1} and the type-3 wavefunction overlaps in \eqs{type3_overlaps} and \eqref{type3_phis} in hand, we now turn to calculating the type-3 contribution to each channel in \eq{DSA}.


\subsubsection{TT channel}

The TT structure entering the numerator of the DSA in \eq{DSA} is
\begin{align}
   z (1-z) \, \frac{1}{2} \! \sum_{S_L, \lambda = \pm 1} \!\! S_L \, \lambda   \,
   \frac{d\sigma_{\lambda \lambda}^{\gamma^* p \to q {\bar q} p'}}{d^2 p_1 \, d^2 p_2 \, d z}.
\end{align}
Employing \eqs{type3_overlaps} in \eq{t3_XS1} and projecting onto $\lambda = \lambda' = \pm 1$, we
obtain
\begin{align}
\notag
 & z (1-z) \, \frac{1}{2} \! \sum_{S_L, \lambda = \pm 1} \!\! S_L \, \lambda\,
 \frac{d\sigma_{\lambda \lambda}^{\gamma^* p \to q {\bar q} p'\, \mathrm{G}[3]}}
 {d^2 p_1 \, d^2 p_2 \, d z}
 = \frac{1}{(2 \pi)^5 \, s} \, \int d^2 x_1 \, d^2 x_{1'} \, d^2 x_2 \, d^2 x_{2'} \,
 e^{- i {\un p}_1 \cdot {\un x}_{11'} - i {\un p}_2 \cdot {\un x}_{22'}}
\notag
\\ &
\times \Bigg\{
\Bigg[ \Phi^{[3]}_{\mathrm{TT}} (\un x_{12}, \un x_{1'2'}, z)
+ \frac{z-\frac{1}{2}}{z(1-z)}  \Phi^{[2]}_{\mathrm{TT}} (\un x_{12}, \un x_{1'2'}, z)
\Bigg] 
\frac{1}{N_c}
\llangle \tord \tr \left[ V^\textrm{G[3]}_{\un 1} \, V_{\un 2}^\dagger
- V_{\un 1} V^{\mathrm{G}[3] \,\dagger}_{\un 2} \right] \rrangle(s) \, 
\Big[ 1- S_{1'2'}^*(s) \Big]
\notag
 \\&
 + \Bigg[ \Phi^{[3]}_{\mathrm{TT}}(\un x_{1'2'}, \un x_{12}, z)
  + \frac{z-\frac{1}{2}}{z(1-z)} \Phi^{[2]}_{\mathrm{TT}}(\un x_{1'2'}, \un x_{12}, z)
 \Bigg] \Big[ 1- S_{12}(s) \Big] \,
 \frac{1}{N_c}  \llangle \atord \tr \left[ V_{\un 2'} V^{\textrm{G[3]} \, \dagger}_{\un 1'} \, 
- V^{\mathrm{G}[3]}_{\un 2'} V^\dagger_{\un 1'}  \right] \rrangle(s)
 \Bigg\} ,
 \label{t3_XS2}
\end{align}
with the unpolarized dipole $S$-matrix $S_{10} (s)$ defined in \eq{Sdef}. 

In the last term on the right-hand side of \eq{t3_XS2}, which contains the sub-eikonal correlators at $\un x_{1'}$ and $\un x_{2'}$, we relabel the integration variables $\un x_{1} \leftrightarrow \un x_{1'}$ and
$\un x_{2} \leftrightarrow \un x_{2'}$, compensating the change of sign of the exponential
by taking $\un p_1 \to -\un p_1$, $\un p_2 \to - \un p_2$. The latter transformation is justified since all the terms in \eq{t3_XS2} are scalar functions of only two transverse vectors, $\un p_1$ and $\un p_2$. We get
\begin{align}\label{t3_XS3}
 & z (1-z) \, \frac{1}{2} \! \sum_{S_L, \lambda = \pm 1} \!\! S_L \, \lambda\,
 \frac{d\sigma_{\lambda \lambda}^{\gamma^* p \to q {\bar q} p'\, \mathrm{G}[3]}}
 {d^2 p_1 \, d^2 p_2 \, d z}
 = \frac{1}{(2 \pi)^5 \, s} \, \int d^2 x_1 \, d^2 x_{1'} \, d^2 x_2 \, d^2 x_{2'} \,
 e^{- i {\un p}_1 \cdot {\un x}_{11'} - i {\un p}_2 \cdot {\un x}_{22'}}
\\ &
\times \Bigg\{ \Bigg[ \Phi^{[3]}_{\mathrm{TT}} (\un x_{12}, \un x_{1'2'}, z)
+ \frac{z-\frac{1}{2}}{z(1-z)}  \Phi^{[2]}_{\mathrm{TT}} (\un x_{12}, \un x_{1'2'}, z)
\Bigg] 
\frac{1}{N_c}
\llangle \tord \tr \left[ V^\textrm{G[3]}_{\un 1} \, V_{\un 2}^\dagger
- V_{\un 1} V^{\mathrm{G}[3] \,\dagger}_{\un 2} \right] \rrangle(s)
\,
\Big[ 1- S_{1'2'}^*(s) \Big] + \mathrm{c.c.}
 \Bigg\} . \notag
\end{align}
Following \cite{Kovchegov:2024wjs}, we assume that quark and antiquark jets are experimentally indistinguishable and symmetrize between the outgoing jets via
\begin{align} \label{symmetric_XS}
 \frac{d\sigma_{\textrm{symm} \, \lambda \lambda'}^{\gamma^* p \to q {\bar q} p'}}{d^2 p_1 \, d^2 p_2 \, d z} \equiv \frac{1}{2} \left[ \frac{d\sigma_{\lambda \lambda'}^{\gamma^* p \to q {\bar q} p'}}{d^2 p_1 \, d^2 p_2 \, d z} + \frac{d\sigma_{\lambda \lambda'}^{\gamma^* p \to q {\bar q} p'}}{d^2 p_2 \, d^2 p_1 \, d (1-z)} \right].
 \end{align}
Applying this symmetrization to \eq{t3_XS3}, we arrive at
\begin{align}
\notag
 & z (1-z) \, \frac{1}{2} \! \sum_{S_L, \lambda = \pm 1} \!\! S_L \, \lambda\,
 \frac{d\sigma_{\mathrm{symm}\,\lambda \lambda}^{\gamma^* p \to q {\bar q} p'\, \mathrm{G}[3]}}
 {d^2 p_1 \, d^2 p_2 \, d z}
 = \frac{2}{ (2 \pi)^5 \, s} \, \int d^2 x_1 \, d^2 x_{1'} \, d^2 x_2 \, d^2 x_{2'} \,
 e^{- i {\un p}_1 \cdot {\un x}_{11'} - i {\un p}_2 \cdot {\un x}_{22'}}
\notag
\\ &
\times \,N_{1'2'}(s)  \Bigg\{ 
\Bigg[ \Phi^{[3]}_{\mathrm{TT}} (\un x_{12}, \un x_{1'2'}, z)
+ \frac{z-\frac{1}{2}}{z(1-z)}  \Phi^{[2]}_{\mathrm{TT}} (\un x_{12}, \un x_{1'2'}, z)
\Bigg]  \,
\Big[ G^{\mathrm{[3]}}_{12}(s) -  G^{[3]}_{21}(s) \Big] 
\Bigg\},
 \label{t3_XS4}
\end{align}
where we have used the definition in \eq{G3_def} and decomposed the unpolarized $S$-matrix as \cite{Hatta:2005as,Kovchegov:2003dm}
\begin{align} \label{unpdas}
    1- S_{10}(s) &= N_{10}(s) - i O_{10}(s),
\end{align}
where 
\begin{subequations}
\begin{align}
& N_{10} (s) = 1 - \frac{1}{2 N_c} \, \left\langle \tord \tr \left[ V_{\un 1} \, V^\dagger_{\un 0} \right] + \tord \tr \left[ V_{\un 0} \, V^\dagger_{\un 1} \right] \right\rangle, \\
& O_{10} (s) = \frac{1}{2 N_c \, i} \, \left\langle \tord \tr \left[ V_{\un 1} \, V^\dagger_{\un 0} \right] - \tord \tr \left[ V_{\un 0} \, V^\dagger_{\un 1} \right] \right\rangle,
\end{align}
\end{subequations}
are the $\mathcal{C}$-even Pomeron and $\mathcal{C}$-odd odderon dipole amplitudes, respectively. Both of these dipole amplitudes have been extensively studied in the literature. See \cite{Mueller:1994rr,Mueller:1994jq,Mueller:1994gb,Balitsky:1995ub,Balitsky:1998ya,Kovchegov:1999yj,Kovchegov:1999ua,Jalilian-Marian:1997ubg,Jalilian-Marian:1997jhx,Weigert:2000gi,Iancu:2001ad,Iancu:2000hn,Ferreiro:2001qy} for the Pomeron and  \cite{Hatta:2005as,Kovchegov:2003dm} for the odderon. To leading order in $\as$ both $N_{1'2'}$ and $O_{1'2'}$ are real. In arriving at \eq{t3_XS4}, we have neglected terms involving the odderon since they are $\alpha_s$-suppressed compared to the terms we keep \cite{Kwiecinski:1980wb, Bartels:1999yt, Ewerz:2003xi, Hatta:2005as,Kovchegov:2003dm}. 

To integrate over impact parameters and isolate the contributions from the moment amplitudes, we first need to expand the phase in \eq{t3_XS4}. To do so, we assume that the net transverse momentum transfer $\Delta_\perp$ is small and write 
\begin{align} \label{dexp}
    e^{-i \un p_1 \cdot \un x_{11'}} e^{-i \un p_2 \cdot x_{22'}} = e^{- i \un p \cdot ( x_{12} - x_{1'2'})} e^{-i \un \Delta \cdot (\un b - \un b')} = e^{- i \un p \cdot ( x_{12} - x_{1'2'})} 
    \Big[ 1 - i \un \Delta \cdot \Big(\un b - \un b' \Big) + \mathcal{O}(\Delta_\perp^2) \Big],
\end{align}
where we have defined the relative transverse momentum and momentum imbalance of the two jets 
\begin{subequations}
    \begin{align}
        \un p &= (1-z) \,\un p_1 - z\, \un p_2, 
        \\ 
        \un \Delta &= \un p_1 + \un p_2,
    \end{align}
\end{subequations}

and the impact parameters in the amplitude and the complex conjugate amplitude
\begin{subequations}
\begin{align}
    \un b \equiv& z\, \un x_{1} + (1-z) \, \un x_{2} = {\un x}_1 - (1-z) \, {\un x}_{12} = {\un x}_2 + z \, {\un x}_{12}, 
    \\ 
    {\un b}' \equiv& z \, {\un x}_{1'} + (1-z) \, {\un x}_{2'} =
     {\un x}_{1'} - (1-z) \, {\un x}_{1'2'} = {\un x}_{2'} + z \, {\un x}_{1'2'}.
\end{align}
\end{subequations}
In expanding \eq{dexp}, we restrict ourselves to the kinematic region where $\Delta_\perp \ll \Lambda_{\mathrm{QCD}} \ll p_\perp, Q$.

The integrals over impact-parameters can be done with the help of \eqref{moment3}:
\begin{subequations} \label{t3_b_ints}
\begin{align} \label{g3_int}
    \int d^2 b \, G^{[3]}_{12}(s) &= \int d^2 b \, G^{[3]}_{21}(s) = G^{[3]}(x_{12}^2, s), \\
    \int d^2 b \, b^i \, G^{[3]}_{12}(s) &= \epsilon^{ij} x_{12}^j \, I_E(x_{12}^2, s)
    + x_{12}^i \, J_E(x_{12}^2, s) - (1-z) \, x_{12}^i \, G^{[3]}(x_{12}^2, s), \\
    \int d^2 b \, b^i \, G^{[3]}_{21}(s) &= - \epsilon^{ij} x_{12}^j \, I_E(x_{12}^2, s)
    - x_{12}^i \, J_E(x_{12}^2, s) + z \, x_{12}^i \, G^{[3]}(x_{12}^2, s),
\end{align}
\end{subequations}
where \eq{g3_int} is equivalent to \eq{G3_int_def}. We also define the impact-parameter integral of $N_{12}$ as 
\begin{align} \label{nb}
    \int d^2 b\, N_{12}(s) &\equiv N(x_{12}^2, s).
\end{align}

Finally, expanding the Fourier phase in \eq{t3_XS4} to the linear order in $\Delta_\perp$ via \eq{dexp} and employing \eqs{t3_b_ints} and \eqref{nb}, along with
\begin{align}
    d^2 x_1 \, d^2 x_{1'} \, d^2 x_2 \, d^2 x_{2'} = d^2 x_{12} \, d^2 x_{1'2'} \, d^2 b \, d^2 b' ,
\end{align}
to integrate over $\un b$ and $\un b'$, we obtain
\begin{align}
\notag
 & z (1-z) \, \frac{1}{2}   \sum_{S_L, \, \lambda = \pm 1}  \!\! S_L \,  \lambda\,
 \frac{d\sigma_{\mathrm{symm}\, \lambda \lambda}^{\gamma^* p \to q {\bar q} p'\, \mathrm{G}[3]}}
 {d^2 p \, d^2 \Delta \, d z}
 = \frac{2}{ (2 \pi)^5 \, s} \, \int d^2 x_{12} \, d^2 x_{1'2'} \,
 e^{- i {\un p} \cdot ({\un x}_{12}  - \un x_{1'2'})}
   \, N(x_{1'2'}^2, s)
\notag
\\ &
\times
\Bigg[ \Phi^{[3]}_{\mathrm{TT}} (\un x_{12}, \un x_{1'2'}, z) 
 + \frac{z- \frac{1}{2}}{z(1-z)} \Phi^{[2]}_{\mathrm{TT}} (\un x_{12}, \un x_{1'2'}, z) 
\Bigg] 
\notag  \\ & \hspace{2cm} \times 
\Big[
 - 2 \, i \, \un \Delta \times \un  x_{12} \, I_E(x_{12}^2, s)
 - 2 \, i \, \un \Delta \cdot \un x_{12} \, J_E(x_{12}^2, s) + i \, \un \Delta \cdot \un x_{12} \, G^{[3]}(x_{12}^2, s)  \Big] 
 + \ord{\Delta_\perp^2}.
 \label{t3_XS5}
\end{align}
The terms containing $I_E$ carry two Levi-Civita tensors, one from $\un \Delta \times \un x_{12}$ and one from the cross products inside $\Phi^{[3]}_{\mathrm{TT}}$ and $\Phi^{[2]}_{\mathrm{TT}}$ (cf. \eqs{type3_phis} above and Eqs.~(108) of \cite{Kovchegov:2024wjs}). After the integrations over $\un x_{12}$ and $\un x_{1'2'}$ are performed, it therefore contributes to the $\un \Delta \cdot \un p$ azimuthal harmonic, exactly like the moment amplitudes $I_3, I_4$ and $I_5$ in \cite{Kovchegov:2024wjs}. Additionally, we will explicitly show below that the $I_E$ term does not vanish after integrating over the angles of $\un x_{12}$ and $\un x_{1'2'}$. This term therefore represents a new contribution to the elastic dijet DSA. By contrast, the $J_E$ and $G^{[3]}$ terms carry a single $\epsilon^{ij}$ each and hence generate the P-odd structure $\un \Delta \times \un p$, which cannot contribute to the P-even DSA. By the same argument used in \cite{Kovchegov:2024wjs} to conclude that $G_1 = J_3 = J_4 = J_5 = 0$ for a longitudinally polarized target, one concludes that $J_E = G^{[3]}= 0$ as well. 


\subsubsection{LT channel}

We turn now to the interference between longitudinally and transversely polarized virtual photons, for which the relevant structure in \eq{DSA} is
\begin{align}
    z(1-z) \, \frac{1}{2} \sum_{S_L, \lambda = \pm 1} S_L \,
    \Bigg[
            e^{i \lambda \phi_k} \frac{d\sigma^{\gamma^*p}_{ 0\lambda }}{d^2 p_1 d^2 p_2 dz}
            + \mathrm{c.c.}
     \Bigg] .
\end{align}
Employing \eqs{type3_overlaps} in \eq{t3_XS1} and projecting out the LT channel, we sum over $\lambda = \pm 1$ and obtain
\begin{align} \label{t3_XS6}
\notag
 & z (1-z) \, \frac{1}{2} \! \sum_{S_L, \lambda = \pm 1} \!\!   S_L \,
 \Bigg[ e^{i\lambda \phi_k} \,
 \frac{d\sigma_{0 \lambda}^{\gamma^* p \to q {\bar q} p'\, \mathrm{G}[3]}}
 {d^2 p_1 \, d^2 p_2 \, d z} + \mathrm{c.c.} \Bigg]
 =  \frac{- i \sqrt{2}}{2 (2 \pi)^5 \, s} \, \int d^2 x_1 \, d^2 x_{1'} \, d^2 x_2 \, d^2 x_{2'}
 \, e^{- i {\un p}_1 \cdot {\un x}_{11'} - i {\un p}_2 \cdot {\un x}_{22'}} \, \frac{\hat k \times \un x_{12}}{x_{12}}
\notag
\\ &
\times 
 \, \Bigg\{
\Bigg[ \Phi^{[3]}_{\mathrm{TL}}(\un x_{12}, \un x_{1'2'}, z)
+ \frac{z-\frac{1}{2}}{z(1-z)} \Phi^{[2]}_{\mathrm{LT}}(\un x_{12}, \un x_{1'2'}, z)
\Bigg]\, \frac{1}{N_c}
\llangle \tord \tr \left[ V^\textrm{G[3]}_{\un 1} \, V_{\un 2}^\dagger
- V_{\un 1} V^{\mathrm{G}[3] \,\dagger}_{\un 2} \right] \rrangle(s) \,
\Big[ 1- S_{1'2'}^*(s) \Big]
 \\ 
& - \Bigg[ \Phi^{[3]}_{\mathrm{LT}}(\un x_{1'2'}, \un x_{12}, z) 
+ \frac{z-\frac{1}{2}}{z(1-z)} \Phi^{[2]}_{\mathrm{LT}}(\un x_{12}, \un x_{1'2'}, z) 
\Bigg] 
\Big[ 1- S_{12}(s) \Big] \, \frac{1}{N_c} 
\llangle \atord \tr \left[ V_{\un 2'} V^{\textrm{G[3]} \, \dagger}_{\un 1'} \, 
- V^{\mathrm{G}[3]}_{\un 2'} V^\dagger_{\un 1'}  \right] \rrangle(s)
 \Bigg\}  + \mathrm{c.c.}, \notag
\end{align}
where $\hat k$ is the unit vector along the transverse momentum of the incoming (and outgoing) electron, $\hat k = (\cos \phi_k, \sin \phi_k)$. In the second term with correlators of Wilson lines at $\un x_{1'}$ or $\un x_{2'}$ 
in the curly brackets of \eq{t3_XS6}, we again exchange $\un x_{1} \leftrightarrow \un x_{1'}$ and
$\un x_{2} \leftrightarrow \un x_{2'}$ while taking $\un p_1, \un p_2, \hat k \to - \un p_1, - \un p_2, -\hat k$, and subsequently exchange that term with its complex conjugate subjected to the same substitution. Once again, this $\un p_1, \un p_2, \hat k \to - \un p_1, - \un p_2, -\hat k$ transformation is justified since the cross section is rotationally invariant in the transverse plane and dependent only on three transverse vectors, $\un p_1, \un p_2$, and $\hat k$. We arrive at 
\begin{align}
\notag
 & z (1-z) \, \frac{1}{2} \! \sum_{S_L, \lambda = \pm 1} \!\! S_L \,
 \Bigg[ e^{i\lambda \phi_k} \,
 \frac{d\sigma_{0\lambda}^{\gamma^* p \to q {\bar q} p'\, \mathrm{G}[3]}}
 {d^2 p_1 \, d^2 p_2 \, d z} + \mathrm{c.c.} \Bigg]
 =  \frac{- i \sqrt{2}}{2 (2 \pi)^5 \, s} \, \int d^2 x_1 \, d^2 x_{1'} \, d^2 x_2 \, d^2 x_{2'}
 \, e^{- i {\un p}_1 \cdot {\un x}_{11'} - i {\un p}_2 \cdot {\un x}_{22'}}
\notag
\\ & \notag
\times \Bigg\{
\Bigg[
\frac{\hat k \times \un x_{12}}{x_{12}} \,
\Bigg( 
\Phi^{[3]}_{\mathrm{TL}}(\un x_{12}, \un x_{1'2'}, z)
  + \frac{z-\frac{1}{2}}{z(1-z)} \Phi^{[2]}_{\mathrm{LT}}(\un x_{12}, \un x_{1'2'}, z)
\Bigg)
\notag \\ & \hspace{1cm}
+ 
\frac{\hat k \times \un x_{1'2'}}{x_{1'2'}} \,
\Bigg( 
\Phi^{[3]}_{\mathrm{LT}}(\un x_{12}, \un x_{1'2'}, z)
  + \frac{z-\frac{1}{2}}{z(1-z)} \Phi^{[2]}_{\mathrm{LT}}(\un x_{1'2'}, \un x_{12}, z)
\Bigg)
\Bigg] \notag 
\\ & \hspace{1cm} \times
\, \frac{1}{N_c}
\Bigg[
\llangle \tord \tr \left[ V^\textrm{G[3]}_{\un 1} \, V_{\un 2}^\dagger
- V_{\un 1} V^{\mathrm{G}[3] \,\dagger}_{\un 2} \right] \rrangle(s)
\Bigg]
\Big[ 1- S_{1'2'}^*(s) \Big]
 \Bigg\}  + \mathrm{c.c.} .
 \label{t3_XS8}
\end{align}
Symmetrizing between the quark and antiquark jets via \eq{symmetric_XS} and neglecting the odderon contribution yields
\begin{align}
\notag
 & z (1-z) \, \frac{1}{2} \! \sum_{S_L, \lambda = \pm 1} \!\! S_L \,
 \Bigg[ e^{i\lambda \phi_k} \,
 \frac{d\sigma_{\mathrm{symm}\, 0 \lambda }^{\gamma^* p \to q {\bar q} p'\, \mathrm{G}[3]}}
 {d^2 p_1 \, d^2 p_2 \, d z} + \mathrm{c.c.} \Bigg]
 =  \frac{- i \sqrt{2}}{(2 \pi)^5 \, s} \, \int d^2 x_1 \, d^2 x_{1'} \, d^2 x_2 \, d^2 x_{2'}
 \, e^{- i {\un p}_1 \cdot {\un x}_{11'} - i {\un p}_2 \cdot {\un x}_{22'}}
\notag
\\ 
& \times  \Bigg\{ \Bigg[
\frac{\hat k \times \un x_{12}}{x_{12}} \,
\Bigg( 
\Phi^{[3]}_{\mathrm{TL}}(\un x_{12}, \un x_{1'2'}, z)
  + \frac{z-\frac{1}{2}}{z(1-z)} \Phi^{[2]}_{\mathrm{LT}}(\un x_{12}, \un x_{1'2'}, z)
\Bigg)
\notag \\ & \hspace{1cm}
+ 
\frac{\hat k \times \un x_{1'2'}}{x_{1'2'}} \,
\Bigg( 
\Phi^{[3]}_{\mathrm{LT}}(\un x_{12}, \un x_{1'2'}, z)
 + \frac{z-\frac{1}{2}}{z(1-z)} \Phi^{[2]}_{\mathrm{LT}}(\un x_{1'2'}, \un x_{12}, z)
\Bigg)
\Bigg] 
\, \Big[ G^{\mathrm{[3]}}_{12}(s) -  G^{[3]}_{21}(s) \Big] \,
N_{1'2'} (s) 
\Bigg\}
.
 \label{t3_XS88}
\end{align}

Expanding \eq{t3_XS88} to the linear order in $\Delta_\perp$ via \eq{dexp} with the help of \eqs{t3_b_ints} and \eqref{nb}, we obtain
\begin{align}
\notag
 & z (1-z) \, \frac{1}{2} \! \sum_{S_L, \lambda = \pm 1} \!\! S_L \,
 \Bigg[ e^{i\lambda \phi_k} \,
 \frac{d\sigma_{\mathrm{symm} \, 0\lambda}^{\gamma^* p \to q {\bar q} p'\, \mathrm{G}[3]}}
 {d^2 p \, d^2 \Delta \, d z} + \mathrm{c.c.} \Bigg]
 =  \frac{- i \sqrt{2}}{(2 \pi)^5 \, s} \, \int d^2 x_{12} \, d^2 x_{1'2'} \,
 e^{- i {\un p} \cdot ({\un x}_{12} - \un x_{1'2'})} \, N(x_{1'2'}^2, s)
\notag
\\ &
\times
\Bigg[
\frac{\hat k \times \un x_{12}}{x_{12}} \,
\Bigg( 
\Phi^{[3]}_{\mathrm{TL}}(\un x_{12}, \un x_{1'2'}, z)
  + \frac{z-\frac{1}{2}}{z(1-z)} \Phi^{[2]}_{\mathrm{LT}}(\un x_{12}, \un x_{1'2'}, z)
\Bigg)
\notag \\ & \hspace{1cm}
+ 
\frac{\hat k \times \un x_{1'2'}}{x_{1'2'}} \,
\Bigg( 
\Phi^{[3]}_{\mathrm{LT}}(\un x_{12}, \un x_{1'2'}, z)
  + \frac{z-\frac{1}{2}}{z(1-z)} \Phi^{[2]}_{\mathrm{LT}}(\un x_{1'2'}, \un x_{12}, z)
\Bigg)
\Bigg] 
\notag \\ & 
\times
\Bigg\{
 - 2 \, i \, \un \Delta \times \un  x_{12} \, I_E(x_{12}^2, s)
 - 2 \, i \, \un \Delta \cdot \un x_{12} \, J_E(x_{12}^2, s) + i \, \un \Delta \cdot \un x_{12} \, G^{[3]}(x_{12}^2, s) \Bigg\} + \ord{\Delta_\perp^2}.
 \label{t3_XS10}
\end{align}
As in the TT channel, the moment amplitude $I_E$ enters \eq{t3_XS10} with two Levi-Civita tensors, one from $\un \Delta \times \un x_{12}$ and one from $\hat k \times \un x$. It therefore contributes to the same two P-even LT harmonics identified in \cite{Kovchegov:2024wjs}, namely $(\un p \cdot \un \Delta)( \un p \cdot \un k)$ and $(\un p \times \un \Delta)(\un p \times \un k)$. The $J_E$ and $G^{[3]}$ terms again come with a single $\epsilon^{ij}$ and generate the P-odd structures $(\un p \times \un \Delta)( \un p \cdot \un k)$ and $(\un p \cdot \un \Delta)(\un p \times \un k)$: therefore, they do not contribute to the DSA.


\subsection{Summary of our results and the angular integrated form}
\label{sec:t3_summary}

We may now assemble the complete numerator of the DSA in elastic dijet production, including the type-3 contributions derived above. Adding \eqs{t3_XS5} and \eqref{t3_XS10} to Eqs.~(107) of \cite{Kovchegov:2024wjs} we obtain our final results for the numerators of the DSA in elastic dijet production in $\vec e+ \vec p$ collisions: 

\begin{tcolorbox}[colback=blue!10!white]
\begin{subequations}\label{integrated_res_t3}
\begin{align}\notag
&z (1-z) \ \frac{1}{2} \sum_{S_L, \lambda = \pm 1} S_L \, \lambda  \,
\frac{d\sigma_{\mathrm{symm.}\, \lambda \lambda}^{\gamma^* p \to q {\bar q} p'}}
{d^2 p \, d^2 \Delta \, d z} =
      -\frac{2}{(2\pi)^5 \, z (1-z) \, s}
        \int d^2 x_{12} \, d^2 x_{1^\prime 2^\prime} \,
        e^{-i \un p \cdot (\un x_{12} - \un x_{1^\prime 2^\prime})} \, N(x_{1^\prime 2^\prime}^2, s)
 \\ \notag &
  \times  \Bigg\{
            \Bigg[
     \left(1-2z + i \un \Delta \cdot \un x_{12} \, \left( z^2 + (1-z)^2 \right)
     - \frac{i}{2} \, \un \Delta \cdot \un x_{1'2'} \, (1-2 z)^2  \right) Q(x_{12}^2, s)
    - i \un \Delta \cdot \un x_{12} \, I_3(x_{12}^2, s) \\
    \notag & \hspace{1cm}
    - i \un \Delta \times \un x_{12} \, J_3(x_{12}^2, s)
    \Bigg] \, \Phi^{[1]}_{\mathrm{TT}}(\un x_{12}, \un x_{1^\prime 2^\prime}, z)
    \\ \notag & \hspace{0.5cm}
     + \Bigg[
     i (1- 2 z) \Big( \Delta^j \epsilon^{ji} x_{12}^2 I_4(x_{12}^2, s)
     + \un \Delta \times \un x_{12} \, x_{12}^i I_5(x_{12}^2, s)
     + \Delta^i  \, x_{12}^2 J_4(x_{12}^2, s)
     + \un \Delta \cdot \un x_{12} \, x_{12}^i  J_5(x_{12}^2, s)
    \Big)
    \\ \notag & \hspace{1cm}
    - \left[ 1+ i \left(1 - 2\, z \right) \un \Delta \cdot
    \left( \un x_{12} - \frac{\un x_{1'2'}}{2} \right) \right]
    \Big( \epsilon^{ik} x_{12}^k G_2(x_{12}^2, s)  + x_{12}^i G_1(x_{12}^2, s)
    \Big) \Bigg]
    \left( \partial^i_{\un 1} - ip^i \right)
     \Phi^{[2]}_{\mathrm{TT}}(\un x_{12}, \un x_{1'2'}, z)
    \\ \notag & \hspace{0.5cm}
    + \, z(1-z) \Bigg[
    2 \, i \, \un \Delta \times \un  x_{12} \, I_E(x_{12}^2, s)
 + 2 \, i \, \un \Delta \cdot \un x_{12} \, J_E(x_{12}^2, s) - i \, \un \Delta \cdot \un x_{12} \, G^{[3]}(x_{12}^2, s) \Bigg] 
 \notag \\ & \hspace{3cm} \times \Bigg[
 \Phi^{[3]}_{\mathrm{TT}}(\un x_{12}, \un x_{1'2'}, z)
  + \frac{z- \frac{1}{2}}{z(1-z)} \Phi^{[2]}_{\mathrm{TT}} (\un x_{12}, \un x_{1'2'}, z)  
 \Bigg]
     \Bigg\} + \ord{\Delta_\perp^2},
      \label{TT_result_t3}
     \\[3mm] \notag
  & z (1-z) \ \frac{1}{2} \sum_{S_L, \lambda = \pm 1} S_L \,
  \left[ e^{i \lambda \phi_k} \,
  \frac{d\sigma_{\mathrm{symm.}\,0\lambda}^{ \gamma^* p \to q {\bar q} p'}}
  {d^2 p \, d^2 \Delta \, d z} + \mathrm{c.c.} \right] =
    -\frac{i \sqrt{2}}{(2\pi)^5 \, z (1-z) \, s}
        \int d^2 x_{12} \, d^2 x_{1^\prime 2^\prime} \,
        e^{-i \un p \cdot (\un x_{12} - \un x_{1^\prime 2^\prime})}
 \\ \notag &
  \times N(x_{1^\prime 2^\prime}^2, s)  \Bigg\{
            \Bigg[
 \left(1-2z + i \, \un \Delta \cdot \un x_{12} \left( z^2 + (1-z)^2 \right)
 - \frac{i}{2} \un \Delta \cdot \un x_{1'2'} \, (1-2 z)^2 \right) Q(x_{12}^2, s)
    - i \un \Delta \cdot \un x_{12} \, I_3(x_{12}^2, s) \\
    & \notag \hspace{1cm}
    - i \un \Delta \times \un x_{12} \, J_3(x_{12}^2, s)
    \Bigg] \,
    \left[ \frac{\hat{k} \cdot \un x_{12}}{x_{12}}
    \Phi^{[1]}_{\mathrm{LT}}(\un x_{12}, \un x_{1^\prime 2^\prime}, z)
- \frac{\hat{k} \cdot \un x_{1^\prime 2^\prime}}{x_{1^\prime 2^\prime}}
\Phi^{[1]}_{\mathrm{LT}}(\un x_{1^\prime 2^\prime}, \un x_{1 2}, z) \right]
    \\ \notag & \hspace{0.5cm}
     + \Bigg[  i (1- 2 z) \Big( \Delta^j \epsilon^{ji} \, x_{12}^2 \, I_4(x_{12}^2, s)
     + \un \Delta \times \un x_{12} \, x_{12}^i \, I_5(x_{12}^2, s)
     + \Delta^i \, x_{12}^2 \, J_4(x_{12}^2, s)
     + \un \Delta \cdot \un x_{12} \, x_{12}^i \, J_5(x_{12}^2, s) \Big)
    \\ \notag & \hspace{1cm}
    - \left[ 1+ i \left(1 - 2\, z \right) \un \Delta \cdot
    \left( \un x_{12} - \frac{\un x_{1'2'}}{2} \right) \right]
    \Big( \epsilon^{ik} x_{12}^k G_2(x_{12}^2, s)  + x_{12}^i G_1(x_{12}^2, s) \Big)  \Bigg]
\\ \notag & \hspace{1cm}
    \times  \left( \partial^i_{\un 1} - ip^i \right)
     \left[ \frac{\hat{k} \times \un x_{12}}{x_{12}}
     \Phi^{[2]}_{\mathrm{LT}}(\un x_{12}, \un x_{1^\prime 2^\prime}, z)
    + \frac{\hat{k} \times \un x_{1^\prime 2^\prime}}{x_{1^\prime 2^\prime}}
    \Phi^{[2]}_{\mathrm{LT}}(\un x_{1^\prime 2^\prime}, \un x_{1 2}, z) \right]
    \\ \notag & \hspace{0.5cm}
    - z(1-z) \Bigg[
    2 \, i \, \un \Delta \times \un  x_{12} \, I_E(x_{12}^2, s)
 + 2 \, i \, \un \Delta \cdot \un x_{12} \, J_E(x_{12}^2, s) - i \, \un \Delta \cdot \un x_{12} \, G^{[3]}(x_{12}^2, s)
    \Bigg]
    \\  & \hspace{1cm} \times
\Bigg[
\frac{\hat k \times \un x_{12}}{x_{12}} \,
\Bigg( 
\Phi^{[3]}_{\mathrm{TL}}(\un x_{12}, \un x_{1'2'}, z)
  + \frac{z-\frac{1}{2}}{z(1-z)} \Phi^{[2]}_{\mathrm{LT}}(\un x_{12}, \un x_{1'2'}, z)
\Bigg)
\notag \\ & \hspace{2cm}
+ 
\frac{\hat k \times \un x_{1'2'}}{x_{1'2'}} \,
\Bigg( 
\Phi^{[3]}_{\mathrm{LT}}(\un x_{12}, \un x_{1'2'}, z)
  + \frac{z-\frac{1}{2}}{z(1-z)} \Phi^{[2]}_{\mathrm{LT}}(\un x_{1'2'}, \un x_{12}, z)
\Bigg)
\Bigg] 
     \Bigg\} + \ord{\Delta_\perp^2},
      \label{TL_result_t3}
\end{align}
\end{subequations}
\end{tcolorbox}

The type-1 and type-2 wave functions overlaps $\Phi$ are given in Eqs.~(108) and (109) of \cite{Kovchegov:2024wjs} and the type-3 overlaps are shown in \eqs{type3_phis} above.

Equations~\eqref{integrated_res_t3} supersede Eqs.~(107) of \cite{Kovchegov:2024wjs}. The structural conclusions of that work are unchanged. Counting Levi-Civita tensors as before, the terms containing $Q, G_2, I_3, I_4, I_5$ and, now, $I_E$ carry either zero or two $\epsilon^{ij}$ and therefore generate the P-even angular structures $\un p \cdot \un \Delta$ in the TT channel and $(\un p \cdot \un \Delta)( \un p \cdot \un k)$, $(\un p \times \un \Delta)(\un p \times \un k)$ in the LT channel. The remaining amplitudes $G_1, J_3, J_4, J_5$ and $J_E$, along with all the $G^{[3]}$ terms, come with a single $\epsilon^{ij}$ and hence generate P-odd structures; by the parity argument of \cite{Kovchegov:2024wjs} they cannot contribute to the DSA and are set to zero. The new type-3 physics therefore enters the observable exclusively through the moment amplitude $I_E$, which is precisely the amplitude whose novel DLA evolution equation we derived in \eq{IE_evol1}.

To explicitly identify which linear combination of amplitudes enters each harmonic, and to prepare \eqs{integrated_res_t3} for the numerical analysis of the companion paper \cite{JAMCollaborationSmall-xAnalysisGroup:inprep}, we now carry out the integrations over the azimuthal angles of $\un x_{12}$ and $\un x_{1'2'}$ analytically. We first rewrite \eq{DSA} with the differential phase space element suitable for elastic dijet production we consider, 
\begin{align} \notag
     E_{k^\prime} z(1-z)\frac{d \sigma^{\mathrm{DSA}}}{d^3 k^\prime\, d^2p \, d^2 \Delta \,dz} &= 
        \frac{\alpha_{EM}}{4\pi^2 \, Q^2}
         \half \, \sum_{S_L} S_L 
        \Bigg\{ 
            (2-y) \, \sum_{\lambda = \pm 1} \lambda \, z(1-z)\frac{d \sigma^{\gamma^*p \to q\bar{q} p'}_{\lambda \lambda}}{d^2p \, d^2 \Delta\, dz} \\[-0.1cm] 
            \label{angles1}
            & \hspace{3cm}
            +  
            \sqrt{2(1-y)} \sum_{\lambda = \pm 1} 
            \Big[
            e^{i \lambda \phi_k} z(1-z)\frac{d \sigma^{\gamma^*p \to q\bar{q} p'}_{0 \lambda}}{ d^2p \, d^2 \Delta\, dz} + \mathrm{c.c.} 
            \Big] 
        \Bigg\}.
\end{align}
The right-hand side of \eq{angles1} consists of quantities we have just calculated, given in \eqs{integrated_res_t3}.

Writing the final-state phase space as\footnote{There is a subtlety here:~see Appendix~\ref{sec:frame}.} 
\begin{align} \label{e_phasespace}
    \frac{d^3 k^\prime}{E_{k^\prime}} \, \frac{dz}{z(1-z)} \, d^2p\, d^2 \Delta
    = \frac{\pi}{2} \, \frac{p_\perp}{z(1-z)} \, dy \, dQ^2 \, dz\, dp_\perp \, dt
     \, d\phi_{kp} \, d \phi_{\Delta p},
\end{align}
with $\phi_{kp} = \phi_k - \phi_p$ and $\phi_{\Delta p} = \phi_\Delta - \phi_p$ the differences between the azimuthal angles of $\un k$ and $\un p$, and of $\un \Delta$ and $\un p$,  \eq{angles1} decomposes into five independent azimuthal harmonics,
\begin{align}
\frac{d \sigma^{\mathrm{DSA}}}{dy \, dQ^2  \, dz \, dp_\perp \, dt\, \, d\phi_{kp} \, d\phi_{\Delta p} }
&=
            (2-y)   \, C_{\mathrm{TT}} +
            \sqrt{2(1-y)} \,
            \cos \phi_{kp}\, \,C_{\mathrm{LT}}^{\cos\phi_{kp}}
+ \frac{\sqrt{-t}}{p_\perp} (2-y) \cos \phi_{\Delta p}\,
            C_{\mathrm{TT}}^{\cos \phi_{\Delta p} }
\notag
\\
& \hspace{-1cm}
+ \frac{\sqrt{-t}}{p_\perp}  \sqrt{2(1-y)}
  \bigg(
              \cos\phi_{kp} \cos\phi_{\Delta p}\,  C_{\mathrm{LT}}^{\cos\phi_{kp} \cos\phi_{\Delta p}}
            +  \sin\phi_{kp} \sin\phi_{\Delta p}\,  C_{\mathrm{LT}}^{\sin\phi_{kp} \sin\phi_{\Delta p}}
  \bigg) .
\label{angles2}
\end{align}
The angular integrals themselves are performed with the help of
\begin{subequations} \label{ang_ints}
    \begin{align}
        \int \displaylimits^{2\pi}_0 d\phi_{12} \,e^{i \un p \cdot \un x_{12}} &= 2\pi \, J_0(p_\perp x_{12}),
        \\
        \int \displaylimits^{2\pi}_0 d\phi_{12} \,e^{i \un p \cdot \un x_{12}} \, \hat{x}_{12}^i &= 2\pi i \,  \frac{p^i}{p_\perp} \,J_1 ( p_\perp x_{12}),
        \\
        \int \displaylimits^{2\pi}_0 d\phi_{12} \,e^{i \un p \cdot \un x_{12}} \, \hat{x}_{12}^i \hat{x}_{12}^j &= 2\pi \, \left[ \frac{\delta^{ij}}{p_\perp \, x_{12}}
        J_1(p_\perp x_{12}) - \frac{p^i p^j}{p_\perp^2} \, J_2 ( p_\perp x_{12}) \right],
                \\ 
        \int \displaylimits^{2\pi}_0 d\phi_{12} \,e^{i \un p \cdot \un x_{12}} \, \hat{x}_{12}^i \hat{x}_{12}^j \hat{x}_{12}^k &= 2\pi i \,  \Bigg[\frac{p^i \delta^{kj} + p^j \delta^{ki} + p^k \delta^{ij}}{p_\perp^2 \, x_{12}}  J_2(p_\perp x_{12}) 
        - \frac{p^i p^j p^k}{p_\perp^3} J_3 ( p_\perp x_{12} ) \Bigg],
    \end{align}
\end{subequations}
where $\phi_{12}$ is the azimuthal angle of $\un x_{12}$ and $\hat x_{12}$ is the corresponding unit vector. The remaining radial integrals are organized using the Fourier-Bessel transform
\begin{align} \label{fb_transform}
    F^{(cd)}_{(ab)}(p_\perp, Q\sqrt{z(1-z)}, s) \equiv \int \displaylimits^\infty_0 dx_{12} \, x_{12}\, \left( p_\perp x_{12} \right)^c \left( Q\sqrt{z(1-z)} x_{12} \right)^d J_a(p_\perp x_{12} ) K_b \left( Q\sqrt{z(1-z)} \, x_{12} \right)\, F(x_{12}^2, s),
\end{align}
where $F$ is any impact-parameter integrated dipole or moment amplitude. Below we suppress the arguments of $F^{(cd)}_{(ab)}$ throughout.

Carrying out the angular integrals in \eq{TT_result_t3}, we obtain the two TT coefficients
\begin{subequations} \label{tt_coeffs}
    \begin{align} \label{tt_coeffs_a}
        C_{\mathrm{TT}} &= -  \frac{\alpha_{EM}^2 \, Z_f^2 \, N_c^2\, p_\perp}{16  \pi^5 \, s} \, N_{(11)}^{(00)}
        \Bigg\{
        (1- 2 z)^2 \, {Q}_{(11)}^{(00)} + 2\Big[z^2 + (1-z)^2 \Big] \,  {G_2}_{(11)}^{(00)}
        \Bigg\},
        \\
         \notag
        C^{\cos\phi_{\Delta p}}_{\mathrm{TT}} &=
       \frac{\alpha_{EM}^2 \, Z_f^2 \, N_c^2\, p_\perp}{16  \pi^5 \, s} \, (1-2z)
         \\ \notag &\times
         \Bigg\{  \frac{(1- 2z)^2}{2} \, N^{(10)}_{(01)} \, {Q}_{(11)}^{(00)}
         + \big[ z^2 + (1-z)^2 \big] \, N^{(10)}_{(01)}\, {G_2}_{(11)}^{(00)}
         + N_{(11)}^{(00)}  \left[ \frac{1}{2} {Q}_{(11)}^{(00)} + {I_{3}}_{(11)}^{(00)} - {I_{3}}^{(10)}_{(01)}\right]
         \\ \notag
         & \hspace{1cm}
         -\Big[z^2 + (1-z)^2 \Big] \, N_{(11)}^{(00)} \Bigg[ {Q}_{(21)}^{(10)} + 2\, {G_2}_{(21)}^{(10)}
         + {I_4}^{(10)}_{(21)}
        +  {I_4}^{(01)}_{(10)}
         - {I_5}_{(11)}^{(00)}
         + {I_5}^{(01)}_{(10)}
         + {I_5}^{(10)}_{(01)}
         \Bigg]
         \\  \label{tt_coeffs_b}
         & \hspace{1cm}
        + \, N_{(11)}^{(00)} \, \Bigg[ 2z(1-z) \, {I_E}_{(11)}^{(00)}
         + \Big[ z^2 + (1-z)^2 \Big] \, {I_E}^{(01)}_{(10)} \Bigg]
         \Bigg\} .
    \end{align}
\end{subequations}
As mentioned above, we see the $I_E$-containing terms are manifestly nonzero, establishing the claim made below \eq{t3_XS5} that the type-3 operator generates a genuinely new contribution to the $\un p \cdot \un \Delta$ harmonic.

Turning next to the LT terms, we again integrate over the angles of $\un x_{12}$ and $\un x_{1'2'}$ to obtain 
\begin{subequations} \label{lt_coeffs}
    \begin{align}
         &C_{\mathrm{LT}}^{\cos\phi_{kp}} = -  \frac{\alpha_{EM}^2  \, Z_f^2 \, N_c^2 \, \sqrt{z(1-z)} \, p_\perp }{ 8 \sqrt{2} \,\pi^5  \, s} \, (1-2z)
        \Bigg\{
        N_{(00)}^{(00)} \Big[ 2 \, {G_2}_{(11)}^{(00)} -  {Q}_{(11)}^{(00)} \Big]
        - N_{(11)}^{(00)} \,  {Q}_{(00)}^{(00)}
        \Bigg\},
        \\
        \notag
        &C_{\mathrm{LT}}^{\cos\phi_{kp} \cos\phi_{\Delta p}} =  -  \frac{\alpha_{EM}^2  \, Z_f^2 \, N_c^2 \, \sqrt{z(1-z)} \, p_\perp }{ 8 \sqrt{2} \,\pi^5  \, s}
         \Bigg\{  N_{(00)}^{(00)} \Big[
         {I_{3}}_{(11)}^{(00)}  - {I_{3}}_{(01)}^{(10)}
         \Big]
         - \frac{1}{2}(1-2z)^2 \Big[N_{(11)}^{(00)} - N_{(01)}^{(10)}  \Big] \, {Q}_{(00)}^{(00)}
         \\ \notag & \hspace{0.5cm}
         + N_{(11)}^{(00)} \, {I_{3}}^{(10)}_{(10)} + \Big[ z^2 + (1-z)^2 \Big] \, N_{(00)}^{(00)} \Big[ {Q}_{(01)}^{(10)} - {Q}_{(11)}^{(00)} \Big]
         - \Big[ z^2 + (1-z)^2 \Big] \, N_{(11)}^{(00)} \,  {Q}_{(10)}^{(10)}
         \\ \notag & \hspace{0.5cm}
         - \frac{1}{2} (1-2z)^2 \, N_{(10)}^{(10)} \, {Q}_{(11)}^{(00)}
         + (1-2z)^2 \, N_{(00)}^{(00)} \Bigg[3 \, {G_2}_{(11)}^{(00)}  - 2 \,{G_2}^{(10)}_{(01)}  + {I_4}^{(10)}_{(21)}  + {I_4}^{(01)}_{(10)}
         - {I_5}_{(11)}^{(00)} + {I_5}^{(10)}_{(01)}  + {I_5}^{(01)}_{(10)}
         \Bigg]
         \\ \notag & \hspace{0.5cm}
         + (1-2z)^2 \, N_{(10)}^{(10)} \,  {G_2}_{(11)}^{(00)}
         \\ \label{lt_coscos}
         & \hspace{0.5cm}
         + \, N_{(00)}^{(00)} \Bigg[  4 z(1-z) \, {I_E}_{(11)}^{(00)}
         + (1-2z)^2 \, {I_E}^{(01)}_{(10)} \Bigg]
         \Bigg\},
         \\
         \notag
        &C_{\mathrm{LT}}^{\sin\phi_{kp} \sin\phi_{\Delta p}} =  - \frac{\alpha_{EM}^2 \, Z_f^2 \, N_c^2 \, \sqrt{z(1-z)} \, p_\perp }{ 8 \sqrt{2} \,\pi^5  \, s}
         \Bigg\{  \Big[ z^2 + (1-z)^2 \Big] \, N_{(00)}^{(00)} \, {Q}_{(11)}^{(00)} - N_{(00)}^{(00)} \, {I_{3}}_{(11)}^{(00)}
        \\ \notag &  \hspace{0.5cm}
        + \frac{1}{2}(1-2z)^2 \, N_{(11)}^{(00)} \, {Q}_{(00)}^{(00)}
         + (1-2z)^2 \, N_{(11)}^{(00)} \Bigg[
        {G_2}^{(10)}_{(10)} + {I_4}^{(11)}_{(11)}
        + {I_4}^{(20)}_{(00)} + {I_5}^{(11)}_{(11)}
        - {I_5}^{(10)}_{(10)} + {I_5}^{(20)}_{(00)}
        \Bigg]
        \\ \notag & \hspace{0.5cm}
        + (1-2z)^2 \, N_{(00)}^{(00)}\Bigg[ {G_2}^{(10)}_{(01)} - 3\, {G_2}_{(11)}^{(00)}
         - 2 \, {I_4}_{(11)}^{(00)} + {I_4}^{(10)}_{(01)} - {I_4}^{(20)}_{(11)} - {I_4}^{(01)}_{(10)}
        + {I_4}^{(11)}_{(00)}
        + {I_5}_{(11)}^{(00)}
        - {I_5}^{(20)}_{(11)} - {I_5}^{(01)}_{(10)} + {I_5}^{(11)}_{(00)}
        \Bigg]
        \\ \notag & \hspace{0.5cm}
        +\, N_{(00)}^{(00)} \Bigg(  4 z(1-z) 
        \Big[ {I_E}_{(11)}^{(00)} - {I_E}^{(10)}_{(21)} \Big]
        + (1-2z)^2 \Big[ {I_E}^{(01)}_{(10)} - {I_E}^{(11)}_{(20)} \Big] \Bigg)
        \\ \label{lt_sinsin}
        & \hspace{0.5cm}
        - (1-2z)^2 \, N_{(11)}^{(00)} \Bigg[ 2 \, {I_E}^{(10)}_{(10)} - {I_E}^{(11)}_{(11)} \Bigg]
       \Bigg\} .
    \end{align}
\end{subequations}
Again we see the $I_E$ terms contribute to the $\cos \phi_{kp} \cos \phi_{\Delta p}$ and $\sin \phi_{kp} \sin \phi_{\Delta p}$ harmonics as expected.

Equations \eqref{tt_coeffs} and \eqref{lt_coeffs}, together with \eq{angles2}, constitute the angular integrated form of the elastic dijet DSA numerator at linear order in $\Delta_\perp$ and are the main phenomenological results of this Section. We emphasize again that, for simplicity, we have assumed that all polarized amplitudes are independent of quark flavor. For phenomenology, one must restore a flavor index in the $\widetilde{Q}, Q, I_3$, and $\widetilde{I}$ amplitudes since they contain quark operators. Furthermore, assuming we are able to experimentally separate the three harmonics $C^{\cos\phi_{\Delta p}}_{\mathrm{TT}}$, $C_{\mathrm{LT}}^{\cos\phi_{kp} \cos\phi_{\Delta p}}$, and $C_{\mathrm{LT}}^{\sin\phi_{kp} \sin\phi_{\Delta p}}$,  which depend on the moment amplitudes in \eq{angles2} by varying the $y$ and azimuthal angles, we still only have 3 observables to constrain 6 moment amplitudes ($I_3, \widetilde{I}_3, \widetilde{I}, I_4, I_5,$ and $I_E$). Fortunately, since the initial condition of $I_E$ is zero at the sub-eikonal order, it appears to be fully constrained via evolution only. Therefore we only have $5$ independent moment amplitudes ($9$ if we account for the fact that $I_3$ and $\widetilde{I}$ depend on the three light quark flavors). Although we will ultimately address this question in the companion feasibility study \cite{JAMCollaborationSmall-xAnalysisGroup:inprep} (including scattering on light nuclei instead of the proton helps), we emphasize here that the multitude of operators which arises at sub-eikonal order and couples to the proton spin poses a challenge for phenomenological extractions of all the independent degrees of freedom. For any potential extraction of OAM distributions, this is especially true given the scarcity of observables coupling to the moment amplitudes (and therefore the OAM distributions). 


\section{Conclusions and outlook}
\label{sec:conclusions}

In this work we revisited the small-$x$ OAM distributions of quarks and gluons in the large-$N_c\&N_f$ limit. On the analytic side, we corrected the quark OAM distribution of \eqs{OAMs} suggested in our earlier work: defining the sub-eikonal operator $\widetilde{L}_{10}$ of \eq{Ltilde}, we showed that $L_{q+\bar{q}}$ is determined by the corresponding moment amplitude $\widetilde{I}$, leading to the compact expressions \eqref{OAMs_corr}. These mirror the small-$x$ helicity PDFs \eqref{pdfs} term by term, so that the OAM distributions are now determined entirely by the moment amplitudes, with the helicity PDFs being given by the impact-parameter integrated amplitudes.

We then derived the large-$N_c\&N_f$ evolution equations for the full set of moment amplitudes $I_3, \widetilde{I}_3, I_4, I_5, \widetilde{I}$, and $I_E$ together with their neighbor counterparts, collected in the matrix form in \eqs{mom_eqn} and \eqref{neigh_mom_eqn}. Because these equations do not close among the moment amplitudes alone, we solved them together with the KPS-CTT-BCL helicity evolution equations 
\cite{Kovchegov:2015pbl, Kovchegov:2016zex, Kovchegov:2017lsr, Kovchegov:2018znm, Chirilli:2021lif, Cougoulic:2022gbk, Borden:2024bxa}.

We solved the coupled system numerically for $N_f = 2, 3, 4, 5, 6$ using a discretization of the recursion relations \eqref{recursions}. From the large-$\eta$ behavior of the amplitudes we extracted the small-$x$ intercepts, finding that all of the moment amplitudes share a common intercept with the helicity dipole amplitudes, as summarized in \eqs{intercepts_res} and Table~\ref{t:intercepts_Nf}. At $N_f = 3$ this yields the OAM distributions asymptotics \eqref{oam_asym}, with an intercept of $3.48\sqrt{\bar{\alpha}_s}$, slightly smaller than the large-$N_c$ value of $3.66\sqrt{\bar{\alpha}_s}$ \cite{Kovchegov:2023yzd} due to the inclusion of the quarks. The intercept values are also in agreement with the large-$N_c\&N_f$ helicity intercepts obtained via the analytic solution in \cite{Borden:2025ehe}.

 We also studied the ratios of the OAM distributions to the helicity PDFs. Fitting our numerical solutions to the asymptotic forms \eqref{m2_ansatz}, we obtained the continuum-limit coefficients in \eqs{m2_res} and Table~\ref{t:ratio_coeffs} for $N_f=2,3,4,5,6$. The leading constants, $L_{q+\bar{q}}/\Delta\Sigma \to -1$ and $L_G/\Delta G \to -2$ at $N_f = 3$ and $Q^2= 10\, \mathrm{GeV}^2$, agree with the predictions $-1$ and $-2$ of \cite{Boussarie:2019icw} and remain numerically consistent with the large-$N_c$ values of \cite{Kovchegov:2023yzd}. Furthermore, we showed that both the quark and gluon hPDF to OAM distribution ratios are independent of $N_f$. We caution, as in \cite{Kovchegov:2023yzd}, that the approach to these asymptotic values is slow and sets in only at extremely small $x$, where saturation effects neglected in our linearized treatment are expected to be important.

 In the future, it would be worthwhile to adapt the method of \cite{Borden:2025ehe} to analytically solve \eqs{mom_eqn} and \eqref{neigh_mom_eqn} in closed form, which would, hopefully, justify the ans\"atze \eqref{m2_ansatz} and sharpen the comparison with the results in \cite{Boussarie:2019icw}; the latter may reveal a discrepancy in the intercepts and hPDF to OAM distribution ratios our present numerical accuracy cannot resolve. 

 Finally, we also revisited the elastic dijet production cross section in longitudinally polarized electron--proton collisions computed in \cite{Kovchegov:2024wjs}, where the $F^{+-}$ operator had been omitted. Treating the minus-momentum dependence of the scattering amplitude carefully, we found that $V^{\mathrm{G}[3]}_{\un x}$ generates a contribution proportional to the derivative of the photon light-cone wave functions with respect to the momentum fraction $z$. These corrections, along with the results from \cite{Kovchegov:2024wjs}, are collected in \eqs{integrated_res_t3}, which supersede Eqs.~(107) of \cite{Kovchegov:2024wjs}. We then carried out the azimuthal integrations over the dipole orientations analytically, arriving at \eqs{tt_coeffs} and \eqref{lt_coeffs} for the five independent harmonics in \eq{angles2}. This angular integrated form is well suited for numerical evaluation, and provides the starting point for the companion study \cite{JAMCollaborationSmall-xAnalysisGroup:inprep}, in which we assess the feasibility of using this observable at the future EIC to extract the moment amplitudes and, through \eqs{OAMs_corr}, the OAM distributions. Combined with the ongoing small-$x$ helicity phenomenology program \cite{Adamiak:2021ppq, Adamiak:2023yhz, JAMCollaborationSmall-xAnalysisGroup:2025tfa}, such a measurement would offer a route to quantify the partonic decomposition of the proton spin at small $x$, and, in particular, to measure, for the first time ever, the orbital contributions that have so far remained unconstrained by the existing experimental data.


\section*{Acknowledgments}

The authors are grateful to Jeremy Borden for his contributions during the early stages of this project, and to Farid Salazar for an informative discussion. We also would like to thank Daniel Adamiak, Wally Melnitchouk, Daniel Pitonyak, and Josh Tawabutr for reading parts of this manuscript and for their suggestions aimed at improving the presentation.

This study was funded in part by the Coordination for the Improvement of Higher Education Personnel – Brazil (CAPES) – Finance Code 001. The work of YK and BM is supported by the U.S. Department of Energy, Office of Science, Office of Nuclear Physics under Award Number DE-SC0004286. BM also gratefully acknowledges support from The Ohio State University via the Presidential Fellowship. BM also acknowledges support from the U.S. National Science Foundation under Grant No.~PHY-2609761.
AT is supported by the U.S. Department of Energy, Office of Science, Office of Nuclear Physics through Contract No. DE-SC0020081. The work of YK, BM, and AT was also supported within the framework of the Saturated Glue (SURGE) Topical Theory Collaboration.

\appendix

\section{The dipole frame}
\label{sec:frame}

The dipole/Trento frame \cite{Bacchetta:2004jz} with $q_\perp =0$ is not a convenient frame to calculate the cross section for the entire process at hand, because the integration over $d^3 k'/E_{k'}$ in this frame results in varying the momentum $k$ of the incoming electron (since $\un k' = \un k$ in that frame). Rather, one can start with \eq{angles1} and argue that due to its left-hand side being Lorentz-invariant, the cross sections multiplied by phase factors and summed over helicities on its right-hand side are also Lorentz-invariant, and, hence, can be calculated in any frame. Further, for the longitudinally polarized proton at hand, the scattering problem possesses azimuthal rotational symmetry, such that nothing in the cross sections depends on $\phi_k$ but only on the differences of azimuthal angles instead. (Note that, up to higher powers of $x$ \cite{Diehl:2005pc}, the longitudinal spin asymmetries are the same when calculated along the electron or virtual photon momentum directions.)  Therefore, the integration measure $d^3 k'/E_{k'}$ can be rewritten in terms of Lorentz invariants $dy \, d Q^2$, while $\phi_k$ can be integrated out while keeping all differences between azimuthal angles fixed. This can be done in any frame that allows one to do so without changing $k$, and is particularly simple in the proton rest frame with $\un k =0$: there, one obtains 
\begin{align}\label{d3kE}
    \frac{d^3 k'}{E_{k'}} = \pi \, dy \, dQ^2 .
\end{align}

Alternatively, consider a more general frame where the  proton momentum is given by \eq{P}, while the incoming ($k$) and outgoing ($k'$) electron momenta are
\begin{subequations}
\begin{align}
    & k^\mu = \left( \frac{k_\perp^2}{2 k^-} , k^- , \un k_\perp \right), \label{k} \\
    & k^{\prime \, \mu} = \left( \frac{k_\perp^{\prime \, 2}}{2 k^{\prime \, -}}, k^{\prime \, -}, \un k' \right).
\end{align} 
\end{subequations}
In such a frame, one can show that
\begin{align}\label{Q^2}
    Q^2 = \frac{\left( \un k' - (1-y) \, \un k \right)^2}{1-y}
\end{align} 
if we neglect the electron's mass. Therefore, we write, using $k^{\prime \, -} = k^- \, (1-y),$
\begin{align}
    \frac{d^3 k'}{E'} = - \frac{d k^{\prime \, -}}{k^{\prime \, -}} \, d^2 k'_\perp = \frac{d y}{1-y} \int dQ^2 \, \delta \left( Q^2 - \frac{\left( \un k' - (1-y) \, \un k \right)^2}{1-y}\right) \, d^2 k' .
\end{align}
Defining a new transverse momentum vector
\begin{align}
    \un k'' \equiv \un k' - (1-y) \, \un k
\end{align}
we can write $d^2 k' = d^2 k'' = \thalf \, d k^{\prime\prime \, 2}_\perp \, d \phi_{k''}$ and integrate out $k^{\prime\prime \, 2}_\perp$ obtaining
\begin{align}
    \frac{d^3 k'}{E'} = \half \, dy \, dQ^2 \,  d \phi_{k''}.
\end{align}
We can use \eq{angles1} to
argue again that the cross sections on the right-hand side of that equation depend only on $Q^2$ and $x$ (out of all the possible ``initial $\gamma^* +p$ state" invariants), and, hence, are independent of $\phi_{k''}$. Therefore, we can integrate out $\phi_{k''}$, obtaining \eq{d3kE} once more.

\providecommand{\href}[2]{#2}\begingroup\raggedright\endgroup

\end{document}